\documentclass[11pt]{article}
\usepackage{epsfig,amsfonts,amssymb}
\usepackage{amsmath,graphics}
\usepackage{slashed}
\usepackage[utf8]{inputenc}
\usepackage{scalerel}
\usepackage{graphicx}
\usepackage{stackengine}
\usepackage{hyperref}
\usepackage{cleveref}
\usepackage{amsmath}
\usepackage{amsmath,mleftright}
\usepackage{xparse}
\usepackage{slashed}
\usepackage{amssymb}
\usepackage{cancel}
\usepackage{multirow}
\usepackage[utf8]{inputenc}
\usepackage{color}                                                      
\usepackage{xcolor}
\usepackage{bbm}
\usepackage{amsmath}
\usepackage{url}
\usepackage{longtable}
\usepackage{array}
\usepackage{bbold}
\numberwithin{equation}{section}

\newcommand{\ads}{\mathrm{AdS_4}}

\newcommand{\abs}[1]{{|#1|}}

\newcommand{\half}{\frac{1}{2}}

\newcommand{\Action}{\mathcal{S}}
\newcommand{\deriD}{\mathcal{D}}
\newcommand{\Tr}{\mathop{\mathrm{tr}}}

\newcommand{\sign}{\mathop{\mathrm{sign}}}
\newcommand{\g}{\dagger}
\newcommand{\NN}{\mathcal{N}}
\newcommand{\WW}{\mathcal{W}}

\newcommand{\ZZ}{\mathcal{Z}}
\newcommand{\VV}{\mathcal{V}}
\newcommand{\Ep}{(\varepsilon_1+i\varepsilon_2)}
\newcommand{\Em}{(\varepsilon_1-i\varepsilon_2)}

\newcommand{\Ev}{\varepsilon}

\newcommand{\comm}[2]{[#1,#2]}
\newcommand{\m}{{{(j)}}}

\newcommand{\n}{{{(j-1)}}}
\newcommand{\p}{{{(j+1)}}}
\newcommand{\1}{{{(1)}}}
\newcommand{\2}{{{(2)}}}
\newcommand{\3}{{{(3)}}}
\newcommand{\5}{{{(5)}}}
\newcommand{\6}{{{(6)}}}
\newcommand{\4}{{(4)}}
\newcommand{\7}{{(7)}}
\newcommand{\nn}{\nonumber}

\newcommand{\A}{\mathcal{A}}
\newcommand{\su}{{SU(2)}}
\newcommand{\ben}{\begin{eqnarray*}}
\newcommand{\en}{\end{eqnarray*}}

\newcommand{\superN}{\mathcal{N}}

\newcommand{\bigbrk}[1]{\bigl(#1\bigr)}
\newcommand{\Bigbrk}[1]{\Bigl(#1\Bigr)}

\newcommand{\Bigsbrk}[1]{\Bigl[#1\Bigr]}

\newcommand{\rep}[1]{{\mathbf{#1}}}
\ifx\href\asklfhas\newcommand{\href}[2]{#2}\fi
\ifx\arxivref\asklfhas\newcommand{\arxivref}[1]{\href{http://arxiv.org/abs/#1}%
{#1}}\fi
\ifx\doiref\asklfhas\newcommand{\doiref}[2]{\href{http://dx.doi.org/#1}{#2}}\fi

\newcommand{\be}{\begin{eqnarray}}
\newcommand{\ee}{\end{eqnarray}}

\usepackage{url}
\usepackage{longtable}

 \usepackage{hyperref}

\def\one{{\hbox{ 1\kern-.8mm l}}}
\def\zero{{\hbox{ 0\kern-1.5mm 0}}}

\begin{document}

\baselineskip 24pt

\begin{center}
{\Large \bf Charges of Monopole Operators in $\widehat{ADE}$ Chern-Simons Quiver Gauge Theories}

\end{center}

\vskip .6cm
\medskip

\vspace*{4.0ex}

\baselineskip=18pt

\centerline{\large \rm   Moumita Patra}

\vspace*{4.0ex}

\centerline{\large \it National Institute of Science Education and Research Bhubaneshwar,}

\centerline{\large \it  P.O. Jatni, Khurda, 752050, Odisha, INDIA}
\vspace*{1.0ex}

\centerline{\large \it Homi Bhabha National Institute, Training School Complex,}

\centerline{\large \it  Anushakti Nagar, Mumbai, India 400085}

\vspace*{4.0ex}
\centerline{E-mail:  mpatra91@niser.ac.in }

\vspace*{5.0ex}

\centerline{\bf Abstract} \bigskip
\thispagestyle{empty}
We compute R-charges of the BPS-monopole operators in   $\mathcal{N} = 3$ $\widehat{ADE}$ Chern-Simons quiver  gauge theories, along the lines of the work of Benna, Klebanov and Klose in \cite{bkk}. These theories have a  weakly coupled UV completion in terms of $\NN=3$ supersymmetric Chern-Simons Yang-Mills theories.  In the UV limit the monopole operators are well approximated by  classical solutions.
We construct classical BPS and anti-BPS monopole solutions to these theories which preserve $\frac{1}{3}$ supersymmetry all along the RG flow. We  compute the $SU(2)_R$ charges   in these  backgrounds and 
 show that the smallest possible value of quantised $SU(2)_R$ charge is zero in each quiver theory.

\vfill \eject

\baselineskip=18pt

\newpage
\setcounter{page}{1}
\renewcommand{\thefootnote}{\arabic{footnote}}
\setcounter{footnote}{0}

\hrule
\tableofcontents
\vspace{8mm}
\hrule
\vspace{4mm}



\section{\label{1}Introduction}

\par
Study of $\ads/\mathrm{CFT}_3$ correspondence\cite{jschwarz}received a lot of interest after the  discovery of a  $\mathcal{N}=6$ superconformal Chern-Simons(CS) matter theory which describes the world volume theory of multiple M2 branes  \cite{abjm}(also see \cite{blg}-\cite{raams}) in the low energy limit. This theory, widely known as ABJM theory, has gauge group $U(N)_k\times U(N)_{-k}$ where $k$ is the 
CS level.  In the matter sector there are  four complex scalar fields  in $(N,\bar{N})$ representation and their  complex conjugate fields in the $(\bar{N},{N})$ representation along with  their  fermionic partners. ABJM theory can be obtained as IR limit of a brane construction \cite{bergman}\cite{kitao} which preserves $\NN=3$ supersymmetry. The field content in the brane construction is similar to ABJM theory but the gauge fields and their superpartners become dynamical in the high energy regime.  One of the important objects in ABJM theory is the monopole operator  which are crucial for the supersymmetry enhancement from $\NN=6$ to $\NN=8$ for special values of the CS levels $k=1,2$. 
\par
\smallskip
 Monopole operator was first studied in the context of   QED and supersymmetric QED in \cite{kapustin}\cite{bor2} where it is defined  as a vortex creating operator with unit vortex charge.  The vortex charge is the  conserved charge of the current $J^\mu=\frac{1}{4\pi}\epsilon^{\mu\nu\rho}F_{\nu\rho}$, which exists in any three dimensional gauge theory and conserved by virtue of Bianchi identity. Monopole operator can also be thought of  as  't Hooft operator which is a topological disorder operator and naturally arises as   dual to an "order" operator  in a topological quantum field theory \cite{kapustin-witten}. Disorder operators can not be expressed as polynomials of basic field variables in the Lagrangian. Therefore they are defined by specifying the singularities of the classical fields in the theory and performing the path integral with a boundary condition that the fields take those specified configurations at the point of singularity. For example  a $U(1)$ Dirac monopole in $\mathbbm{R}^3$ is defined by specifying the classical gauge field configuration,
\be
\vec{A}= \frac{q}{2r\sin\theta}(\pm 1 - \cos\theta)\,\hat{e}_\varphi
\ee
where $\hat{e}_\varphi$  is the unit vector in spherical polar co-ordinate system $(r, \theta, \varphi)$, the upper and lower sign is for northern and southern hemisphere respectively, $q$ is the magnetic charge. Observe that the gauge field has a real singularity at $r=0$. Therefore the statement  that there exists such a monopole operator in a theory  implies that we are inserting a singular gauge field at a point in space-time. Insertion of such a monopole operator  at a point $p$  amounts to  integrating over the  gauge fields which have a singularity at $x = p$ such that the magnetic flux through a $2$-sphere surrounding $x = p$ is $q$. In  $U(N)$ gauge theory a monopole operator is obtained by defining a homomorphism, 
    \be
      \rho:\, U(1)\longmapsto U(N)
     \ee
One such mapping  takes a $U(1)$ element $e^{i\alpha}$ to $\mathrm{diag}( e^{i\,q_1\alpha},  e^{i\,q_2\alpha},...,  e^{i\,q_N\alpha})\in U(N)$. An algebra element gets mapped to
$H=\mathrm{diag}(q_1, q_2,...,  q_N)\in u(N)$. 
 It is shown by Goddard, Nuyts and Olive in \cite{gno}  that when $H$ is written as a linear combination of Cartan generators then $(q_1 \,\, q_2,\, ...,q_N)$ are the weights of the dual of $U(N)$.  The monopole operator will transform in the $U(N)$ representation with highest weight state labelled by $(q_1, q_2,...,q_N)$\cite{kkm}.
\smallskip
\par
 Monopole operators  play crucial role in establishing   various   non-perturbative dualities in three  dimensional quantum field theories \cite{int-mirror}\cite{bash-aharony}  and  are often useful in condensed matter systems \cite{Pufu:2013eda}\cite{Dyer:2015zha}\cite{Dyer:2013fja}. In AdS/CFT studies 
it has been found that monopole operators with zero conformal dimensions are the ones important for matching the spectra with supergravity. Such monopole operators in ABJM theory  which are singlets under the global symmetries were first studied by Benna, Klebanov and Klose(BKK) in \cite{bkk}. Since ABJM theory is strongly coupled for small values of $k$, which is the only coupling in the theory, it is difficult to  compute the conformal dimension of the monopole operators.  To overcome this difficulty BKK introduces a method which goes as follows:

\par   \begin{itemize}
\item A small coupling $g$ is introduced through  Yang-Mills deformation of the action  which provides a weakly coupled UV completion of the theory. The UV completion is a $\mathcal{N}=3$ CS Yang-Mills  theory\footnote{ There are other UV completions that have been studied in the literature  can be  found in \cite{Kim:2009wb},  \cite{Bashkirov:2010kz}, where a smaller amount of supersymmetry is preserved along the RG flow. }.
\item  To retain $\NN=3$ supersymmetry one has to  add dynamical fields in the adjoint representation. Thus   the R-symmetry group of the theory is $SU(2)$. 
\item In the UV limit the monopole operators  are described by  classical BPS(anti-BPS) monopole solutions of the  CS-Yang-Mills theory.
\item Since the final aim is to compute the spectrum of conformal dimension of monopole operator which is valid at the IR fixed point, they compute a quantity in the UV which is related to the conformal dimension   and does not change along the renormalisation group(RG) flow.  This quantity is the quantised $\su_R$ charge of the monopole operator.
 Due to the non-abelian nature of the R-symmetry group the quantised charges will take  discrete values  and  will not be changed under the continuous RG flow in the parameter $g$. 
\item At the conformal fixed point these  charges are related to the   conformal dimensions of the BPS monopole operators by state operator correspondence.
\end{itemize}

\smallbreak
\par Monopole operators in ABJM theory have been extensively studied over many years now. In the light of the important role played by monopole operators in ABJM theory we study monopole operators in a wide class of $\widehat{ADE}$ CS quiver gauge theories. 
  ABJM theory is the low energy world volume description of   M2 branes probing a transverse toric hyperK\"{a}hler manifold which has singularities of the form $\mathbbm{C}^4/\mathbbm{Z}_k$. Therefore it is natural to construct theories which can arise as world volume theories of multiple M2 branes probing other hyperK\"{a}hler singularities. In \cite{jaff} a $\NN=3$ superconformal quiver CS matter theory was constructed which  is a world volume theory of M2 branes placed at singularities in the transverse eight dimensional hyper-K\"{a}hler manifold and are dual by the AdS/CFT conjecture \cite{Maldacena:1997re}\cite{Witten:1998qj}\cite{Gubser:1998bc} to
M-theory on $\ads$ times the seven dimensional tri-Sasakian manifold whose cone is the eight
dimensional hyper-K\"{a}hler manifold. The corresponding quiver of the field theory is a Dynkin diagram of $\widehat{A}$ algebra. It is known that
 one can always construct a three dimensional $\NN=3$ superconformal CS matter theory whose field content can be summarised by  a $\widehat{ADE}$ quiver diagram \cite{gulotta}.

\smallbreak
\par
In this note we find classical monopole solutions and compute   R-charges of the BPS monopole operators in $\NN=3$ CS quiver gauge theories with $\widehat{A}_{n-1}, \widehat{D}_n$ and $\widehat{E}_6$ quiver diagrams along the lines of the work of BKK. 
 The rest of the paper is organised as follows. In section \ref{sec:action} we construct $\NN=3$ action of $\widehat{ADE}$ theories and write down   supersymmetry variation equations. In section \ref{sec:monopole soln} we obtain  classical BPS and anti-BPS monopole solutions in these theories. In section \ref{sec:u1_charge} we compute the quantum corrections to the $U(1)_R$ charges and  in section \ref{sec:su2_charge} 
 computation of $SU(2)_R$ charges is presented.  Appendix \ref{appendix:A} contains all notations and conventions used for calculations. Appendix \ref{appendix:B} contains the superfield expressions and explicit expression of the component action. In appendix \ref{app:susy_variation_check} we present the computation to check supersymmetry in the $\widehat{D}$ case. In appendix \ref{app:d4_monopole_soln} we give a small example of monopole solution in $\widehat{D}_4$.


\bigskip
 
 \section{\label{sec:action}Action of Yang-Mills deformed $\NN=3$ Chern-Simons theory}
 One way to obtain a three dimensional CS matter theory with $\NN=3$ supersymmetry is to first construct a theory with $\NN=4$ supersymmetry without a CS term. A $\NN=4$ theory in three dimension is a dimensional reduction of $\superN=2$ theory in four space-time dimension which has the following supersymmetric multiplets, \\
(i)\,\,\,$\superN=2$ gauge/vector multiplet = ($\superN=1$ gauge $\VV$) $\oplus$ ($\superN=1$ chiral $\Phi$ ) multiplet in adjoint rep. of the gauge group of the theory.\\
(ii)\,\,\,$\superN=2$ hypermultiplet = ($\superN=1$ chiral $\ZZ$ in rep. $R_i$ ) $\oplus$ ($\superN=1$ chiral $\WW$  in rep. $R_i^*$ ). where, $R_i$ can be fundamental or bi-fundamental under the gauge group of the theory.
\smallskip
\par The above multiplet is used to construct  the $\NN=3$ CS matter theory in three dimensions after dimensional reduction.  The component expansions of all the superfields are given in Appendix \ref{appendix:B}. 
Keeping in mind that in three dimension the R-symmetry group of $\NN=3$ theory is $SO(3)$ which is isomorphic to $SU(2)$ at the algebra level, the on-shell component fields in the superfield expansion are  arranged in  R-symmetry representations in BKK as follows:
\smallskip
\par The   non-auxiliary scalars in the vector multiplet are arranged as\footnote{The subscript $\m$ was not present in BKK and will be explained shortly}, 
\be
\phi^a_{b\m} = (\phi_\m)_i (\sigma_i)^{~a}_b=\left( \begin{array}{cc}
 -\sigma_\m & \phi^\g_\m \\
     \phi_\m & \sigma_\m
\end{array} \right) 
\ee
where, the lower/upper index is the row/column index, $i=1,2,3$ is $so(3)$ vector index. $\phi^a_{b\m}$ forms a 3-dimensional representation of $SU(2)$ algebra.
Fermions in the vector multiplet are written in terms of a  $2\times 2$ matrix,
\[\lambda^{ab}_\m=\left( \begin{array}{cc}
 {\chi_{\sigma}}_\m e^{-i\pi/4} & {\chi_{\phi}^\dagger}_\m \, e^{-i\pi/4} \\
    {\chi_{\phi}}_\m  e^{+i\pi/4} & -{\chi_{\sigma}^\dagger}_\m \, e^{+i\pi/4}
\end{array}  \right) \] 

 Therefore  $\lambda^{ab}_\m$ transforms in the reducible representation $\rep{2}\times\rep{2} = \rep{3}+\rep{1}$ of $SU(2)_R$.
The $SU(2)_R$ indices are raised and lowered by the $SU(2)$ metric $\epsilon^{ab}$ with,  $\epsilon^{12}=\epsilon_{21}=+1$ and the following relations hold.
\begin{eqnarray}
(\lambda^{ab}_\m)^* = -\lambda_{ab_\m} = -\epsilon_{ac} \epsilon_{bd}\,\lambda^{cd}_\m\\
(\phi^a_{b\m})^* = {\phi^b_a}_\m = \epsilon_{ac} \epsilon^{bd}\, {\phi^c_d}_\m
\end{eqnarray}

The bifundamental matter fields are written as   $SU(2)_R$ doublets as follows,

\begin{equation}
\label{eqn:X-SU2R} 
X^{a}_\m =\left( \begin{array}{cc}
Z_\m \\[2mm] W^{\dagger }_\m
\end{array}\right),\hspace{1cm}
{X^\dagger_{a}}_\m = \left( \begin{array}{cc} 
Z^\dagger_\m \\[2mm] W_\m
\end{array} \right)
\end{equation}

and
\begin{equation}
\label{eqn:xi-SU2R}
\xi^{a}_\m =\left( \begin{array}{cc}
\omega^\dagger_\m  \, e^{i\pi/4} \\[2mm] \zeta_\m \, e^{-i\pi/4} 
\end{array}\right) ,\hspace{1cm}
{\xi^\dagger_{a}}_\m =\left(\begin{array}{cc}  \omega_\m \, e^{-i\pi/4} \\[2mm] \zeta^\dagger_\m \, e^{i\pi/4} 
\end{array}\right)
\end{equation}
The component action and the supersymmetry variations will be written in terms of the above R-symmetry representations.
\medskip
 
\subsection{$\widehat{A}$-type quiver}
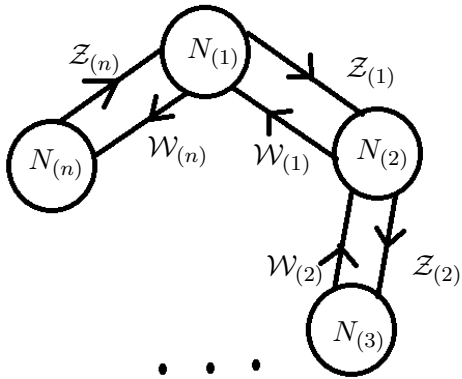
\begin{figure}[h]
\centering
\includegraphics[width=2.5in]{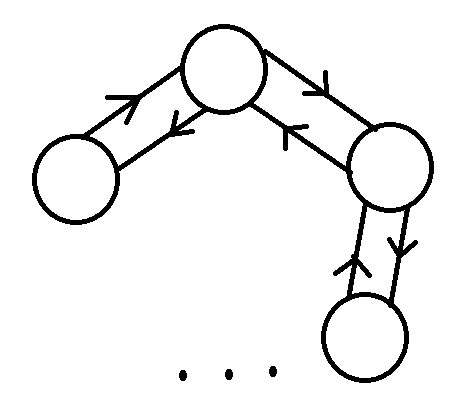}
\put (-100,132) {$N_\1$} \put (-40,125) {$\mathcal{Z}_{(1)}$}\put (-75,92) {$\mathcal{W}_{(1)}$}\put (-36,94) {$N_\2$} \put (-160,90) {$N_{(n)}$} 
\put (-15,50) {$\mathcal{Z}_{(2)}$}\put (-70,50) {$\mathcal{W}_{(2)}$}\put (-45,25) {$N_\3$}
\put (-145,130) {$\mathcal{Z}_{(n)}$}\put (-115,95) {$\mathcal{W}_{(n)}$}
\caption{$\widehat A_{n-1}$ quiver diagram.}
\label{fig:an}
\end{figure}

Before writing down the superspace action for quiver gauge theories let us first set up the notation to express the above field content via a quiver diagram. Figure \ref{fig:an} is a Dynkin diagram of affine A-algebra and can be used to represent the $\NN=3$ field content discussed above. Each of the circles(nodes) are associated with a gauge group factor $U(N_\m)$, a CS level $k_\m$ and contains a gauge multiplet $(\VV_\m, \Phi_\m)$.
The subscript in parentheses $(j)$ is used to label the nodes and the edges in the quiver which runs from $1$ to $n$.   The $(j)$-th edge is the one that
joins the  $(j)$-th node to the  $(j+1)$-th node and we should take  $(n+1)$-th edge to be  $(1)$-st
edge.
The arrows represent the bi-fundamental hypermultiplets $\ZZ_\m$ and $\WW_\m$, in the representation $(\rep{N}_\m,\rep{\overline{N}}_\p)$ and $(\rep{\overline{N}}_\m,\rep{N}_\p)$ respectively.
The gauge group of the theory  is $U(N_\1)\times U(N_\2)\times ...\times U(N_{(n)})$.  The CS levels satisfy $\sum_{\m=1}^n \tilde{n}_\m k_\m=0$\cite{gulotta}, where $\tilde{n}_\m$ is the co-mark of the $\m$-th node which is $1$ for all $\m$ in this case, making $\sum_{\m=1}^n k_\m=0$. 
\smallskip
\par The above described field content can be used to write down a $\NN=3$ superconformal CS matter  theory. Such a theory was constructed by Jafferis and Tomasiello in \cite{jaff}. 

The superspace action of Yang-Mills deformed  CS theory consists of five parts, 
\be
\label{eqn:totalaction}
\Action=\Action_{\mathrm{CS}}+\Action_{\mathrm{YM}}+\Action_{\mathrm{adj}}+\Action_{\mathrm{mat}}+\Action_{\mathrm{pot}}
\ee
 The first three parts involve only vector multiplet
fields and the last two parts involve the hypermultiplet fields together with their minimal couplings
to the vector multiplet fields. 
The CS term,

\be
\label{cs_an}
\Action_{\mathrm{CS}}=-\frac{i}{8\pi}\int d^3x \hspace{.1cm} d^4\theta\int_0^1 ds\sum_{\m=1}^n \Tr\Big[k_\m \VV_\m \bar{D}^\alpha(e^{s\VV_\m} D_\alpha e^{-s\VV_\m)}\Big]
\ee
 The Yang-Mills term,
\be
  \Action_{\mathrm{YM}} = \frac{1}{4g^2} \int d^3x\, d^2\theta  \sum_{\m=1}^n \Tr\Bigsbrk{ \mathcal{U}^{\alpha}_\m \mathcal{U}_{\alpha\m}}  \; 
\ee
  $g$ is a coupling of mass dimension $\frac{1}{2}$ which is responsible for the RG flow. ${\mathcal{U}_{\alpha}}_\m = \frac{1}{4} \bar{D}^2 e^{\VV_\m} D_\alpha e^{-\VV_\m}$ is the super field strength. 

The kinetic terms of the adjoint scalar and fermionic fields arise from,
\be
  \Action_{\mathrm{adj}} = \frac{1}{g^2} \int d^3x\,d^4\theta\: \sum_{\m=1}^n\Tr \Bigsbrk{
      - \bar{\Phi}_\m e^{-\VV_\m} \Phi_\m e^{\VV_\m}
        } 
\ee
After introducing the dimensionful coupling $g$ the theory is not  conformal any more. 
At the IR fixed point $g\to \infty$ which sets  $\Action_{\mathrm{YM}}$ and $\Action_{\mathrm{adj}}$ to zero thus making the gauge fields and the adjoint fields non-dynamical. We can then integrate out the $\Phi_\m$’s and recover $\NN=3$ superconformal theory.

 For the bifundamental matter fields we have  minimally coupled action,
\be
\Action_{\mathrm{mat}}=\int d^3x \hspace{.1cm} d^4\theta \sum_{\m=1}^n\Tr\Big[- \bar{\ZZ}_\m e^{-\VV_\m}\ZZ_\m e^{\VV_\p}-\bar{\WW}_\m e^{-\VV_\p}\WW_\m e^{\VV_\m}\Big]
\ee

The last part of the action is a superpotential term,
\be
\Action_{\mathrm{pot}}=\int d^3x \hspace{.1cm} d^2\theta \sum_{\m=1}^n W_\m-\int d^3x \hspace{.1cm} d^2\bar{\theta} \sum_{\m=1}^n \bar{W}_\m
\ee
where,
\be
W_\m=\Tr (\Phi_\m \ZZ_\m \WW_\m - \Phi_\m \WW_\n\ZZ_\n)+\frac{k _\m}{8\pi}\Tr(\Phi_\m \Phi_\m)\nn\\
\bar{W}_\m=\Tr (\bar{\Phi}_\m  \bar{\WW}_\m \bar{\ZZ}_\m - \bar{\Phi}_\m  \bar{\ZZ}_\n \bar{\WW}_\n)+\frac{k _\m}{8\pi}\Tr(\bar{\Phi}_\m \bar{\Phi}_\m)
\ee
We can write the gauge transformations of the fields under which  the action is invariant as,
\begin{eqnarray}
\label{eqn:gauge transformation}
\Phi_\m\longrightarrow   e^{i\Lambda_\m}\Phi_\m e^{-i\Lambda_\m},\quad
e^{\VV_\m}&\longrightarrow & e^{i\Lambda_\m}e^{\VV_\m} e^{-i\Lambda^\g_\m}\nn\\ 
\ZZ_\m\longrightarrow e^{i\Lambda_\m}\ZZ_\m e^{-i\Lambda_\p},\quad
\WW_\m &\longrightarrow &e^{i\Lambda_\p}\WW_\m e^{-i\Lambda_\m} 
\end{eqnarray}
where, $\Lambda_\m \in U(N_\m)$ is a chiral superfield.
\smallskip
\par   
 We do Grassmann integration and integrate out the auxiliary fields to get the   component  action. Since the R-symmetry group $SU(2)$ is preserved all along the RG flow it is desirable to write the component action as follows where the $SU(2)_R$ symmetry is manifest.

The component action on $\mathbbm{R}^{1,2}$ with signature $(-++)$,  written in terms of the above $SU(2)_R$ multiplets is,
\begin{eqnarray}
\label{min_action_kin}
\mathcal{S}_\mathrm{kin} &=& \int d^3x\sum_{\m=1}^{n} \Tr \Big[- \frac{1}{2g^2} F^{\mu\nu}_\m {F_{\mu\nu}}_\m+ \kappa_\m  \epsilon^{\mu\nu\lambda} \big({ {A_\mu}_\m \partial_\nu {A_\lambda}_\m + \frac{2i}{3} {A_\mu}_\m {A_\nu}_\m {A_\lambda}_\m }\big)\nn\\[1mm] \hspace{17mm}
   &-& \frac{1}{2g^2} \mathcal{D}_\mu {\phi^a_b}_\m \mathcal{D}^\mu {\phi^b_a}_\m
   - \frac{1}{2} \kappa_\m^2 g^2 \, {\phi^a_b}_\m {\phi^b_a}_\m
   - \frac{i}{2g^2} \lambda^{ab}_\m \slashed{\mathcal{D}} \lambda_{ab\m}
   - \frac{i\kappa_\m}{2} \,  \lambda^{ab}_\m \lambda_{ba\m}\nn
   \\[1mm]\hspace{17mm}
      &-& \mathcal{D}_\mu X^\dagger_\m \mathcal{D}^\mu X_\m
   + i \xi^\dagger_\m \slashed{\mathcal{D}} \xi_\m
   \Big]
\end{eqnarray}
\begin{eqnarray}
\label{min_action_int}
\mathcal{S}_{\mathrm{int}} &=& \int d^3x \sum_{\m=1}^{n}\Tr \Big[- \kappa_\m g^2 \, {X^\dagger_a}_\m {\phi^a_b}_\m X^b_\m + \kappa_\m g^2 \, X^a_\n {\phi}^b_{a\m} X^\dagger_{b\n}\nn\\[1mm]\hspace{17mm} &-& i {\xi^\dagger_a}_\m {\phi^a_b}_\m \xi^b_\m + i \xi^a_\n {\phi^b_a}_\m \xi^\g_{b\n}  + \epsilon_{ac} \lambda^{cb}_\m X^a_\m {\xi^\dagger_b}_\m- \epsilon^{ac} {\lambda_{cb}}_\m \xi^b_\m {X^\dagger_a}_\m\nn \\ [1mm]\hspace{17mm}& -&\epsilon_{ac} \lambda^{cb}_\m {\xi^\dagger_b}_\n X^a_\n +\epsilon^{ac} {\lambda_{cb}}_\m {X^\g_a}_\n \xi^b_\n  - \frac{\kappa_\m}{6} {\phi^a_b}_\m \comm{{\phi^b_c}_\m}{{\phi^c_a}_\m}\nn\\[1mm]\hspace{17mm} &-& \frac{i}{2g^2}  {\lambda_{ab}}_\m \comm{{\phi^b_c}_\m}{\lambda^{ac}_\m}- \frac{g^2}{4} (X_\m \sigma_i X^\g_\m) (X_\m \sigma_i X^\g_\m)- \frac{g^2}{4} (X^\dagger_\m \sigma_i X_\m) (X^\dagger_\m \sigma_i X_\m)\nn \\ &+& \frac{g^2}{2} (X_\m \sigma_i X^\dagger_\m) (X^\dagger_\n \sigma_i X_\n)-\frac{1}{2} (X_\m X^\dagger_\m) {\phi^a_b}_\m {\phi^b_a}_\m+ {X^\dagger_a}_\m {\phi^b_c}_\m X^{a}_\m {\phi^c_b}_\p\nn\\[1mm]\hspace{17mm} &-&  \frac{1}{2} (X^\dagger_\n X_\n) {\phi^a_b}_\m {\phi^b_a}_\m+ \frac{1}{8g^2} \comm{{\phi^a_b}_\m}{{\phi^c_d}_\m} \comm{{\phi^b_a}_\m}{{\phi^d_c}_\m}
\end{eqnarray}

where $\kappa_{{(j)}} \equiv \frac{k_{{(j)}}}{4\pi}$,  $X_{(j)} \sigma_i X^\dagger_{(j)} \equiv X^{a}_{(j)} (\sigma_i)_a{}^b {X^\dagger_{b}}_{(j)}$ and $X^\dagger_{(j)} \sigma_i X_{(j)} \equiv {X^\dagger_{a}}_{(j)} (\sigma_i)^a{}_b X^{b}_{(j)}$. The $(\sigma_i)_a{}^b = \sigma_i$ are the usual Pauli matrices and the $(\sigma_i)^a{}_b = \sigma_i^{\scriptscriptstyle\mathrm{T}}$ are the transpose of the Pauli matrices. The various gauge covariant derivatives above are,
\begin{align}
{F_{\mu\nu}}_{(j)}&=\partial_\mu {A_\nu}_{(j)}-\partial_\nu {A_\mu}_{(j)}+i[{A_\mu}_{(j)},{A_\nu}_{(j)}] 
,\qquad
\mathcal{D}_\mu {\phi^{a}_b}_{{(j)}} = \partial_\mu{\phi^{a}_b}_{{(j)}} + i [{A_\mu}_{(j)},{\phi^{a}_b}_{{(j)}}]&\; \nn\\
\mathcal{D}_\mu\lambda^{ab}_\m &= \partial_\mu{\lambda^{ab}}_{{(j)}}+i[{A_\mu}_{(j)}, \lambda^{ab}_\m]&\;\nn\\
{\mathcal{D}_\mu} X^{b}_\m &=\partial_\mu X^{b}_\m +i  A_{\mu\m} X^{b}_\m-i X^{b}_\m  A_{\mu\p},\qquad
{\mathcal{D}_\mu} {X^\g_b}_\m =\partial_\mu {X^\g_b}_\m +i A_{\mu\p} {X^\g_b}_\m-i {X^\g_b}_\m A_{\mu\m}&\;\nn\\
\mathcal{D}_\mu \xi^{b}_\m &= \partial_\mu \xi^{b}_\m +i  A_{\mu\m} \xi^{b}_\m-i \xi^{b}_\m A_{\mu\p}\qquad
\qquad\mathcal{D}_\mu {\xi^\g_b}_\m =\partial_\mu {\xi^\g_b}_\m +i A_{\mu\p} {\xi^\g_b}_\m-i {\xi^\g_b}_\m A_{\mu\m}&\;\nn
\end{align}
 It can be  checked that when the above action is unpacked following the notation introduced in \ref{sec:action}  reproduces the expressions given in appendix  \ref{appendix:an_action}. The supersymmetry transformation parameter $\Ev$ in  a $\mathcal{N}=3$ theory is in the  $\mathbf{3}$ of $SU(2)_R$: $ \varepsilon_{ab} = \varepsilon_i\,(\sigma_i)_{ab}$. The supersymmetry transformations of the non-auxiliary component fields that leave the action \eqref{min_action_kin}+\eqref{min_action_int} invariant  are, 
\begin{align}
  \delta {A_\mu}_\m&= -\frac{i}{2} \Ev_{ab} \gamma_\mu\, \lambda^{ab}_\m &\nonumber \\
  \delta \lambda^{ab}_\m &= \half \epsilon^{\mu\nu\lambda} {F_{\mu\nu}}_\m \gamma_\lambda \Ev^{ab}- i \slashed{\deriD} {\phi^b_c}_\m \Ev^{ac}+ \frac{i}{2} [{\phi^b_c}_\m,{\phi^c_d}_\m] \Ev^{ad}+i \kappa_\m g^2 \,  {\phi^b_c}_\m \Ev^{ac}      \nonumber \\&+ ig^2 \,  \big(X^a_\m {X^\g_c}_\m \Ev^{cb} - \Ev^{bc}{X^\dagger_c}_\n X^a_\n \big) - \frac{i g^2}{2} (X_{(j)}\,X^\dagger_{(j)} - X^\dagger_{(j-1)}X_{(j-1)} ) \varepsilon^{ab} &  \nonumber \\
  \delta \phi^a_{b\m}     &= - \Ev_{cb} \lambda^{ca}_\m
+ \half \delta^a_b \Ev_{cd} \lambda^{cd}_\m&  
\end{align}
\begin{align}
  \delta X^{a}_\m & = - i \Ev^a_b\, \xi^{b}_\m\nonumber \;  &
  \delta \xi^{a}_\m & = \slashed{\mathcal{D}} X^{b}_\m \Ev^a_b + {\phi^a_b}_\m \Ev^b_c\, X^{c}_\m - X^{c}_\m \Ev^b_c\, {{\phi}^a_b}_\p \;  \\
  \delta X^\dagger_{{a}_\m}   & = - i \xi^\dagger_{{b}_\m} \Ev^b_a \;  &
  \delta {\xi^\dagger_{a}}_\m & = \slashed{\mathcal{D}} X^\dagger_{{b}_\m} \Ev^b_a - {{\phi}^b_a}_\p \Ev^c_b\, X^\dagger_{{c}_\m}+ {X^\dagger_c}_\m \Ev^c_b\, {\phi^b_a}_\m \;
\end{align}
\medskip
\subsection{$\widehat{D}$-type quiver}

\begin{figure}[h]
\centering
\includegraphics[width=5in]{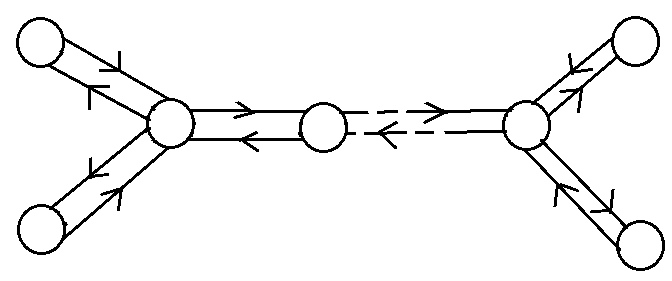}
\put (-27,15) {$N_\3$} \put (-29,125) {$N_\4$}\put (-349,125) {$N_\1$} \put (-315,127) {$\mathcal{Z}_\1$}\put (-315,85) {$\mathcal{W}_\1$} 
\put (-305,28) {$\mathcal{Z}_\2$}\put (-348,50) {$\mathcal{W}_\2$}
\put (-221,96) {$\mathcal{Z}_\5$}\put (-279,82) {$N_\5$}
\put (-60,127) {$\mathcal{Z}_\4$}\put (-35,100) {$\mathcal{W}_\4$}
\put (-60,20) {$\mathcal{Z}_\3$}\put (-35,50) {$\mathcal{W}_\3$}
\put (-349,25) {$N_\2$} \put (-221,65) {$\WW_\5$} \put (-198,80) {$N_\6$} \put (-160,60) {$\cdots$} \put (-87,83) {${\scaleto{N}{6pt}}_{\scaleto{(n+1)}{5pt}}$ } \put (-87,83) {${\scaleto{N}{6pt}}_{\scaleto{(n+1)}{5pt}}$ }
\caption{$\widehat D_n$ quiver diagram.}
\label{fig:Dn}
\end{figure}
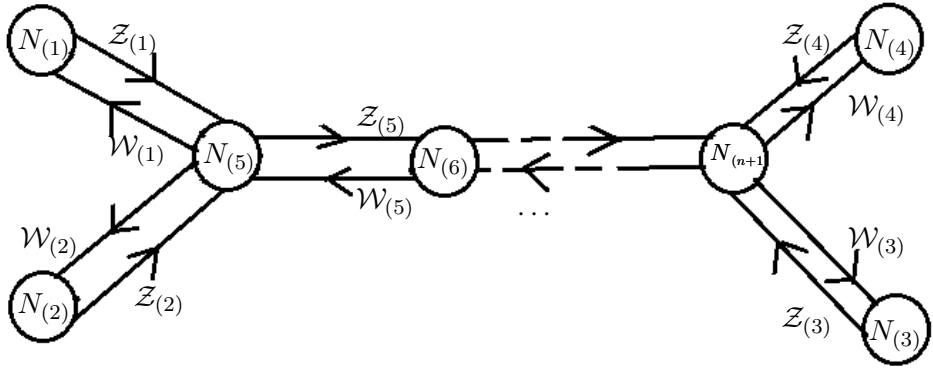

Figure \ref{fig:Dn} is a Dynkin diagram of $\widehat{D}_n$ algebra.
The  main differences in $\widehat{D}$ quiver diagram  from the $\widehat{A}$-type quiver are the external nodes $\m=1,2,3,4$ which have only one edge attached to them and the nodes labelled by $\5$ and $(n+1)$ which have three edges attached to them. The hypermultiplet corresponding to an external  edge say $\1$  is  $(\ZZ_ \1 , \WW _\1)$, which   are  in the gauge representations $(\rep{N_\1},\rep{\overline{N}}_\5)$ and $(\rep{\overline{N}}_\1,\rep{N}_\5)$ respectively. The representations of the hypermultiplets for   other external edges can be written similarly. Rest of the quiver, i.e from $\m=6,..., n$ is similar to the previous case.  The CS levels satisfy $\sum_\m \tilde{n}_\m k_\m=0$. $\tilde{n}_\m=1$ for $\m=1,2,3,4$ and $\tilde{n}_\m=2$ for $\m=5,6,..., n+1$ which implies,
\be 
k_\1+k_\2+k_\3+k_\4+2\big(k_\5+...+ k_{(n+1)}\big)=0
\ee 
 Such quiver gauge theory is dual to M-theory on $AdS_4\times Y$, where, $Y$ is the base of the hyperK\"{a}hler cone $\mathbbm{H}^{2n-8}///U(1)^n\times SU(2)^{n-3}$ \cite{non-toric}.

\par The superspace action for $\mathcal{S}_{\mathrm{CS}}, \mathcal{S}_{\mathrm{YM}}$ and $\mathcal{S}_{\mathrm{adj}}$ remains same except now the limit of summation  runs form $\m=1$ to $n+1$. The rest of the action is as follows:

The minimally coupled action in the  matter sector,
\be
\Action_{\mathrm{mat}}=\int d^3x \hspace{.1cm} d^4\theta\,\Tr\Big[ \sum_{\m=1}^2\Big(- \bar{\ZZ}_{\m} e^{-\VV_{\m}}\ZZ_{\m} e^{\VV_{(5)}}-\bar{\WW}_{\m} e^{-\VV_{(5)}}\WW_{\m} e^{\VV_{\m}}\Big) \nn\\
+  \sum_{\m=3}^4\Big(- \bar{\ZZ}_{\m} e^{-\VV_{\m}}\ZZ_{\m} e^{\VV_{(n+1)}}-\bar{\WW}_{\m} e^{-\VV_{(n+1)}}\WW_{\m} e^{\VV_{\m}}\Big)\nn\\
+  \sum_{\m=5}^n\Big(- \bar{\ZZ}_{\m} e^{-\VV_{\m}}\ZZ_{\m} e^{\VV_{\p}}-\bar{\WW}_{\m} e^{-\VV_{\p}}\WW_{\m} e^{\VV_{\m}}\Big)\Big]
\ee
and
\begin{eqnarray}
\mathcal{S}_\mathrm{pot}=\int d^3x\hspace{.1cm} d^2\theta\hspace{.1cm} W_\m-\int d^3x\hspace{.1cm} d^2\bar{\theta}\hspace{.1cm} \bar{W}_\m\nonumber
\end{eqnarray}
with,
\begin{eqnarray}
W_\m&=& \Tr\Big[\sum_{\m=1}^n\Phi_\m \mathcal{Z}_\m\mathcal{W}_\m +\sum_{\m=1}^{n+1}\frac{k_\m}{8\pi}[\Phi_\m\Phi_\m]\nn\\&-&\sum_{\m=1}^2 {\Phi}_{\5} \mathcal{W}_\m\mathcal{Z}_\m  - \sum_{\m=3}^4 {\Phi}_{(n+1)} \mathcal{W}_\m\mathcal{Z}_\m -\sum_{\m=5}^n {\Phi}_{\p} \mathcal{W}_\m\mathcal{Z}_\m\Big]\nn\\[2mm]
\bar{W}_\m &=&\Tr\Big[\sum_{\m=1}^n\bar{\Phi}_\m \bar{\mathcal{W}}_\m \bar{\mathcal{Z}}_\m +\sum_{\m=1}^{n+1}\frac{k_\m}{8\pi}[\bar{\Phi}_\m\bar{\Phi}_\m]\nn\\&-&\sum_{\m=1}^2 {\bar{\Phi}}_{\5}\bar{\mathcal{Z}}_\m \bar{\mathcal{W}}_\m  - \sum_{\m=3}^4 \bar{{\Phi}}_{(n+1)} \bar{\mathcal{Z}}_\m\bar{\mathcal{W}}_\m -\sum_{\m=5}^n \bar{{\Phi}}_{\p} \bar{\mathcal{Z}}_\m\bar{\mathcal{W}}_\m\Big]
\ee

which gives the following component action after doing the Grassmann integral and eliminating the auxiliary fields.
\par The interaction part of the action,
\begin{eqnarray}
\label{eqn:min_action_int_dn}
\mathcal{S_{\mathrm{int}}}&=& \int d^3x\,\, \Tr\Bigg( 
  \sum_{\m=1}^{n}\Big[
 -i {\xi^\g_a}_\m {\phi^a_b}_\m {\xi^b}_\m+\epsilon_{ac}\lambda^{cb}_\m X^a_\m{\xi_b^\g}_\m -\epsilon^{ac}{\lambda_{cb}}_\m\xi^b_\m{X^\g_a}_\m\nn\\
&-&g^2\kappa_\m X^\g_{a\m} {\phi^a_b}_\m X^b_\m
-\frac{1}{2} X_\m X^\g_\m{\phi^a_b}_\m {\phi^b_a}_\m
 -\frac{g^2}{4} (X_\m \sigma_i X^\g_\m) (X_\m \sigma_i X_\m^\g)\nn\\[1mm]\hspace{17mm}
 &-& \frac{g^2}{4} (X^\g_\m \sigma_i X_\m)(X^\g_\m \sigma_i X_\m)    \Big]
+ \sum_{\m=1}^{2}\Big[- \epsilon_{ac}\lambda^{cb}_\5\,{\xi^\g_b}_\m X^a_\m + \epsilon^{ac}\lambda_{cb\5}{X^\g_a}_\m\xi^b_\m + i \xi^a_\m\, {\phi^b_a}_\5\, \xi^\g_{b\m}\nn\\
&+& 
-\frac{1}{2}(X^\g_\m X_\m){\phi^a_b}_\5 {\phi^b_a}_\5+ {X^\g_\m}_a{\phi^b_c}_\m X^a_\m {\phi^c_b}_\5 + \kappa_{\5} g^2 X^a_\m {\phi^b_a}_\5 {X^\g_b}_\m\nn\\
&+& \frac{g^2}{2}(X_\5 \sigma_i X^\g_\5)(X^\g_{\m}\sigma_i X_{\m})\Big]
+\sum_{\m=3}^{4}\Big[- \epsilon_{ac}\lambda^{cb}_{(n+1)}{\xi^\g_b}_\m X^a_\m + \epsilon^{ac}{\lambda_{cb}}_{(n+1)}{X^\g_a}_\m\xi^b_\m\nn\\
&+& i {\xi^a}_\m {\phi^b_a}_{(n+1)} {\xi^\g_b}_\m
-\frac{1}{2}(X^\g_\m X_\m){\phi^a_b}_{(n+1)} {\phi^b_a}_{(n+1)}+ {X^\g_\m}_a {\phi^b_c}_\m X^a_\m{\phi^c_b}_{(n+1)}\nn\\
&+& \kappa_{(n+1)} g^2 X^a_\m\, {\phi^b_a}_{(n+1)} {X^\g_b}_\m - \frac{g^2}{2}(X^\g_\m \sigma_i X_\m)(X^\g_{(n)} \sigma_i X_{(n)}) \Big]+\sum_{\m=5}^{n}\Big[- \epsilon_{ac}\lambda^{cb}_{\p}{\xi^\g_b}_\m X^a_\m \nn\\
&+&  \epsilon^{ac}{\lambda_{cb}}_{\p}{X^\g_a}_\m\xi^b_\m +i \xi^a_\m {\phi^b_a}_{\p} {\xi^\g_b}_\m +g^2\kappa_{\p} {X^a_\m} {\phi^b_a}_{\p} {X^\g_b}_\m-\frac{1}{2}(X^\g_\m X_\m){\phi^a_b}_{\p}{\phi^b_a}_{\p}\nn\\
&+& {X^\g_\m}_a {\phi^b_c}_\m X^a_\m{\phi^c_b}_{\p}   \Big]
+ \sum_{\m=1}^{n+1}\Big[\frac{1}{8g^2}[{\phi^a_b}_\m,{\phi^c_d}_\m][{\phi^b_a}_\m,{\phi^d_c}_\m]-\frac{\kappa_\m}{6}{\phi^a_b}_\m[{\phi^b_c}_\m{\phi^c_a}_\m]\nn\\&-&\frac{i}{2g^2}{\lambda_{ab}}_\m[{\phi^b_c}_\m,{\lambda^{ac}}_\m]\Big]+ \frac{g^2}{2}\sum_{\m=6}^n (X_\m \sigma_i X^\g_\m)(X^\g_{\n}\sigma_i X_{\n})\nn\\[1mm]\hspace{17mm}
& -& \frac{g^2}{2}(X^\g_\1 \sigma_i X_\1)(X^\g_\2 \sigma_i X_\2)-\frac{g^2}{2}(X^\g_\3 \sigma_i X_\3)(X^\g_\4 \sigma_i X_\4)\Bigg)
\end{eqnarray}

The kinetic part of the action,
\begin{eqnarray}
\label{eqn:min_action_kin_dn}
\mathcal{S}_{\mathrm{kin}}&=& \int d^3x\,\, \Tr \Bigg[ \sum_{\m=1}^{n+1}\Big(- \frac{1}{2g^2} F^{\mu\nu}_\m {F_{\mu\nu}}_\m + \kappa_\m  \epsilon^{\mu\nu\lambda} \big({ {A_\mu}_\m \partial_\nu {A_\lambda}_\m + \frac{2i}{3} {A_\mu}_\m {A_\nu}_\m {A_\lambda}_\m }\big)\nn\\ [1mm]\hspace{17mm}
  &-& \frac{1}{2g^2} \mathcal{D}_\mu {\phi^a_b}_\m \mathcal{D}^\mu {\phi^b_a}_\m
     - \frac{i}{2g^2} \lambda^{ab}_\m\, \slashed{\mathcal{D}} {\lambda_{ab}}_\m
   - \frac{i\kappa_\m}{2} \,  \lambda^{ab}_\m {\lambda_{ba}}_{\m}- \frac{1}{2}\kappa_\m^2 g^2{\phi^a_b}_\m {\phi^b_a}_\m \Big)\nn\\[1mm]\hspace{17mm}
   &+& \Tr \sum_{\m=1}^{n}\Big(- \mathcal{D}_\mu X^\g_\m \mathcal{D}^\mu X_\m
   + i \xi^\g_\m \slashed{\mathcal{D}} \xi_\m \Big)\Bigg]
\end{eqnarray}
where the covariant derivatives of the bi-fundamental fields are,\\
 for $\m=1,2$:
$\mathcal{D}_\mu X^{b}_\m=\partial_\mu X^{b}_\m +i A_{\mu\m} X^{b}_\m-i X^{b}_\m A_{\mu\5},\quad
{\mathcal{D}_\mu} \xi^{b}_\m=\partial_\mu \xi^{b}_\m +i A_{\mu\m} \xi^{b}_\m-i \xi^{b}_\m A_{\mu\5}$ and  for $\m=3,4$:
$
\mathcal{D}_\mu X^{b}_\m=\partial_\mu X^{b}_\m +i A_{\mu\m} X^{b}_\m-i X^{b}_\m A_{\mu(n+1)},\quad
\mathcal{D}_\mu \xi^{b}_\m=\partial_\mu \xi^{b}_\m +i A_{\mu\m} \xi^{b}_\m-i \xi^{b}_\m  A_{\mu(n+1)}.
$ Rest of the covariant derivatives are same as $\widehat{A}$ case.

The supersymmetry transformations of the vector multiplet fields are,
\begin{align}
\label{eqn:min_var_vec_dn}
  \delta {A_\mu}_\m&= -\frac{i}{2} \Ev_{ab} \gamma_\mu \lambda^{ab}_\m,\qquad \delta \phi^a_{b\m}     = - \Ev_{cb} \lambda^{ca}_\m
+ \half\, \delta^a_b\, \Ev_{cd}\, \lambda^{cd}_\m,\qquad \m=1,..., n+1 &\nonumber \\[4mm]
  \delta \lambda^{ab}_\m &= \half \epsilon^{\mu\nu\lambda} {F_{\mu\nu}}_\m \gamma_\lambda \Ev^{ab}- i \slashed{\deriD} \phi^b_{c\m} \Ev^{ac}+ \frac{i}{2} [\phi^b_{c\m},\phi^c_{d\m}] \Ev^{ad}+ i\kappa_\m g^2 \,  \phi^b_{c\m} \Ev^{ac}      \nonumber \\[1mm]&+i g^2 \,  \big(X^a_\m {X^\g_c}_\m \Ev^{cb} - \Ev^{bc}{X^\g_c}_{\n} X^a_{\n} \big) + \frac{ig^2}{2} \big((X^\g X)_{\n} -(X X^\g)_\m\big) \Ev^{ab},\quad \m=6,..., n &  \nonumber \\[4mm]
    \delta \lambda^{ab}_\5 &= \half \epsilon^{\mu\nu\lambda} F_{\mu\nu_\5} \gamma_\lambda \Ev^{ab}- i \slashed{\deriD} \phi^b_{c\5} \Ev^{ac}+ \frac{i}{2} [\phi^b_{c\5},\phi^c_{d\5}] \Ev^{ad}+ i\kappa_\5 g^2 \,  \phi^b_{c\5} \Ev^{ac}      \nonumber \\[1mm]&+i g^2 \,  \Big(X^a_\5 X^\g_{c\5} \Ev^{cb} - \Ev^{bc}\sum_{\m=1}^2{X^\g_c}_\m X^a_\m \Big) + \frac{ig^2}{2} \Big(\sum_{\m=1}^2(X^\g X)_\m -(X X^\g)_\5\Big) \Ev^{ab} &  \nonumber \\[4mm]
  \delta \lambda^{ab}_{(n+1)} &= \half \epsilon^{\mu\nu\lambda} {F_{\mu\nu}}_{(n+1)} \gamma_\lambda \Ev^{ab}- i \slashed{\deriD} \phi^b_{c (n+1)} \Ev^{ac}+ \frac{i}{2} [\phi^b_{c(n+1)},\phi^c_{d(n+1)}] \Ev^{ad}+ i\kappa_{(n+1)} g^2 \,  \phi^b_{c (n+1)} \Ev^{ac}      \nonumber \\[1mm]&- ig^2 \,    \Ev^{bc}\sum_{\m=3,4,n}{X^\g_c}_\m X^a_\m + \frac{ig^2}{2} \sum_{\m=3,4,n}(X^\g X)_\m  \Ev^{ab} &  \nonumber\\[4mm]
  \delta \lambda^{ab}_\m &= \half \epsilon^{\mu\nu\lambda} {F_{\mu\nu}}_\m \gamma_\lambda \Ev^{ab}- i \slashed{\deriD} \phi^b_{c\m} \Ev^{ac}+ \frac{i}{2} [\phi^b_{c\m},\phi^c_{d\m}] \Ev^{ad}+i \kappa_\m g^2 \,  \phi^b_{c\m} \Ev^{ac}      \nonumber \\[1mm]&+i g^2 \, X^a_\m {X^\g_c}_\m \Ev^{cb}  - \frac{ig^2}{2} (X X^\g)_\m \Ev^{ab},\quad  \m=1,2,3,4 &   
\end{align}

Note: The reason that the transformations of $\lambda^{ab}_{(n+1)}$ and $\lambda^{ab}_\5$ are different in spite of their symmetry in the quiver diagram is because of our convention of labelling the $X_\m$'s. There is no arrow coming out of the  $n+1$-th node like $X_\5$ was coming out of the 5-th node. Therefore the terms   $ g^2 \, i X^a_\5 X^\g_{c\5} \Ev^{cb}$ and $\frac{ig^2}{2} (\sum_{\m=1}^2 -(X X^\g)_\5) \Ev^{ab} $
are not present in $\delta\lambda^{ab}_{(n+1)}$. One other way to check this is, the term  $ g^2 \, i   \sum_{\m=3,4,n} X^a_\m X^\g_{c\m} \Ev^{cb}$ is  a matrix of size $3\times 3$ where the L.H.S $\delta\lambda^{ab}_{(n+1)}$,  is an $(n+1)\times (n+1)$ matrix.

The supersymmetry transformations of the hypermultiplet fields are,
\begin{align}
\label{eqn:min_var_hyper_dn}
 \m=1,..., n,\qquad  \delta X^{a}_\m  &= - i \Ev^a_b\, \xi^b_\m, \quad
  \delta X^\g_{{a}_\m}    = - i \xi^\g_{b\m}\, \Ev^b_a&\nn\\[2mm]
\m=1,2\qquad\delta \xi^{a}_\m  &= \slashed{\mathcal{D}} X^{b}_\m \Ev^a_b + \phi^a_{b\m} \Ev^b_c X^c_\m - X^c_\m \Ev^b_c \phi^a_{b\5}& \;\;\;\nn \\[2mm]
\delta \xi^\g_{a\m} &= \slashed{\mathcal{D}} X^\g_{b\m} \Ev^b_a - \phi^b_{a\5} \Ev^c_b X^\g_{c\m}+ X^\g_{c\m} \Ev^c_b \phi^b_{a\m}& \;\; \nn\\[4mm]
\m=3,4 \qquad\delta \xi^{a}_\m  &= \slashed{\mathcal{D}} X^{b}_\m \Ev^a_b + \phi^a_{b\m} \Ev^b_c X^c_\m - X^c_\m \Ev^b_c \phi^a_{b(n+1)}& \;\nn \\[2mm]
\delta \xi^\g_{a\m}  &= \slashed{\mathcal{D}} X^\g_{b\m} \Ev^b_a - \phi^b_{a{(n+1)}} \Ev^c_b X^\g_{c\m}+ X^\g_{c\m} \Ev^c_b \phi^b_{a\m}& \; \quad\nn\\[4mm]
\m=5, 6,..., n\quad\delta \xi^a_\m  &= \slashed{\mathcal{D}} X^b_\m \Ev^a_b + \phi^a_{b\m} \Ev^b_c X^c_\m - X^c_\m \Ev^b_c \phi^a_{b\p}& \;\nn \\[2mm]
\delta \xi^\g_{a\m} & = \slashed{\mathcal{D}} X^\g_{b\m} \Ev^b_a - \phi^b_{a\p} \Ev^c_b X^\g_{c\m}+ X^\g_{c\m} \Ev^c_b \phi^b_{a\m}& \; 
\end{align}
\subsection{$\widehat{E}_6$ quiver}
\begin{figure}[h]
\centering
\includegraphics[width=5in]{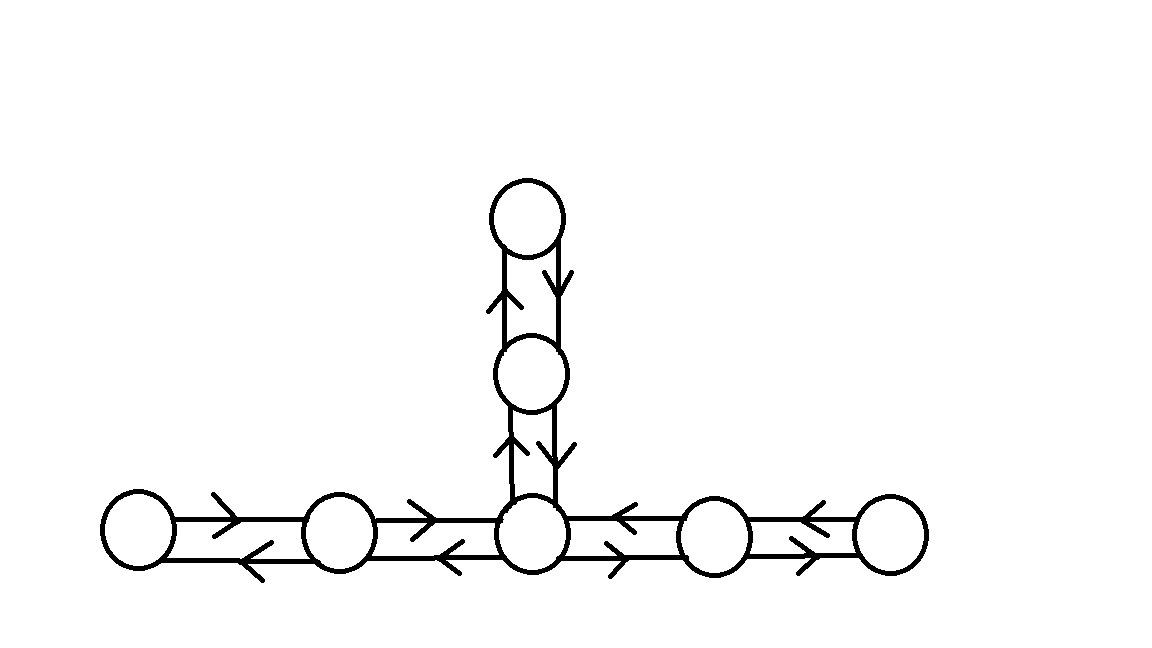}
\put (-327,34) {${\scaleto{N_\1}{10pt}}$}\put (-263,33) {${\scaleto{N_\2}{10pt}}$}  \put (-202,33) {${\scaleto{N_\7}{10pt}}$}\put (-203,132) {$ {\scaleto{N_\5}{10pt}}$} \put (-180,110) {$\mathcal{Z}_{(5)}$}\put (-228,112) {$\mathcal{W}_{(5)}$}\put (-203,82) {${\scaleto{N_\6}{10pt}}$} 
\put (-300,50) {$\mathcal{Z}_\1$} \put (-300,15) {$\mathcal{W}_\1$}
\put (-240,50) {$\mathcal{Z}_\2$} \put (-240,15) {$\mathcal{W}_\2$}
\put (-110,50) {$\mathcal{Z}_\3$} \put (-110,15) {$\mathcal{W}_\3$}
\put (-170,48) {$\mathcal{Z}_\4$} \put (-170,15) {$\mathcal{W}_\4$}
\put (-180,65) {$\mathcal{Z}_{(6)}$}\put (-225,65) {$\mathcal{W}_{(6)}$}
\put (-144,33) {${\scaleto{N_\4}{10pt}}$}  \put (-88,33) {${\scaleto{N_\3}{10pt}}$}
\caption{$\widehat E_6$ quiver diagram.}
\label{e6}
\end{figure}
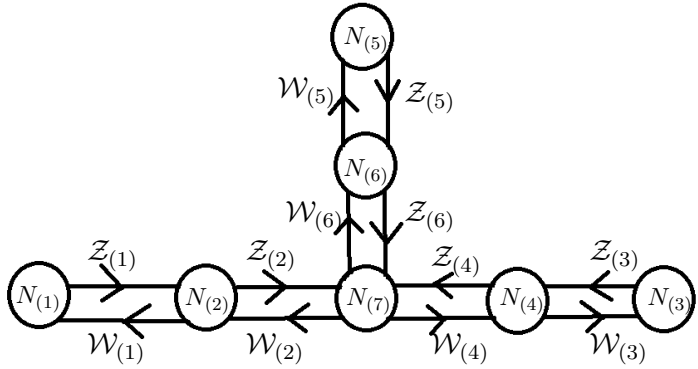

The $\widehat{E}_6$ quiver is similar to $\widehat{D}_n$ quiver diagram after adding some extra nodes and mapping $7$-th node of the former to $n+1$-th node of the latter.  Therefore all the notions can be generalised easily from the $\widehat{D}_n$ case.  The comarks in this case are, $\tilde{n}_\m=1\quad for\,\, \m=1,3,5$,  $\tilde{n}_\m=2\quad for\,\, \m=2,4,6$ and $\tilde{n}_\m=3\quad for\,\, \m=7$ which implies that the CS levels satisfy,
\be 
\kappa_\1+\kappa_\3+\kappa_\5 +2\big(\kappa_\2+\kappa_\4+ \kappa_\6\big)+ 3\kappa_\7=0
\ee 

The first three parts  of the action \eqref{eqn:totalaction} remains same as  except the summation limit changes for $1$ to $7$.
We write down the last two parts of the action,
\be
\Action_{\mathrm{mat}}=\int d^3x \hspace{.1cm} d^4\theta\Tr\Big[ \sum_{\m=2,4,6}\Big(- \bar{\ZZ}_{\m} e^{-\VV_{\m}}\ZZ_{\m} e^{\VV_{(7)}}-\bar{\WW}_{\m} e^{-\VV_{(7)}}\WW_{\m} e^{\VV_{\m}}\Big) \nn\\
+  \sum_{\m=1,3,5}\Big(- \bar{\ZZ}_{\m} e^{-\VV_{\m}}\ZZ_{\m} e^{\VV_{\p}}-\bar{\WW}_{\m} e^{-\VV_{\p}}\WW_{\m} e^{\VV_{\m}}\Big)\Big]
\ee
and
\begin{eqnarray}
\mathcal{S}_\mathrm{pot}=\int d^3x\hspace{.1cm} d^2\theta\hspace{.1cm} W_\m-\int d^3x\hspace{.1cm} d^2\bar{\theta}\hspace{.1cm} \bar{W}_\m\nonumber
\end{eqnarray}
with,
\begin{eqnarray}
W_\m &=&\Tr\Big[\sum_{\m=1}^6\Phi_\m \mathcal{Z}_\m\mathcal{W}_\m +\sum_{\m=1}^{7}\frac{k_\m}{8\pi}[\Phi_\m\Phi_\m]\nn\\&-&\sum_{\m=2,4,6} {\Phi}_{(7)} \mathcal{W}_\m\mathcal{Z}_\m -\sum_{\m=1,3,5} {\Phi}_{\p} \mathcal{W}_\m\mathcal{Z}_\m\Big]
\ee
We write down the interaction part of the action since the kinetic part is straight forward from the previous cases.

\begin{eqnarray}
\mathcal{S_{\mathrm{int}}}&=& \int d^3x\,\, \Tr\Bigg( 
  \sum_{\m=1}^{6}\Big[
 -i {\xi^\g_a}_\m {\phi^a_b}_\m {\xi^b}_\m+\epsilon_{ac}{\lambda^{cb}}_\m X^a_\m{\xi_b^\g}_\m
 -\epsilon^{ac}{\lambda_{cb}}_\m\xi^b_\m{X^\g_a}_\m- g^2\kappa_\m X^\g_{a\m} {\phi^a_b}_\m X^b_\m\nn
\\[1mm]
&-&\frac{1}{2} X_\m X^\g_\m{\phi^a_b}_\m {\phi^b_a}_\m- \frac{g^2}{4} (X_\m \sigma_i X^\g_\m) (X_\m \sigma_i X_\m^\g)-\frac{g^2}{4} (X^\g_\m \sigma_i X_\m)(X^\g_\m \sigma_i X_\m)    \Big]
\nn \\[1mm]
 &+& \sum_{\m=2,4,6}\Big[- \epsilon_{ac}\lambda^{cb}_{(7)}\,{\xi^\g_b}_\m X^a_\m+ \epsilon^{ac}\lambda_{cb(7)}{X^\g_a}_\m\xi^b_\m+ i \xi^a_\m {\phi^b_a}_{(7)} {\xi^\g_b}_\m-\frac{1}{2}(X^\g_\m X_\m){\phi^a_b}_{(7)} {\phi^b_a}_{(7)}  \nn\\[1mm]
   &+&X^\g_{a\m} {\phi^b_c}_\m X^a_\m{\phi^c_b}_{(7)}+\kappa_{(7)} g^2 X^a_\m {\phi^b_a}_{(7)} {X^\g_b}_\m- \epsilon_{ac}\lambda^{cb}_\m\,{\xi^\g_b}_{\n} X^a_{\n}+ \epsilon^{ac}{\lambda_{cb}}_{\m}{X^\g_a}_{\n}\xi^b_{\n} \nn\\[1mm]
   &+&i {\xi^a}_{\n} {\phi^b_a}_{\m} {\xi^\g_b}_{\n}
 + g^2\kappa_{\m} X^a_{\n} {\phi^b_a}_{\m} {X^\g_b}_{\n}-\frac{1}{2}(X^\g_{\n} X_{\n})\,{\phi^a_b}_{\m}{\phi^b_a}_{\m}\nn\\[1mm]
   &+&\frac{g^2}{2} (X_\m \sigma_i X^\g_\m)(X^\g_{\n}\sigma_i X_{\n})\Big]
+  \sum_{\m=1,3,5} X^\g_{a\m}\, {\phi^b_c}_{\m} X^a_{\m}{\phi^c_b}_{\p} \nn\\[1mm]
   &+& \sum_{\m=1}^{7}\Big(\frac{1}{8g^2}[{\phi^a_b}_\m,{\phi^c_d}_\m][{\phi^b_a}_\m,{\phi^d_c}_\m]-\frac{\kappa_\m}{6}{\phi^a_b}_\m[{\phi^b_c}_\m{\phi^c_a}_\m]-\frac{i}{2g^2}{\lambda_{ab}}_\m[{\phi^b_c}_\m,{\lambda^{ac}}_\m]\Big)\nn\\ [1mm]
   &-&\frac{g^2}{2}(X^\g_\2 \sigma_i X_\2)(X^\g_\4 \sigma_i X_\4)
- \sum_{\m=2,4} \frac{g^2}{2}(X^\g_\m \sigma_i X_\m)(X^\g_{(6)} \sigma_i X_{(6)})\Bigg)
\end{eqnarray}
The supersymmetry variations are,
\begin{align}
  \delta {A_\mu}_\m&= -\frac{i}{2} \Ev_{ab} \gamma_\mu \lambda^{ab}_\m,\qquad  \delta \phi^a_{b\m}     = - \Ev_{cb} \lambda^{ca}_\m
+ \half\, \delta^a_b\, \Ev_{cd}\, \lambda^{cd}_\m,\qquad   \quad \m=1,..., 7&\nonumber \\[2mm]
  \delta \lambda^{ab}_\m &= \half \epsilon^{\mu\nu\lambda} {F_{\mu\nu}}_\m \gamma_\lambda \Ev^{ab}- i \slashed{\deriD} \phi^b_{c\m} \Ev^{ac}+ \frac{i}{2} [\phi^b_{c\m},\phi^c_{d\m}] \Ev^{ad}+i \kappa_\m g^2 \phi^b_{c\m} \Ev^{ac}      \nonumber \\&+i g^2  \big(X^a_\m {X^\g_c}_\m \Ev^{cb} - \Ev^{bc}{X^\g_c}_{\n} X^a_{\n} \big) + \frac{ig^2}{2} \big((X^\g X)_{\n} -(X X^\g)_\m\big) \Ev^{ab},\quad  \m=2,4,6\nn & 
\end{align}
\be
  \delta \lambda^{ab}_{(7)} &=& \half \epsilon^{\mu\nu\lambda} {F_{\mu\nu}}_{(7)} \gamma_\lambda \Ev^{ab}- i \slashed{\deriD} \phi^b_{c (7)} \Ev^{ac}+ \frac{i}{2} [\phi^b_{c(7)},\phi^c_{d(7)}] \Ev^{ad}+ i\kappa_{(7)} g^2 \,  \phi^b_{c (7)} \Ev^{ac}  - ig^2 \,    \Ev^{bc}\sum_{\m=2,4,6}{X^\g_c}_\m X^a_\m     \nonumber \\&+& \frac{ig^2}{2} \sum_{\m=2,4,6}(X^\g X)_\m  \Ev^{ab}   \nonumber\\[4mm]
  \delta \lambda^{ab}_\m &=& \half \epsilon^{\mu\nu\lambda} {F_{\mu\nu}}_\m \gamma_\lambda \Ev^{ab}- i \slashed{\deriD} \phi^b_{c\m} \Ev^{ac}+ \frac{i}{2} [\phi^b_{c\m},\phi^c_{d\m}] \Ev^{ad}+i \kappa_\m g^2  \phi^b_{c\m} \Ev^{ac} +i g^2 \, X^a_\m {X^\g_c}_\m \Ev^{cb}\nn\\ &-& \frac{ig^2}{2} (X X^\g)_\m \Ev^{ab},\qquad \m=1,3,5
\ee

\begin{align}
\m=1,...6 ,\quad  \delta X^{a}_\m  &= - i \Ev^a_b\, \xi^b_\m, \quad
  \delta X^\g_{{a}_\m}    = - i \xi^\g_{b\m}\, \Ev^b_a &\nn\\[2mm]
\m=2,4,6 \quad\delta \xi^{a}_\m  &= \slashed{\mathcal{D}} X^{b}_\m \Ev^a_b + \phi^a_{b\m} \,\Ev^b_c X^c_\m - X^c_\m \Ev^b_c\, \phi^a_{b(7)}& \;\nn \\[2mm]
\delta \xi^\g_{a\m}  &= \slashed{\mathcal{D}} X^\g_{b\m} \Ev^b_a - \phi^b_{a{(7)}}\, \Ev^c_b X^\g_{c\m}+ X^\g_{c\m} \Ev^c_b\, \phi^b_{a\m}& \; \quad\nn\\[4mm]
\m=1,3,5\quad\delta \xi^a_\m  &= \slashed{\mathcal{D}} X^b_\m \Ev^a_b + \phi^a_{b\m} \,\Ev^b_c X^c_\m - X^c_\m \Ev^b_c \,\phi^a_{b\p}& \;\nn \\[2mm]
\delta \xi^\g_{a\m} & = \slashed{\mathcal{D}} X^\g_{b\m} \Ev^b_a - \phi^b_{a\p}\, \Ev^c_b X^\g_{c\m}+ X^\g_{c\m} \Ev^c_b\, \phi^b_{a\m}& \; 
\end{align}

\medskip

\subsection{Action and supersymmetry variations on $\mathbbm{R}\times S^2$}
Our main goal is to compute the scaling dimensions of the BPS monopole operators in the theory which is related to its charges by operator state correspondence  in a radially quantized system. Following the method of BKK, to obtain the action on $\mathbbm{R}\times S^2$ from the  action on $\mathbbm{R}^{1,2}$ one has to carry out the following steps. First we do a Wick rotation to go from $\mathbbm{R}^{1,2}$ to $\mathbbm{R}^3$ in the following way,
\be
\label{eqn:wick}
x^0\longrightarrow-ix^3,\quad
A^0\longrightarrow-iA^3,\quad
\gamma^0\longrightarrow-i\gamma^3
\ee
\smallskip
\par The next step is to do a co-ordinate transformation   from $(x^1,x^2,x^3)$ to the polar  co-ordinates $(r,\theta,\varphi)$.  Under this co-ordinate transformation
all the terms in the action except the measure is invariant since they are written in  a co-ordinate independent form. Then we  translate the action to $\mathbbm{R}\times S^2$ by introducing a radial variable $\tau$ by the relation,
\be
r=e^ \tau
\ee
where $\tau\in\mathbbm{R}$, runs from $-\infty$ to $+\infty$, is the Euclidean time.
Now one has to perform a Weyl rescaling of the metric by a factor of $e^{-2\tau}$ so that the  theory  lands on $\mathbbm{R}\times S^2$ described by $(\tau,\theta,\varphi)$ co-ordinate system. The metric after Weyl rescaling is,
$g_{mn}=\mathrm{diag}(1,1, \sin^2\theta)$. $m,n=\tau,\theta,\varphi$ are used to denote the space-time indices on $\mathbbm{R}\times S^2$. One has to also 
rescale the $\mathbbm{R}^3$  fields to obtain $\mathbbm{R}\times S^2$ fields in the following way,
\be
\mathcal{X}_{\mathbbm{R}^3} = e^{-\mathrm{dim}{(\mathcal{X})}\tau}\mathcal{X}_{\mathbbm{R}\times S^2}
\ee
where, $\mathcal{X}$ is any generic field. The coupling $g$ is also rescaled by
\be
g\longrightarrow e^{-\frac{\tau}{2}}\tilde{g}
\ee
$\tilde{g}$ is the Yang-Mills coupling on $\mathbbm{R}\times S^2$ (for the component fields we don't use any tilde).
 
The vector multiplet fermions remains auxiliary in the IR and drop out form the theory. But since they were dynamical in the UV and can effect the charges of the monopole operators by quantum fluctuation one has to rescale them before doing the Weyl rescaling in the following way,

\be
\lambda_{ab\m}\longrightarrow g\,\lambda_{ab\m}
\ee
\subsubsection{$\widehat{A}$-type quiver}

Carrying out  all the steps described above we convert the action on $\mathbbm{R}^{1,2}$ to  $\mathbbm{R}\times S^2$.  The kinetic part of the action,
\begin{eqnarray}
\label{eqn:radial_action_an}
\mathcal{S}^E_\mathrm{kin} &=& \int d\tau d\Omega\,\,\sum_{\m=1}^n \mathrm{tr} \Bigg(\, \frac{1}{2\tilde{g}^2} F^{mn}_\m {F_{mn}}_\m - i \kappa_\m  \epsilon^{mnk} \big( {A_m}_\m \partial_n {A_k}_\m + \frac{2i}{3} {A_m}_\m {A_n}_\m {A_k}_\m \big)\nn\\&+&
    \mathcal{D}_m X^\dagger_\m \mathcal{D}^m X_\m+\frac{1}{4}X^\g_\m X_\m
   - i \xi^\dagger_\m \slashed{\mathcal{D}} \xi_\m
+ \frac{1}{2\tilde{g}^2} \mathcal{D}_m {\phi^a_b}_\m \mathcal{D}^m {\phi^b_a}_\m
   + \frac{1}{2} \kappa^2_\m \tilde{g}^2 \, {\phi^a_b}_\m {\phi^b_a}_\m
     \nn \\[1mm]\hspace{17mm}
  & +& \frac{i}{2} {\lambda^{ab}}_\m \slashed{\mathcal{D}} {\lambda_{ab}}_\m
   + \frac{i\kappa_\m \tilde{g}^2}{2} \, \lambda^{ab}_\m {\lambda_{ba}}_\m
   \Bigg)
\end{eqnarray}
The interaction part is,
\begin{eqnarray}
\label{eqn:radial_action_int_an}
\mathcal{S}^E_{\mathrm{int}}&=& \int d\tau d\Omega\,\,\sum_{\m=1}^n \mathrm{tr} \Bigg(\,\, \kappa_\m \tilde{g}^2 \, {X^\dagger_a}_\m {\phi^a_b}_\m X^b_\m - \kappa_\m \tilde{g}^2 \, X^a_\n {\phi}^b_{a\m} X^\dagger_{b\n}+i {\xi^\dagger_a}_\m {\phi^a_b}_\m \xi^b_\m\nn\\[1mm]\hspace{17mm}
& -&   i \xi^a_\n {\phi^b_a}_\m \xi^\dagger_{b\n}  - \tilde{g}\, \epsilon_{ac} \lambda^{cb}_\m X^a_\m {\xi^\dagger_b}_\m+\tilde{g}\epsilon^{ac} {\lambda_{cb}}_\m \xi^b_\m {X^\dagger_a}_\m + \tilde{g}\epsilon_{ac} {\lambda}^{cb}_\m {\xi^\dagger_b}_\n X^a_\n\nn\\[1mm]\hspace{17mm}
& -&  \tilde{g}\epsilon^{ac} {\lambda_{cb}}_\m {X^\dagger_a}_\n \xi^b_\n  - \frac{\kappa_\m}{6} {\phi^a_b}_\m \comm{{\phi^b_c}_\m}{{\phi^c_a}_\m}+ \frac{i}{2}  {\lambda_{ab}}_\m \comm{{\phi^b_c}_\m}{{\lambda^{ac}}_\m}\nn\\[1mm]\hspace{17mm}
&+& \frac{\tilde{g}^2}{4} (X_\m \sigma_i X^\dagger_\m) (X_\m \sigma_i X^\dagger_\m)+ \frac{\tilde{g}^2}{4} (X^\dagger_\m \sigma_i X_\m) (X^\dagger_\m \sigma_i X_\m)+ \frac{1}{2} (X_\m X^\dagger_\m) {\phi^a_b}_\m {\phi^b_a}_\m\nn \\[1mm]\hspace{17mm} &-& \frac{\tilde{g}^2}{2} (X_\m \sigma_i X^\dagger_\m) (X^\dagger_\n \sigma_i X_\n) + \frac{1}{2} (X^\dagger_\n X_\n) {\phi^a_b}_\m {\phi^b_a}_\m-{X^\dagger_a}_\m {\phi^b_c}_\m X^{a}_\m {\phi^c_b}_\p  \nn
\\[1mm]\hspace{17mm} &-&  \frac{1}{8\tilde{g}^2} \comm{{\phi^a_b}_\m}{{\phi^c_d}_\m} \comm{{\phi^b_a}_\m}{{\phi^d_c}_\m}\Bigg)
\end{eqnarray}
The supersymmetry variations obtained for the  action on $\mathbbm{R}^{1,2}$ can also be converted to $\mathbbm{R}\times S^2$ by following same steps as stated above.
The supersymmetry variation for the action on $\mathbbm{R}\times S^2$ with the rescaled variation parameter $\tilde{\varepsilon}_{ab}=e^{-\frac{\tau}{2}}\,\varepsilon_{ab}$ are as follows,
\begin{align}
\label{eqn:variation}
  \delta {A_m}_\m&= -\frac{i\tilde{g}}{2} \Ev_{ab} \gamma_m \lambda^{ab}_\m,\qquad \delta {\phi^a_b}_\m     = - \tilde{g}\,\Ev_{cb} \lambda^{ca}_\m
+ \frac{\tilde{g}}{2} \delta^a_b \Ev_{cd} \lambda^{cd}_\m &\nn \\[2mm]
  \delta \lambda^{ab}_\m &= \frac{i}{2\tilde{g}} \epsilon^{mnk} {F_{mn}}_\m \gamma_k \Ev^{ab}- \frac{i}{\tilde{g}} \slashed{\deriD} {(\phi^b_c)}_\m \Ev^{ac} -\frac{2i}{3\tilde{g}}{{\phi}^b_c}_\m\slashed{\nabla}\varepsilon^{ac} + \frac{i}{2\tilde{g}} [{\phi^b_c}_\m,{\phi^c_d}_\m] \Ev^{ad}+i \kappa_\m \tilde{g} \,  {\phi^b_c}_\m \Ev^{ac}   \nonumber \\& + i\,\tilde{g}  \big(X^a_\m {X^\dagger_c}_\m \Ev^{cb} - \Ev^{bc}{X^\dagger_c}_\n X^a_\n \big)+ \frac{i\tilde{g}}{2} ((X^\dagger X)_\n -(X X^\dagger)_\m) \Ev^{ab} &   
\end{align}
\begin{align}
\label{hypervariation}
  \delta X^{a}_\m & = - i \Ev^a_b\, \xi^{b}_\m\nonumber \;  &
  \delta \xi^{a}_\m & = \slashed{\mathcal{D}} X^{b}_\m \Ev^a_b +\frac{1}{3}X^b_\m\slashed{\nabla}\varepsilon^a_b  + {\phi^a_b}_\m \Ev^b_c X^{c}_\m - X^{c}_\m \Ev^b_c {{\phi}^a_b}_\p \;  \\
  \delta X^\dagger_{{a}_\m}   & = - i \xi^\dagger_{{b}_\m} \Ev^b_a \;  &
  \delta {\xi^\dagger_{a}}_\m & = \slashed{\mathcal{D}} X^\dagger_{{b}_\m} \Ev^b_a +\frac{1}{3}
  {X^\g_b}_\m\slashed{\nabla}\varepsilon^b_a - {{\phi}^b_a}_\p \Ev^c_b X^\dagger_{{c}_\m}+ {X^\dagger_c}_\m \Ev^c_b {\phi^b_a}_\m  
\end{align}
\subsubsection{$\widehat{D}$-type quiver}
The kinetic part remains similar as the  previous case. We write down the interaction part of the action on $\mathbbm{R}\times S^2$,
\begin{eqnarray}
\mathcal{S}^E_{\mathrm{int}}&=& \int d\tau\, d\Omega\,\, \Tr\Bigg( 
  \sum_{\m=1}^{n}\Big(\kappa_\m\tilde{g}^2 {X^\g_\m}_a {\phi^a_b}_\m X^b_\m
 +i {\xi^\g_a}_\m {\phi^a_b}_\m {\xi^b}_\m-\tilde{g}\,\epsilon_{ac}\lambda^{cb}_\m X^a_\m{\xi_b^\g}_\m\nn\\
 & +&\tilde{g}\epsilon^{ac}{\lambda_{cb}}_\m\xi^b_\m{X^\g_a}_\m 
+\frac{1}{2} X_\m X^\g_\m{\phi^a_b}_\m {\phi^b_a}_\m
 +\frac{\tilde{g}^2}{4} (X_\m \sigma_i X^\g_\m) (X_\m \sigma_i X_\m^\g)\nn\\
 &+& \frac{\tilde{g}^2}{4} (X^\g_\m \sigma_i X_\m)(X^\g_\m \sigma_i X_\m)    \Big)
 + \sum_{\m=1}^{2}\Big(\tilde{g} \epsilon_{ac}\lambda^{cb}_\5\xi^\g_{b\m} X^a_\m -\tilde{g}\, \epsilon^{ac}{\lambda_{cb}}_\5{X^\g_a}_\m\xi^b_\m\nn \\ 
 &-&i \xi^a_\m {\phi^b_a}_\5 {\xi^\g_b}_\m+\frac{1}{2}(X^\g_\m X_\m){\phi^a_b}_\5 {\phi^b_a}_\5
- {X^\g_\m}_a{\phi^b_c}_\m X^a_\m {\phi^c_b}_\5 - \kappa_{\5} \tilde{g}^2 X^a_\m {\phi^b_a}_\5 {X^\g_b}_\m \nn\\
&-& \frac{\tilde{g}^2}{2}(X_5 \sigma_i X^\g_5)(X^\g_{\m}\sigma_i X_{\m})\Big)
 +\sum_{\m=3}^{4}\Big(\tilde{g}\, \epsilon_{ac}\lambda^{cb}_{(n+1)}{\xi^\g_b}_\m X^a_\m -\tilde{g} \epsilon^{ac}\lambda_{cb(n+1)}{X^\g_a}_\m\xi^b_\m\nn \\
 &-&i \xi^a_\m {\phi^b_a}_{(n+1)} {\xi^\g_b}_\m
+\frac{1}{2}(X^\g_\m X_\m){\phi^a_b}_{(n+1)} {\phi^b_a}_{(n+1)}- {X^\g_\m}_a {\phi^b_c}_\m X^a_\m{\phi^c_b}_{(n+1)}\nn\\  
&-& \kappa_{(n+1)} \tilde{g}^2 X^a_\m {\phi^b_a}_{(n+1)} {X^\g_b}_\m
+ \frac{\tilde{g}^2}{2}(X^\g_\m \sigma_i X_\m)(X^\g_n \sigma_i X_n) \Big)+ \sum_{\m=5}^{n}\Big(\tilde{g} \epsilon_{ac}\lambda^{cb}_{\p}{\xi^\g_b}_\m X^a_\m
   \nn\\ 
   & -& \tilde{g}\, \epsilon^{ac}{\lambda_{cb}}_{\p}{X^\g_a}_\m\xi^b_\m  -i \xi^a_\m {\phi^b_a}_{\p} {\xi^\g_b}_\m-\tilde{g}^2\kappa_{\p} {X^a_\m} {\phi^b_a}_{\p} {X^\g_b}_\m  -{X^\g_\m}_a {\phi^b_c}_\m X^a_\m{\phi^c_b}_{\p}  \nn\\ 
   &+& \frac{1}{2}(X^\g_\m X_\m){\phi^a_b}_{\p}{\phi^b_a}_{\p}  \Big]+\sum_{\m=1}^{n+1}\Big(-\frac{1}{8\tilde{g}^2}[{\phi^a_b}_\m,{\phi^c_d}_\m][{\phi^b_a}_\m,{\phi^d_c}_\m]-\frac{\kappa_\m}{6}{\phi^a_b}_\m[{\phi^b_c}_\m{,\phi^c_a}_\m]\nn\\
  &+& \frac{i}{2}{\lambda_{ab}}_\m[{\phi^b_c}_\m,\lambda^{ac}_\m]\Big)- \frac{\tilde{g}^2}{2}\sum_{\m=6}^n (X_\m \sigma_i X^\g_\m)(X^\g_{\n}\sigma_i X_{\n})\nn\\
&+& \frac{\tilde{g}^2}{2}(X^\g_\1 \sigma_i X_\1)(X^\g_\2 \sigma_i X_\2)+\frac{\tilde{g}^2}{2}(X^\g_\3 \sigma_i X_\3)(X^\g_\4 \sigma_i X_\4)\Bigg)
\end{eqnarray}

Supersymmetry variations on $\mathbbm{R}\times S^2$ of the vector multiplet fields,
\begin{align}
\m=1,.., n+1,\quad  \delta {A_m}_\m&= -\frac{i\tilde{g}}{2} \Ev_{ab} \gamma_m \lambda^{ab}_\m,\qquad
  \delta \phi^a_{b\m}     = - \tilde{g}\Ev_{cb} \lambda^{ca}_\m
+\frac{\tilde{g}}{2}\, \delta^a_b\, \Ev_{cd}\, \lambda^{cd}_\m&\\[4mm]
\label{vec_fermion_variation_dn_internal}
\m=6,..., n,\hspace{.4cm}\qquad\delta \lambda^{ab}_\m &= \frac{i}{2\tilde{g}} \epsilon^{mnk} {F_{mn}}_\m \gamma_k \Ev^{ab} - \frac{i}{\tilde{g}} \slashed{\deriD} \phi^b_{c\m} \Ev^{ac} - \frac{2i}{3\tilde{g}}\phi^b_{c\m}\slashed{\nabla}\Ev^{ac}+ \frac{i}{2\tilde{g}} [\phi^b_{c\m},\phi^c_{d\m}] \Ev^{ad}\nonumber \\[2mm]&+i \kappa_\m \tilde{g} \, \phi^b_{c\m} \Ev^{ac}      +i\tilde{g} \,  \big(X^a_\m {X^\g_c}_\m \Ev^{cb} - \Ev^{bc}{X^\g_c}_{\n} X^a_{\n} \big)\nn\\[2mm]& + \frac{i\tilde{g}}{2} \big((X^\g X)_{\n} -(X X^\g)_\m\big) \Ev^{ab} &
\end{align}

\begin{align}
\label{vec_fermion_variation_dn}
    \delta \lambda^{ab}_\5 &= \frac{i}{2\tilde{g}} \epsilon^{mnk} {F_{mn}}_\5 \gamma_k \Ev^{ab}- \frac{i}{\tilde{g}} \slashed{\deriD} \phi^b_{c\5} \Ev^{ac} - \frac{2i}{3\tilde{g}}\phi^b_{c\5}\slashed{\nabla}\Ev^{ac}+ \frac{i}{2\tilde{g}} [\phi^b_{c\5},\phi^c_{d\5}] \Ev^{ad}+ i\kappa_\5 \tilde{g} \,  \phi^b_{c\5} \Ev^{ac}      \nonumber \\[2mm]&+i \tilde{g} \,  \big(X^a_\5 X^\g_{c\5} \Ev^{cb} - \Ev^{bc}\sum_{\m=1}^2{X^\g_c}_\m X^a_\m \big) + \frac{i\tilde{g}}{2} (\sum_{\m=1}^2(X^\g X)_\m -(X X^\g)_\5) \Ev^{ab} &   \\[4mm]
  \delta \lambda^{ab}_{(n+1)} &= \frac{i}{2\tilde{g}} \epsilon^{mnk} {F_{mn}}_{(n+1)} \gamma_k \Ev^{ab}- \frac{i}{\tilde{g}} \slashed{\deriD} \phi^b_{c{(n+1)}} \Ev^{ac} - \frac{2i}{3\tilde{g}}\phi^b_{c{(n+1)}}\slashed{\nabla}\Ev^{ac}+ \frac{i}{2\tilde{g}} [\phi^b_{c(n+1)},\phi^c_{d(n+1)}] \Ev^{ad} \nonumber \\[2mm]&+ i\kappa_{(n+1)} \tilde{g}  \, \phi^b_{c (n+1)} \Ev^{ac}     - i\tilde{g} \,    \Ev^{bc}\sum_{\m=3,4,n}{X^\g_c}_\m X^a_\m + \frac{i\tilde{g}}{2} \sum_{\m=3,4,n}(X^\g X)_\m  \Ev^{ab} & \\[4mm]
  \label{vec_fermion_variation_dn_end}
  \delta \lambda^{ab}_\m &=\frac{i}{2\tilde{g}} \epsilon^{mnk} {F_{mn}}_\m \gamma_k \Ev^{ab}- \frac{i}{\tilde{g}} \slashed{\deriD} \phi^b_{c\m} \Ev^{ac}- \frac{2i}{3\tilde{g}}\phi^b_{c\m}\slashed{\nabla}\Ev^{ac}+ \frac{i}{2\tilde{g}} [\phi^b_{c\m},\phi^c_{d\m}] \Ev^{ad}+ i\kappa_\m \tilde{g} \,  \phi^b_{c\m} \Ev^{ac}      \nonumber \\&+i \tilde{g} \, X^a_\m {X^\g_c}_\m \Ev^{cb}  - \frac{i \tilde{g}}{2} (X X^\g)_\m \Ev^{ab},\quad \textit{for $\m=1,2,3,4$ } &  
\end{align}
Variations of the hypermultiplet fields,
\begin{align}
\m=1,..n\quad \delta X^{a}_\m  = - i \Ev^a_b\, \xi^b_\m \; \quad
  \delta X^\g_{{a}_\m}    = - i \xi^\g_{b\m} \Ev^b_a
\end{align}
\be
\label{hyperino_variation_dn}
\m=1,2\quad &&\delta \xi^{a}_\m  = \slashed{\mathcal{D}} X^{b}_\m \Ev^a_b +\frac{1}{3}X^b_\m\slashed{\nabla}\Ev^a_b + \phi^a_{b\m} \Ev^b_c X^c_\m - X^c_\m \Ev^b_c \phi^a_{b\5} \;\nn \\[1mm]
 &&\delta \xi^\g_{a\m}  = \slashed{\mathcal{D}} X^\g_{b\m} \Ev^b_a +\frac{1}{3} X^\g_{b\m} \slashed{\nabla}\Ev^b_a - \phi^b_{a\5} \Ev^c_b X^\g_{c\m}+ X^\g_{c\m} \Ev^c_b \phi^b_{a\m} \; 
\ee 
\be 
\label{hyperino_variation_dn}
\m=3,4\quad&&\delta \xi^{a}_\m  = \slashed{\mathcal{D}} X^{b}_\m \Ev^a_b +\frac{1}{3}X^b_\m\slashed{\nabla}\Ev^a_b + \phi^a_{b\m} \Ev^b_c X^c_\m - X^c_\m \Ev^b_c \phi^a_{b(n+1)} \;\nn \\[1mm]
 &&\delta \xi^\g_{a\m}  = \slashed{\mathcal{D}} X^\g_{b\m} \Ev^b_a +\frac{1}{3} X^\g_{b\m} \slashed{\nabla}\Ev^b_a - \phi^b_{a(n+1)} \Ev^c_b X^\g_{c\m}+ X^\g_{c\m} \Ev^c_b \phi^b_{a\m} \; 
 \ee
 \be
\label{hyperino_variation_dn}
\m=5, 6,..., n \quad&&\delta \xi^a_\m  = \slashed{\mathcal{D}} X^b_\m \Ev^a_b +\frac{1}{3}X^b_\m\slashed{\nabla}\Ev^a_b + \phi^a_{b\m} \Ev^b_c X^c_\m - X^c_\m \Ev^b_c \phi^a_{b\p} \;\nn \\[1mm]
 &&\delta \xi^\g_{a\m}  = \slashed{\mathcal{D}} X^\g_{b\m} \Ev^b_a+\frac{1}{3} X^\g_{b\m} \slashed{\nabla}\Ev^b_a - \phi^b_{a\p} \Ev^c_b X^\g_{c\m}+ X^\g_{c\m} \Ev^c_b \phi^b_{a\m} \; 
\ee
\subsubsection{$\widehat{E}_6$ quiver}
The $\hat{E}_6$ action on $\mathbbm{R}\times S^2$ is calculated following the same steps. We do not write the expression of the action here since it is straightforward from the $\widehat{D}$-type quiver. Let us write down the supersymmetry variations which are needed to find the monopole solutions later.
\begin{align}
\label{vec_fermion_variation_e6}
\delta \lambda^{ab}_\m &= \frac{i}{2\tilde{g}} \epsilon^{mnk} {F_{mn}}_\m \gamma_k \Ev^{ab} - \frac{i}{\tilde{g}} \slashed{\deriD} \phi^b_{c\m} \Ev^{ac} - \frac{2i}{3\tilde{g}}\phi^b_{c\m}\slashed{\nabla}\Ev^{ac}+ \frac{i}{2\tilde{g}} [\phi^b_{c\m},\phi^c_{d\m}] \Ev^{ad}\nonumber \\[2mm]&+ i\kappa_\m \tilde{g} \,  \phi^b_{c\m} \Ev^{ac}      +i\tilde{g} \,  \big(X^a_\m {X^\g_c}_\m \Ev^{cb} - \Ev^{bc}{X^\g_c}_{\n} X^a_{\n} \big) + \frac{i\tilde{g}}{2} \big((X^\g X)_{\n} -(X X^\g)_\m\big) \Ev^{ab} &\nn\\
&\hspace{12cm}\m=2,4,6&\nn\\[4mm]
  \delta \lambda^{ab}_{(7)} &= \frac{i}{2\tilde{g}} \epsilon^{mnk} {F_{mn}}_\7 \gamma_k \Ev^{ab}- \frac{i}{\tilde{g}} \slashed{\deriD} \phi^b_{c\7} \Ev^{ac} - \frac{2i}{3\tilde{g}}\phi^b_{c\7}\slashed{\nabla}\Ev^{ac}+ \frac{i}{2\tilde{g}} [\phi^b_{c(7)},\phi^c_{d(7)}] \Ev^{ad} \nonumber \\[2mm]&+ i\kappa_{(7)} \tilde{g}   \phi^b_{c (7)} \Ev^{ac}     - i\tilde{g} \,    \Ev^{bc}\sum_{\m=2,4,6}{X^\g_c}_\m X^a_\m + \frac{i\tilde{g}}{2} \sum_{\m=2,4,6}(X^\g X)_\m  \Ev^{ab} &\nn \\[4mm]
  \delta \lambda^{ab}_\m &=\frac{i}{2\tilde{g}} \epsilon^{mnk} {F_{mn}}_\m \gamma_k \Ev^{ab}- \frac{i}{\tilde{g}} \slashed{\deriD} \phi^b_{c\m} \Ev^{ac}- \frac{2i}{3\tilde{g}}\phi^b_{c\m}\slashed{\nabla}\Ev^{ac}+ \frac{i}{2\tilde{g}} [\phi^b_{c\m},\phi^c_{d\m}] \Ev^{ad}+i \kappa_\m \tilde{g} \,  \phi^b_{c\m} \Ev^{ac}      \nonumber \\[2mm]&+i \tilde{g} \, X^a_\m {X^\g_c}_\m \Ev^{cb}  - \frac{i \tilde{g}}{2} (X X^\g)_\m \Ev^{ab}\hspace{7cm} \m=1,3,5 &  
\end{align}
Variations of the hypermultiplet fermions,
\be
\label{hyperino_variation_e6}
\m=2,4,6\quad&&\delta \xi^{a}_\m  = \slashed{\mathcal{D}} X^{b}_\m \Ev^a_b +\frac{1}{3}X^b_\m\slashed{\nabla}\Ev^a_b + \phi^a_{b\m} \Ev^b_c X^c_\m - X^c_\m \Ev^b_c \phi^a_{b(7)} \;\nn \\[1mm]
 &&\delta \xi^\g_{a\m}  = \slashed{\mathcal{D}} X^\g_{b\m} \Ev^b_a +\frac{1}{3} X^\g_{b\m} \slashed{\nabla}\Ev^b_a - \phi^b_{a(7)} \Ev^c_b X^\g_{c\m}+ X^\g_{c\m} \Ev^c_b \phi^b_{a\m} \; \\[2mm]
\m=1,3,5 \quad&&\delta \xi^a_\m  = \slashed{\mathcal{D}} X^b_\m \Ev^a_b +\frac{1}{3}X^b_\m\slashed{\nabla}\Ev^a_b + \phi^a_{b\m} \Ev^b_c X^c_\m - X^c_\m \Ev^b_c \phi^a_{b\p} \;\nn \\[1mm]
 &&\delta \xi^\g_{a\m}  = \slashed{\mathcal{D}} X^\g_{b\m} \Ev^b_a+\frac{1}{3} X^\g_{b\m} \slashed{\nabla}\Ev^b_a - \phi^b_{a\p} \Ev^c_b X^\g_{c\m}+ X^\g_{c\m} \Ev^c_b \phi^b_{a\m} \; 
\ee
\bigskip

\section{\label{sec:monopole soln} Classical monopole solution}
\subsection{$\widehat{A}$-type quiver}

In this section we compute the classical bosonic BPS and anti-BPS monopole solution to the  theories under consideration. To find a monopole solution we start  with a Dirac monopole solution on $\mathbbm{R}\times S^2$,
\be
\label{eqn:gauge}
A_\m=\frac{H}{2}\,\,(\pm1-\cos\theta)d\varphi
\ee
where, $H=\mathrm{diag}(q_{\scaleto{1}{4pt}},q_{\scaleto{2}{4pt}},...q_{\scaleto{N}{4pt}})$ and $q_i\in \mathbbm{Z}$ are the magnetic charges. The  upper sign is for the northern hemisphere and lower sign for the southern one. The the dual of the field strength of the above gauge field, 
\be
\epsilon^{\theta\varphi\tau}F_{\theta\varphi}=\frac{H}{2} d\tau
\ee
 is a constant. If there exists a BPS monopole solution with such a gauge potential \eqref{eqn:gauge}, then corresponding fermionic variation $\delta\lambda^{ab}_\m, \delta\xi^{a}_\m, \delta\xi^\g_{a\m}$ should be equal to zero for a non trivial supersymmetry variation parameter. As we can see form \eqref{eqn:variation} that $\delta\lambda^{ab}_\m$ contains terms of order $\tilde{g}$ and $\tilde{g}^{-1}$.  The goal of finding solutions all along the flow
means we have to set them separately to zero.  In our following calculation we are also assuming that the background fields ($A,\phi$) are  in the Cartan subalgebra of the gauge group factors, which makes all commutator vanish. 
\smallskip
\par 
\underline{$\delta \lambda^{ab}_{(j)}$ at order $\frac{1}{\tilde{g}}$:}
Equating order $\tilde{g}^{-1}$ terms of \eqref{eqn:variation} to zero,
\be
\label{eqn:var_vec_fermion_an}
\frac{i}{2g} \epsilon^{mnk} {F_{mn}}_\m \gamma_k \Ev^{ab}- \frac{i}{g} \slashed{\deriD} {\phi^b_c}_\m \Ev^{ac} -\frac{2i}{3g}{{\phi}^b_c}_\m\slashed{\nabla}\varepsilon^{ac} =0
\ee
Recalling that $\delta \lambda^{ab}_{\,(j)}$ is in the reducible representation $\mathbf{2}\times\mathbf{2} = \mathbf{1} + \mathbf{3}$ of $SU(2)_R$ we isolate the $\mathbf{1}$ part i.e. the $SU(2)_R$ trace, by computing $\delta \lambda^{ab}_{\,(j)}\,\epsilon_{b a}$. 
The first term is zero since $\Ev_a^a=0$. Therefore only the trace of the second term contributes to the $SU(2)_R$ trace of $\delta \lambda^{ab}_{\,(j)}$ at order $\frac{1}{\tilde{g}}$. Since, $\epsilon^{mnk}F_{mn}$ is a constant, we can have  a simple situation where $\phi$'s are also constant.  Then the second term of the above equation is zero.  The third term in \eqref{eqn:var_vec_fermion_an}, after using the  the Killing spinor equation on $\mathbbm{R}\times S^2$,  
\be
\label{killing_eqn_s2}
\nabla_m \varepsilon=-\frac{1}{2}\gamma_m \gamma^\tau\varepsilon
\ee 
gives,
\be
\label{eqn:con1}
{\phi}_{i\m }\,\Ev_i =0
\ee
i.e, the supersymmetry variation parameter is orthogonal to the background scalar. We can solve the above by choosing, 
\begin{equation}
\varepsilon_3 = 0, \qquad \phi_{i(j)} \sim \delta_{i3}
\end{equation}

Then we isolate the $\mathbf{3}$ part by  computing $\delta \lambda^{ab}_{\,(j)}\,(\sigma_i)_{b a}$ and we get,
\be
\label{eqn:con2}
&&\frac{i}{2} \epsilon^{mnk} {F_{mn}}_\m \gamma_k \Ev_i+ \epsilon_{ijk}{{\phi}_j}_\m\gamma^\tau\varepsilon_k =0
\ee

easily solvable in two ways:
\begin{eqnarray}
\label{eqn:phi_an}
(i)\qquad \phi_{3\,(j)} = - \frac{{H}}{2}, \quad \varepsilon_1 - i \varepsilon_2 = 0 \\
(ii)\qquad \phi_{3\,(j)} =  \frac{{H}}{2}, \quad \varepsilon_1 + i \varepsilon_2 = 0.
\end{eqnarray}
In the first case, the preserved supersymmetry is $\varepsilon_1 + i \varepsilon_2$ and is the BPS solution and the second case, the preserved supersymmetry is $\varepsilon_1 - i \varepsilon_2$ and is the anti-BPS solution. In both cases, since one of the three supersymmetry parameters are preserved by the solution we have $\frac13$-BPS solutions. Now let us examine the order $\tilde{g}$ terms in $\delta \lambda^{ab}_{(j)}$.
\smallskip
\par
\underline{$\delta \lambda^{ab}_{(j)}$ at order $\tilde{g}$:}  We obtain from \eqref{eqn:variation}
\begin{eqnarray}\label{orderg}
\kappa_{(j)}   \,\phi^b_{c\,(j)}\varepsilon^{ac}   +   \big(X^a_{(j)}\,X^\dagger_{c\,(j)} \varepsilon^{cb} - \varepsilon^{bc} X^\dagger_{c\,(j-1)}\,X^a_{(j-1)}\big) - \frac{1}{2} \big(X_{(j)}\,X^\dagger_{(j)} - X^\dagger_{(j-1)}X_{(j-1)} \big) \varepsilon^{ab} = 0\nn\\
\end{eqnarray}
We analyse the above as before by considering the $\mathbf{1}$ and the $\mathbf{3}$ parts separately. The $SU(2)_R$ trace of the first term is  $\kappa_{(j)}\,\varepsilon_i\,\phi_{i(j)}$ which vanishes on using \eqref{eqn:con1} and the third term is traceless. The  second term gives
\begin{eqnarray}\label{+1trace}
(i) &\eta =1\qquad  (-W^\dagger_{(j)}Z^\dagger_{(j)} + Z^\dagger_{(j-1)} W^\dagger_{(j-1)})(\varepsilon_1 + i \varepsilon_2 ) = 0\ \nonumber \\
(ii) &\eta =-1\qquad (-Z_{(j)}W_{(j)} + W_{(j-1)} Z_{(j-1)})(\varepsilon_1 - i \varepsilon_2 ) = 0.
\end{eqnarray}
By isolating the $\mathbf{3}$ part of \eqref{orderg} one obtains the following equation,
\begin{equation}\label{+1three}
2 \kappa_{(j)}\,\phi_{i\,(j)} = - X_{(j)}\sigma_i X^\dagger_{(j)} + X^\dagger_{(j-1)}\sigma_i X_{(j-1)}
\end{equation}
by which we see that the hypermultiplet scalars are already constrained by the CS levels and the magnetic charges in addition to other constraints that are yet to come from analysing the hypermultiplet fermions variation equations.
\smallskip
\par Moving on to the hypermultiplet fermion variations,\\
\underline{$\mathbf{\delta \xi^{a}_{(j)}}$:} We obtain from \eqref{hypervariation}
\begin{eqnarray}
\delta \xi^{a}_\m & = \slashed{\mathcal{D}} X^{b}_\m \Ev^a_b +\frac{1}{3}X^b_\m\slashed{\nabla}\varepsilon^a_b  + {\phi^a_b}_\m \Ev^b_c X^{c}_\m - X^{c}_\m \Ev^b_c {{\phi}^a_b}_\m\nn
\end{eqnarray}
The last two terms cancel\footnote{There won't be a straightforward cancellation in the more general case  where the magnetic charges are different at each node, $A_{(j)} \sim H_{(j)}, ~ \phi_{i\,(j)} = - \eta \frac{{H_{(j)}}}{2}\,\delta_{i3}$ .} because in the special case we are working in, all the $\phi_{(j)}$'s are equal. Using Killing- spinor equation we get,
\be
\delta\xi^a_\m = \Big(\gamma^m\mathcal{D}_m X^b_\m - \frac{1}{2} X^b_\m \gamma^\tau\Big)\Ev^a_b \nonumber
\ee
Expanding the above,
\be
\label{eqn:con4}
 \delta\xi^1_\m =\Ep \Big(\gamma^m\mathcal{D}_m X^2_\m - \frac{1}{2} X^2_\m \gamma^\tau\Big)\qquad \textrm{for $a=1$}
\ee
and
\be\label{eqn:con5}
 \delta\xi^2_\m = \Em \Big(\gamma^m\mathcal{D}_m X^1_\m - \frac{1}{2} X^1_\m \gamma^\tau\Big)\qquad \textrm{for $a=2$}
\ee
Demanding $\delta\xi^a_\m=0$, fixes the functional dependence of $X^a_\m$ to,
$X^1_\m\sim \exp(-\eta\tau/2)$ and $X^2_\m\sim \exp(\eta\tau/2)$. Then we can make all $\delta\xi^a_\m$
's  zero either by the functional dependence or by the condition derived above $\Ev_1- i\eta\Ev_2=0$. The solution chosen for the X's also satisfy their classical equation of motion.
The only other equations of motion that remain to be satisfied by the so-far obtained background are for the gauge fields:
\begin{equation}\label{eomgauge}
\kappa_\m \epsilon^{mnp}F_{np\m}=   X_\m\,\mathcal{D}^m X^\g_{\m}-\mathcal{D}^m X_\m. X^\g_{\m} - \mathcal{D}^m X^\g_{\n}\, .X_{\n}+X^\g_{\n}\,\mathcal{D}^m X_{\n}
\end{equation}

To summarise, we have  obtained the background gauge and adjoint scalar fields,
\begin{eqnarray}
\label{monopole_solution}
A_{(j)} = \frac{{H}}{2} (\pm 1 - \cos \theta)\,d\phi, \quad \phi_{i\,(j)} = - \eta \frac{{H}}{2}\,\delta_{i3}
\end{eqnarray}
 with $\varepsilon_1 + i \eta\,\varepsilon_2$ the preserved supersymmetry. 
 And we have several constraints on the hypermultiplet scalar fields: (i) equation \eqref{+1trace}, (ii) equation \eqref{+1three} (iii) the $\tau$ dependence that follow from the hypermultiplet fermion variations and (iv) the equation \eqref{eomgauge}. We will analyse these constraints and find solutions first for the three node quiver and with the experience gained thus, we can then generalize for a generic $n$-node quiver. 
 
\subsubsection{Example of a  three node case}
In the previous section we saw how to choose the fields in the vector multiplet to make the fermionic variations(both vector and hypermultiplet) zero. In this section we will consider a simple example of a  three node case and
\begin{figure}[hbtp]
\centering
\includegraphics[scale=.5]{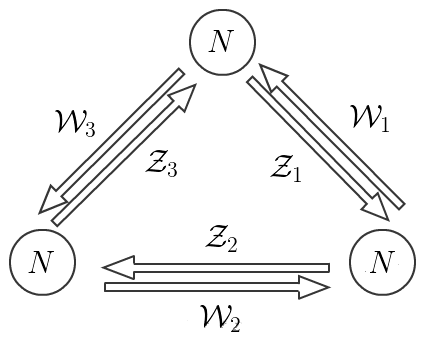}
\caption{$\widehat{A}$ quiver diagram with three gauge groups}
\end{figure}
see how to completely fix  the hypermultiplet scalar fields to obtain full monopole solution. For that we need to solve,
\be
\label{tosolve}
 {\phi_l}_\m  =      -\frac{1}{2\kappa_\m}\Big(   X_\m {(\sigma_l)} {X^\dagger}_\m  - X^\dagger_\n {(\sigma_l)} X_\n\Big)
\ee
where, $l=1,2,3$. 

\eqref{tosolve}  when expanded in terms of components gives the following set of equations,
\be
\label{eqn:phi1}
{\phi_1}_\1\longrightarrow  Z_\1 W_\1+ W^\g_\1 Z^\g_\1 - Z^\g_\3 W^\g_\3 - W_\3 Z_\3 =0\nonumber\\
{\phi_1}_\2\longrightarrow  Z_\2 W_\2+ W^\g_\2 Z^\g_\2 - Z^\g_\1 W^\g_\1 - W_\1 Z_\1 =0\nonumber\\
{\phi_1}_\3\longrightarrow  Z_\3 W_\3+ W^\g_\3 Z^\g_\3 - Z^\g_\2 W^\g_\2 - W_\2 Z_\2 =0
\ee
Similarly,
\be
\label{eqn:phi2}
{\phi_2}_\1\longrightarrow   W^\g_\1 Z^\g_\1 - Z_\1 W_\1 - Z^\g_\3 W^\g_\3 + W_\3 Z_\3 =0\nonumber\\
{\phi_2}_\2\longrightarrow   W^\g_\2 Z^\g_\2 - Z_\2 W_\2 - Z^\g_\1 W^\g_\1 + W_\1 Z_\1 =0\nonumber\\
{\phi_2}_\3\longrightarrow   W^\g_\3 Z^\g_\3 - Z_\3 W_\3 - Z^\g_\2 W^\g_\2 + W_\2 Z_\2                                                 =0
\ee
For a simple case, there can be three possibilities which solve the aove six equations.\\
(i)$Z_\1=Z_\2=Z_\3=0$\\
(ii)$W_\1=W_\2=W_\3=0$\\
(iii)$W_\1=Z_\2=W_\3=0$ or $Z_\1=W_\2=Z_\3=0$\\
Now expanding $\phi_3$,  in terms of $X$  and $X^\g$ from \eqref{tosolve} we get the following equations,
\be
\label{eqn:phi3con}
{\phi_3}_\1\longrightarrow Z_\1 Z^\g_1 - W^\g_\1 W_\1 - Z^\g_\3 Z_\3 + W_\3 W^\g_\3 =\eta H\kappa_\1 \nn\\
{\phi_3}_\2\longrightarrow Z_\2 Z^\g_\2 - W^\g_\2 W_\2 - Z^\g_\1 Z_\1 + W_\1 W^\g_\1 = \eta H\kappa_\2\nn\\
{\phi_3}_\3\longrightarrow Z_\3 Z^\g_\3 - W^\g_\3 W_\3 - Z^\g_\2 Z_\2 + W_\2 W^\g_\2 =\eta H\kappa_\3
\ee
 We need to solve the above three equations for the three cases written above. Since the Chern-Simons levels satisfy,
\be
\kappa_\1+\kappa_\2+\kappa_\3=0\nonumber
\ee 
either one or two of them have to be negative.
For a positive semi-definite $H$, we can solve  several constraints on the hypermultiplet scalars  in the following ways. These solutions for each case are consistent with the equation of motion of the gauge fields \eqref{eomgauge}. For example,
\par \textbf{(i)$\boldsymbol{Z_\1=Z_\2=Z_\3=0}$}
\par \underline{\textbf{BPS $\eta=+1$}}
\eqref{eqn:phi3con}reduces to,
\be
 - W^\g_\1 W_\1  + W_\3 W^\g_\3 = H\kappa_\1,\quad
 - W^\g_\2 W_\2  + W_\1 W^\g_\1 =  H\kappa_\2,\quad
 - W^\g_\3 W_\3  + W_\2 W^\g_\2 = H\kappa_\3\nn\\
\ee
are solved by,
\be
 W_\1 = \sqrt{A}\, e^{ -\tau/2} \ , \hspace{5mm}  W^\g_\1 = \sqrt{A}\, e^{\tau/2} \nn  \\
 W_\2 = \sqrt{ A-H \kappa_\2}\, e^{- \tau/2} \ , \hspace{3mm}  W^\g_\2 = \sqrt{A -H \kappa_\2}\, e^{\tau/2}  \nn \\
 W_\3 = \sqrt{A + H\kappa_\1 }\, e^{- \tau/2} \ , \hspace{5mm} W^\g_\3 = \sqrt{A + H \kappa_\1}\, e^{\tau/2} 
\ee
The above equations are satisfied for any diagonal matrix $A$ with  entries $\geq 0$ assuming, $\kappa_\1,\kappa_\3>0$ and $\kappa_\2<0$.  \\
\underline{\textbf{anti-BPS $\eta=-1$}:}
Similarly the anti-BPS case is solved by,
\be
 W_\1 = \sqrt{A}\, e^{\tau/2} \ , \hspace{5mm}  W^\g_\1 = \sqrt{A}\, e^{-\tau/2} \nn  \\
 W_\2 = \sqrt{ A+H \kappa_\2}\, e^{ \tau/2} \ , \hspace{3mm}  W^\g_\2 = \sqrt{A +H \kappa_\2}\, e^{-\tau/2}  \nn \\
 W_\3 = \sqrt{A - H\kappa_\1 }\, e^{ \tau/2} \ , \hspace{5mm} W^\g_\3 = \sqrt{A - H \kappa_\1}\, e^{-\tau/2} 
\ee
where,
\be
A_\alpha\geq q_\alpha|\kappa_\2|,\quad A_\alpha\geq q_\alpha\kappa_\1\nn
\ee
where $A_\alpha$ and $q_\alpha$ denote the $\alpha$-th entry of the matrix $A$ and $H$ respectively.
Similarly the rest of the cases can be solved.
\subsubsection{$n$ node case}
Now we can move on  to more general case with gauge group $U(N)\times U(N)\times...\times U(N)$ and apply similar steps as above to obtain the full monopole solution in this theory.
From \eqref{eqn:phi1} and \eqref{eqn:phi2} we can see that,
\be
Z_\m W_\m = Z_\n W_\n\nn
\ee
Because all $Z_\m$'s and $W_\m$'s are diagonal, if we take $d_\alpha$ to be the $\alpha$-th entry of the matrix $Z_\m W_\m$, then
\be
\label{eqn:1}
{Z_\m}_\alpha {W_\m}_\alpha = {Z_\n}_\alpha {W_\n}_\alpha = d_\alpha(\mathrm{=constant})
\ee 
Also from  \eqref{eqn:phi3con} we can write,
\be
{|Z_\m|}^2_\alpha - {|W_\m|}^2_\alpha = {|Z_\n|}^2_\alpha - {|W_\n|}^2_\alpha + \eta q_\alpha \kappa_\m\nn
\ee
Using the above equation again and again we get,
\be
\label{eqn:2}
{|Z_\m|}^2_\alpha - {|W_\m|}^2_\alpha = {|Z_\1|}^2_\alpha - {|W_\1|}^2_\alpha + \eta q_\alpha K_\m 
\ee
where,
\be
K_\m=\sum_{\m=2}^{\m}\kappa_\m
\ee
Let ${|Z_\m|}^2_\alpha = u_\m, {|W_\m|}^2_\alpha= v_\m$.
Using \eqref{eqn:1} and substituting in \eqref{eqn:2}
\be
u^2_\m u_\1 + u_\m (- u^2_\1  + {|d_\alpha|^2} - \eta q_\alpha K_\m u_\1 )- {|d_\alpha|^2}u_\1 =  0 
\ee
Above is a quadratic equation in $u_\m$ with a solution,
\be
u_\m = \frac{ u^2_\1  - {d_\alpha^2} + \eta q_\alpha K_\m u_\1 \pm\sqrt{(u^2_\1  - {|d_\alpha|^2} + \eta q_\alpha K_\m u_\1)^2 - 4 d_\alpha^2 u^2_\1} }{2 u_\1}
\ee
These seem to exist for any value of $\kappa_\m$. These form a moduli space of solutions classified by
$Z_{\1\alpha}, d_\alpha, \mathrm{arg}(Z_{\m\alpha})$.
The above is a general solution. 
\smallskip
\par Let us examine the case if we choose $d_\alpha=0$. Then from \eqref{eqn:1} either $v_\1$ or $u_\1$ is zero. Also from \eqref{eqn:2}
\be
\label{eqn3}
u_\m- v_\m = u_\1-v_\1 +\eta q K_\m
\ee
Now, we have a constraint that $ \kappa_\1 +\kappa_\2+...+\kappa_{(n)}=0$.
Therefore some of the $\kappa_\m$'s are positive and the rest are negative. In other words there will be some positive $K_\m$'s , let us call that set $[K_+]$ and rest of them will be negative which are denoted by $[K_-]$. Now let us solve case by case as we did in the three node example. \\
\textbf{CASE 1: $v_\1=v_\2=...=v_{(n)}=0$}\\
From \eqref{eqn3}
\be
u_\m = u_\1 +\eta q_\alpha K_\m\nn
\ee
For,\\
(i) \textbf{BPS $\eta=1$} \\
Since $u_\1$ is positive, positivity of the LHS implies solutions exist for $K_\m>0$. For negative $K_\m$'s 
\be
u_\1-q[|K_-|]>0\implies u_\1>q_\alpha\,\mathrm{max}([|K_-|])
\ee
(ii) \textbf{anti-BPS $\eta=-1$}\\
There is no restriction on $u_\1$ for $[K_-]$. But for $[K_+]$ we get,
\be
u_\1-q\,[K_+]>0\implies u_\1>q\,\mathrm{max}([K_+])
\ee
\textbf{CASE 2: $u_\1=u_\2=...=u_{(n)}=0$}\\
From \eqref{eqn3}
\be
v_\m = v_\1 -\eta q_\alpha K_\m\nn
\ee
For,\\
(i) \textbf{BPS $\eta=1$} \\
positivity of LHS implies solutions exist for $K_\m<0$. For positive $K_\m$'s 
\be
v_\1-q_\alpha [K_+]>0\implies v_\1>q_\alpha \,\mathrm{max}([K_+])
\ee
(ii) \textbf{anti-BPS $\eta=-1$}\\
There is no restriction on $v_\1$ for $[K_+]$. But for $[K_-]$ we get,
\be
v_\1-q_\alpha \,[|K_-|]>0\implies v_\1>q_\alpha \,\mathrm{max}([|K_-|])
\ee
\textbf{CASE 3}: $v_\1=0$, some $u_\m=0$ and complementary $v_\m=0$\\
Let us first consider the nodes where $u_\m=0$. For this case there will be a subset of $[K]$ for which $u_\m=0$ and K's are positive or negative. Lets us denote the first case by $[K_{u_+}]$ and the second by $[K_{u_-}]$. Similar analysis for the rest of the nodes will bring two more subsets of $[K]$ i.e  $[K_{v_+}]$ and $[K_{v_-}]$. Hence,
\be
[K]=[K_+]\cup [K_-]\nn
\ee
and
\be
[K_+]=[K_{u_+}]\cup [K_{v_+}]\nn
\ee
Similarly,
\be
[K_-]=[K_{u_-}]\cup [K_{v_-}]\nn
\ee
Now, let us look at the solutions. For the nodes where $v_\m=0$,  from \eqref{eqn3}
\be
u_\m = u_\1 +\eta q_\alpha K_\m\nn
\ee
For,\\
(i) \textbf{BPS $\eta=1$} \\
\be
u_\m = u_\1 + q_\alpha K_\m\nn
\ee
Now this $K_\m$ will either belong to $[K_{v_+}]$ or $[K_{v_-}]$
For $[K_{v_+}]$ solutions exist without any restriction on $u_\1$. For  $[K_{v_-}]$ 
\be
u_\1-q[|K_{v_-}|]>0\implies u_\1>q\,\mathrm{max}([|K_{v_-}|])
\ee
(ii) \textbf{anti-BPS $\eta=-1$}\\
There is no restriction on $u_\1$ for $[K_{v_-}]$. But for $[K_{v_+}]$ we get,
\be
u_\1-q_\alpha \,[K_{v_+}]>0\implies u_\1>q_\alpha\,\mathrm{max}([K_{v_+}])
\ee
For the nodes where $u_\m=0$, we have,
\be
v_\m =-( u_\1 +\eta q_\alpha K_\m)\nn
\ee
For,\\
(i) \textbf{BPS $\eta=1$} \\
\be
v_\m =-( u_\1 + q_\alpha K_\m)\nn
\ee
For the set $[K_{u_+}]$ there is no  solution.  For  $[K_{u_-}]$ 
\be
u_\1-q_\alpha [|K_{u_-}|]<0\implies u_\1<q_\alpha\,\mathrm{min}([|K_{u_-}|])
\ee
(ii) \textbf{anti-BPS $\eta=-1$}\\
\be
v_\m =-( u_\1 - q_\alpha K_\m)\nn
\ee
There is no solution for  $[K_{u_-}]$. But for $[K_{u_+}]$ we get,
\be
u_\1-q_\alpha\,[K_{u_+}]<0\implies u_\1<q_\alpha\,\mathrm{min}([K_{u_+}])
\ee
The above four cases exhaust all the possibilities. To summarize,
\\
\textbf{BPS solutions}
\be
q_\alpha.\,\mathrm{min}([|K_{u_-}|])>  u_\1> q_\alpha.\,\mathrm{max}([|K_{v_-}|])
\ee
\textbf{anti-BPS solutions}
\be
q_\alpha.\,\mathrm{min}([|K_{u_+}|])>  u_\1> q_\alpha.\,\mathrm{max}([|K_{v_+}|])
\ee

Therefore we can conclude that we have obtained full monopole solution in the CS Yang-Mills  $\widehat{A}$ quiver theory with $\mathcal{N}=3$ supersymmetry and with gauge group $U(N)\times U(N)\times... U(N)$.

\subsection{$\widehat{D}$-type quiver}
\begin{figure}[h]
\centering
\includegraphics[width=4in]{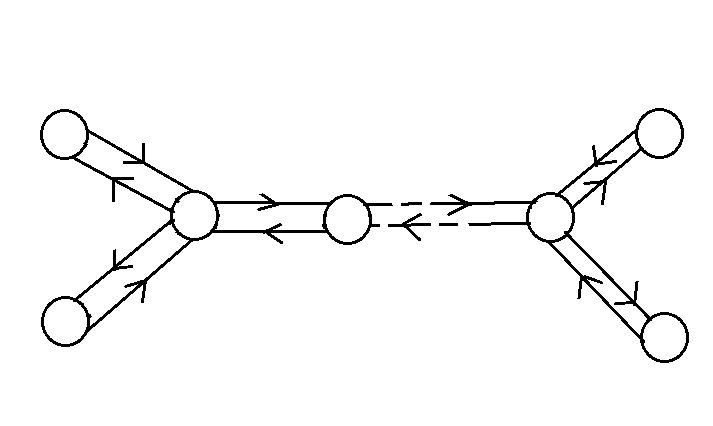}
\put (-23,32) {$\scaleto{N}{5pt}$} \put (-24,116) {$\scaleto{N}{5pt}$}\put (-265,116) {$\scaleto{N}{5pt}$}
\put (-265,40) {$\scaleto{N}{5pt}$} \put (-214,82) {$\scaleto{2N}{5pt}$} \put (-152,80) {$\scaleto{2N}{5pt}$}  \put (-70,82) {$\scaleto{2N}{5pt}$}
\caption{$\widehat D_n$ quiver diagram with gauge group $U(2N)^{n-3}\times U(N)^4$}
\label{Dn_ranks}
\end{figure}
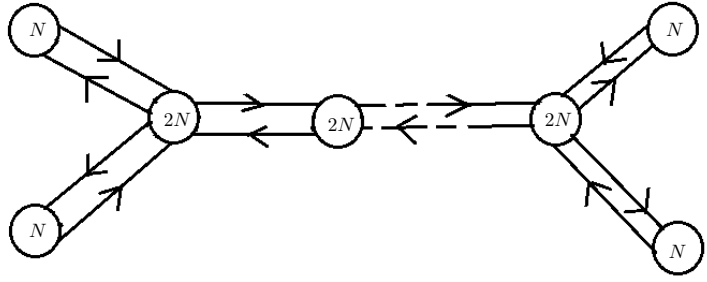
We consider the gauge group of the theory to be $U(2N)^{n-3}\times U(N)^4$\cite{nishioka}\cite{gulotta}\cite{non-toric} since these are the ones whose gravity dual is $\ads\times M_7$ vacua of M-theory, where $M_7$ is a tri-Sasakian manifold. Also this class of  theories have been considered in the context of $\widehat{ADE}$ matrix models since they have a nice large $N$ limit.
Since the ranks of all gauge groups are not equal like the previous case, we need to choose two ansatz for the gauge fields as following,
\be
\label{eqn:gauge_dn}
A_\m &=&\frac{H^{(1)}}{2}\,\,(\pm1-\cos\theta)\,d\varphi,\quad for\,\,\, \m=1,2,3,4\nn\\
A_\m &=&\frac{H^{(2)}}{2}\,\,(\pm1-\cos\theta)\,d\varphi,\quad for\,\,\, \m=5,..., n+1
\ee
where, $H^{(1)}=\mathrm{diag}(q_{\scaleto{1}{4pt}},q_{\scaleto{2}{4pt}},...q_{\scaleto{N}{4pt}}), H^{(2)}=\mathrm{diag}(q'_{\scaleto{1}{4pt}},q'_{\scaleto{2}{4pt}},...q'_{\scaleto{2N}{4pt}})$.  The superscript in $H$ is the co-mark of the associated gauge group.

The logic here is same as the $\widehat{A}$ quiver case, i.e we need to make all the supersymmetry variations of the theory zero for such gauge ansatz with a non trivial variation parameter.  Since the order $\tilde{g}^{-1}$ terms  in $\delta\lambda^{ab}_\m$ is same for $\m=1$ to $n+1$. The analysis is same as before and we  get analogues of \eqref{eqn:con1}\eqref{eqn:phi_an},
\be
\label{monopolesolution_dn}
\phi_{i\m}\Ev_i&=&0,\quad \Ev_3=0,\quad \Ev_1-i\eta \Ev_2=0\nn\\
\phi_{i\m} &=&  -\eta \frac{H^{(1)}}{2}\delta_{i3},\quad for\,\,\, \m=1,2,3,4\nn\\
\phi_{i\m} &=&  -\eta \frac{H^{(2)}}{2}\delta_{i3}, \quad for\,\,\, \m=5,..., n+1
\ee

 The difference occurs in order $\tilde{g}$ terms in $\delta\lambda^{ab}_\m$. Let us analyse them from  \eqref{vec_fermion_variation_dn_internal}-\eqref{vec_fermion_variation_dn_end}
\medskip
\par \underline{$\delta\lambda^{ab}_\m|_{\tilde{g}}=0$, for $\m=1,...4$,}
\be
 \kappa_\m \, i {\phi^b_c}_\m \Ev^{ac}      + i  \big(X^a_\m {X^\g_c}_\m \Ev^{cb} \big)- \frac{i}{2}  (X X^\g)_\m \Ev^{ab}=0\nn
\ee
Contracting with $\sigma_{ab}$ to isolate the $\mathbf{3}$ part we get,
\be
\label{phi_corners}
 {\phi_i}_\m  =      -\frac{1}{2\kappa_\m}\Big(   X_\m \sigma_i {X^\dagger}_\m \Big), \quad for\,\, \m=1,...4
\ee
Similarly,
\par
\underline{${\delta\lambda^{ab}_\5}|_{\tilde{g}}=0$ gives,}

\be
\label{phi_5}
 {\phi_i}_\5  =      -\frac{1}{2\kappa_\5}\Big(   X_\5 \sigma_i {X^\dagger}_\5  - \sum_{\m=1}^2 X^\dagger_\m  \sigma_i X_\m\Big)
\ee
 
 \underline{${\delta\lambda^{ab}_\m}|_{\tilde{g}}=0$, for $\m=6,...n$ gives,}

\be
\label{phi_p}
 {\phi_i}_\m  =      -\frac{1}{2\kappa_\m}\Big(   X_\m \sigma_i {X^\dagger}_\m -{X^\dagger}_{\n} \sigma_i X_{\m-1}\Big), \quad for\,\, \m=6,...n
\ee

\par 
\underline{\textbf{$\delta\lambda^{ab}_{(n+1)}|_{\tilde{g}}=0$ } gives,}

\be
\label{phi_(n+1)}
 {\phi_i}_{(n+1)}  =\frac{1}{2\kappa_{(n+1)}} \sum_{\m=3,4,n}{X^\dagger}_\m  \sigma_i X_\m
\ee

\medskip
\par 
At this stage we have obtained solutions for $A_\m$ and ${\phi_i}_\m$ explicitly.  But to call it a  monopole solution  we need to check if these solutions make all $\delta\xi^a_\m$'s and its complex complex conjugates zero. Let us verify it,

For ${\m=1,2}$ we have,
\be
\delta \xi^{a}_\m  = \slashed{\mathcal{D}} X^{b}_\m \Ev^a_b +\frac{1}{3}X^b_\m\slashed{\nabla}\Ev^a_b + \phi^a_{b\p} \Ev^b_c X^c_\m - X^c_\m \Ev^b_c \phi^a_{b\5} \nn 
\ee
Using Killing spinor equation  \eqref{killing_eqn_s2} above,  we get,

\underline{for $a=1$}
\begin{eqnarray*}
\delta \xi^{1}_\m  
=\Big( \slashed{\mathcal{D}} X^{2}_\m -\frac{1}{2}X^2_\m \gamma^\tau   -\eta \frac{H^{(1)}}{2}  X^2_\m + \eta X^2_\m  \,  \frac{H^{(2)}}{2}\Big)\Ep
\end{eqnarray*}
In the anti-BPS case $\Ep=0$ hence the above supersymmetry variation vanishes. But in the BPS case the following has to be zero,
\begin{eqnarray*}
\Big( \slashed{\mathcal{D}} X^{2}_\m -\frac{1}{2}X^2_\m \gamma^\tau   -\frac{H^{(1)}}{2}  X^2_\m + X^2_\m  \,  \frac{H^{(2)}}{2}\Big)=0
\end{eqnarray*}
To cancel the first two terms we fix the functional dependence of $X^2_\m\sim e^{\eta \tau/2}$.
To cancel the last two terms for  the simplest case, one has to fix the diagonal entries of $H^{(1)}$ and $H^{(2)}$ to be all equal, say $q$ (of course there can be non-trivial choices)
\be
\label{gauge_ansatz_dn}
H^{(1)}&=&\mathrm{diag}(q_1,q_2,...,q_{\scaleto{N}{4pt}}),\quad q_i=q, \quad i=1,..., N\nn\\
\quad H^{(2)}&=&\mathrm{diag}(q_1,q_2,...,q_{\scaleto{2N}{4pt}}), \quad q_i=q, \quad i=1,...2N
\ee
By taking the above ansatz one can check that all the supersymmetry variations $\delta \xi^{a}_\m $ and their complex conjugates $\delta \xi^\g_{a\m} $ will vanish and we take the functional dependence of the hypermultiplet scalars to be $X^1_\m\sim e^{-\eta\tau/2}, X^\g_{1\m}\sim e^{\eta\tau/2},  X^\g_{2\m}\sim e^{-\eta\tau/2}$.
\smallskip

\par Now we can generalise same steps used in the $\widehat{A}$ case to solve for the hypermultiplet scalars.
\par \underline{For $\m=1,...,4$}:
\be
{\phi_1}_\m=0\implies  Z_\m W_\m+ W^\g_\m Z^\g_\m &=& 0\nn\\
{\phi_2}_\m=0 \implies  Z_\m W_\m - W^\g_\m Z^\g_\m&=& 0\nn\\
{\phi_3}_\m=-\eta \frac{H^{(1)}}{2}\implies Z_\m Z^\g_\m - W^\g_\m W_\m &=&\eta H^{(1)}\kappa_\m 
\ee
First two  equations imply that $Z_\m W_\m=0$. Therefore we have the following two cases.
\smallskip
\par  Case 1: $Z_\m=0$
\be
 W^\g_\m W_\m &=&-\eta H^{(1)}\kappa_\m 
\ee
For a positive semi-definite solution we get more constrains on the CS levels associated to the external nodes of the quiver, i.e $\eta\kappa_\m\leq 0 $.
\smallskip
\par Case 2: $W_\m=0$
\be
  Z_\m Z^\g_\m &=&\eta H^{(1)}\kappa_\m 
\ee
For a positive semi-definite solution we get, $\eta\kappa_\m\geq 0 $.

\smallskip

\par \underline{For ${\m=5}$:}
\be
{\phi_1}_\5=0\implies Z_\5 W_\5+ W^\g_\5 Z^\g_\5 -\sum_{\m=1}^2\big(Z_\m^\g W^\g_\m+ W_\m Z_\m\big) &=& 0\nn\\
{\phi_2}_\5=0\implies -iZ_\5 W_\5 +iW^\g_\5 Z_\5^\g -\sum_{\m=1}^2 \big(iZ^\g_\m W^\g_\m -i W_\m Z_\m\big) &=& 0\nn\\
 {\phi_3}_\5=-\eta \frac{H^{(2)}}{2}\implies       Z_\5 Z_\5^\g -W_\5^\g W_\5 - \sum_{\m=1}^2 \big(Z^\g_\m Z_\m- W_\m W^\g_\m\big)&=& \eta\kappa_\5 H^{(2)}\nn
\ee
Using the solutions from the previous the last summed over terms vanish and the remaining can be solved exactly as before.
\smallskip
\par \underline{For ${\m=6,...,n}$:}
This analysis is similar to the $\widehat{A}$ case.
\smallskip
\par \underline{For ${\m=n+1}$:} This case is same as $\m=5$.
\par
While fixing the coefficients of $X$ and $X^\g$'one has to keep in mind that they should satisfy the equations of motions of the gauge fields,
\be
\kappa_{(j)} \epsilon^{mnp}F_{np\m} &=&  X_{(j)}\,\mathcal{D}^m X^\g_{(j)} -\mathcal{D}^m X_{(j)}. X^\g_{(j)},\qquad\qquad \m=1,2,3,4\nn\\
\kappa_{(5)} \epsilon^{mnp}F_{np(5)}&=&   
 \sum_{\m=1}^2 \big( X^\g_{\m}\mathcal{D}^m X_{\m }-\mathcal{D}^m X^\g_{\m }. X_{\m }  \big)
+ \big( X_{\5 }\mathcal{D}^m X^\g_{\5 }  - \mathcal{D}^m X_{\5}.X^\g_{\5 }\big)\nn\\
\kappa_{(n+1)} \epsilon^{mnp}F_{np(n+1)}&=&   
 \sum_{\m=3,4,n} \big( X^\g_{\m}\mathcal{D}^m X_{\m }-\mathcal{D}^m X^\g_{\m }. X_{\m }  \big)\nn\\
\kappa_\m \epsilon^{mnp}F_{np\m}&=& X_\m\,\mathcal{D}^m X^\g_{\m}-\mathcal{D}^m X_\m. X^\g_{\m} - \mathcal{D}^m X^\g_{\n}\, .X_{\n}+X^\g_{\n}\,\mathcal{D}^m X_{\n},\quad \m=6,..., n\nn\\
\ee

\subsection{$\widehat{E}_6$ quiver}
\smallskip
\par Solving for the hypermultuplet scalars for $\hat{E}_6$ is straight forward from the $\widehat{D}$ case.   The  theory we analyse has gauge group $U(2N)^3\times U(N)^3\times U(3N)$. 
\begin{figure}[h]
\centering
\includegraphics[width=5in]{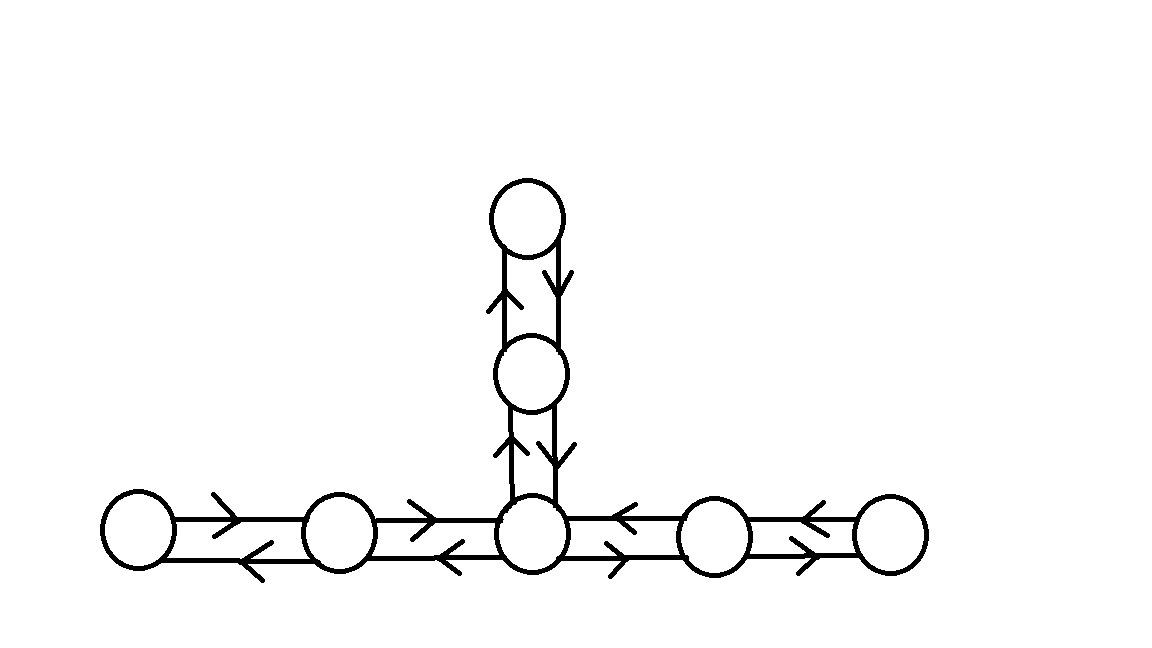}
\put (-324,32) {$N$}\put (-261,32) {$2N$}  \put (-200,32) {$3N$}\put (-199,130) {$N$} \put (-200,82) {$2N$}  
 \put (-144,30) {$2N$}  \put (-84,30) {$N$}
\caption{$\widehat{E}_6$ quiver diagram with gauge group $U(2N)^3\times U(N)^3\times U(3N)$.}
\label{e6_rank}
\end{figure}
In this case we have to start with three different gauge potentials on $\mathbbm{R}\times S^2$, since there are three distinct co-marks associated to the nodes,  such as,
\be
\label{eqn:gauge_e6}
A_\m &=&\frac{H^{(1)}}{2}\,\,(\pm1-\cos\theta)\,d\varphi,\quad for\,\,\, \m=1,3,5\nn\\
A_\m &=&\frac{H^{(2)}}{2}\,\,(\pm1-\cos\theta)\,d\varphi,\quad for\,\,\, \m=2,4,6\nn\\
A_\7 &=&\frac{H^{(3)}}{2}\,\,(\pm1-\cos\theta)\,d\varphi,\quad for\,\,\, \m=7
\ee
where, $H^{(1)}=\mathrm{diag}(q_1,q_2,...q_{\scaleto{N}{4pt}}), H^{(2)}=\mathrm{diag}(q^\prime_1,q^\prime_2,...q^\prime_{\scaleto{2N}{4pt}}), H^{(3)}=\mathrm{diag}(q^{\prime\prime}_1,q^{\prime\prime}_2,...q^{\prime\prime}_{\scaleto{3N}{4pt}})$.

\smallskip
\par  Equating order ${\tilde{g}^{-1}}$ of $\delta\lambda^{ab}_\m$ one chooses the  background scalars to be,
\be
\label{monopolesolution_e6}
\phi_{i\m} &=&  -\eta \frac{H^{(1)}}{2}\delta_{i3},\quad for\,\,\, \m=1,3,5\nn\\
\phi_{i\m} &=&  -\eta \frac{H^{(2)}}{2}\delta_{i3}, \quad for\,\,\, \m=2,4,6\nn\\
\phi_{i\7} &=&  -\eta \frac{H^{(3)}}{2}\delta_{i3}, \quad for\,\,\, \m=7
\ee
From \eqref{vec_fermion_variation_e6},
\par \underline{$\delta\lambda^{ab}_\m|_{\tilde{g}}=0$ for $\m=1,3,5$}

\be
\label{phi_corners_e6}
 {\phi_i}_\m  =      -\frac{1}{2\kappa_\m}\Big(   X_\m \sigma_i {X^\dagger}_\m \Big) 
\ee

\smallskip
\par 
 \underline{$\delta\lambda^{ab}_\m|_{\tilde{g}}=0$ for $\m=2,4,6$}

\be
\label{phi_p_e6}
 {\phi_i}_\m  =      -\frac{1}{2\kappa_\m}\Big(   X_\m \sigma_i {X^\dagger}_\m -{X^\dagger}_{\n} \sigma_i X_{\n}\Big)
\ee
\smallskip
\par 
\underline{$\delta\lambda^{ab}_\7|_{\tilde{g}}=0$ }

\be
\label{phi_(7)_e6}
 {\phi_i}_\7  =\frac{1}{2\kappa_\7} \sum_{\m=2,4,6}{X^\dagger}_\m  \sigma_i X_\m
\ee
To make the variations of the hypermultiplet fermions vanish, lets say 
for ${\m=2,4,6}$, one has to make the following zero,\\
\underline{for $a=1$}
\begin{eqnarray*}
\delta \xi^{1}_\m  
=\Big( \slashed{\mathcal{D}} X^{2}_\m -\frac{1}{2}X^2_\m \gamma^\tau   -\eta \frac{H^{(2)}}{2}  X^2_\m + \eta X^2_\m  \,  \frac{H^{(3)}}{2}\Big)\Ep
\end{eqnarray*}
Equating this to zero we get the functional dependence of the $X$'s  same as before
and,
\be
H^{(2)}=\mathrm{diag}(q,...,q),\quad H^{(3)}=\mathrm{diag}(q,...,q)\nn
\ee
Now looking at other cases, for $\m=1,3,5$ we fix the allowed entries of $H^{(1)}$. To summarize,
 \be
H^{(1)}&=&\mathrm{diag}(q_1,..., q_{\scaleto{N}{4pt}}),\quad q_i=q, \quad i=1,...N\nn\\
H^{(2)}&=&\mathrm{diag}(q_1,..., q_{\scaleto{2N}{4pt}}),\quad q_i=q, \quad i=1,...2N\nn\\
\quad H^{(3)}&=&\mathrm{diag}(q_1,..., q_{\scaleto{3N}{4pt}}), \quad q_i=q, \quad i=1,...3N
\ee
Now it is straight forward to solve \eqref{phi_corners_e6}, \eqref{phi_p_e6}, \eqref{phi_(7)_e6} from the $\widehat{D}$ case.

\bigskip

\section{\label{sec:u1_charge}$\mathrm{U}(1)_R$ charge}\label{sec:u(1) charge}
\smallskip
The quantity that is preserved  by the static background monopole solution in \eqref{monopole_solution}(and their $\widehat{DE}$ analogs)   is the $\mathrm{U}(1)_R$ charge. This charge is not exact since it is abelian and can receive quantum correction under RG flow.
In this section we compute the quantum corrections to the $\mathrm{U}(1)_R$ charge, following method of BKK, who do it by computing the normal ordering constant of the $\mathrm{U}(1)_R$ charge operator. The $\mathrm{U}(1)_R$ charge for ABJM theory is computed in BKK as a consistency check of results found in  \cite{gaiotto2008s}\cite{gaiotto2012notes}. BKK start this computation by considering a simple toy model and then generalising it to ABJM.  Let us consider a single fermion  $\psi(\tau,\Omega)$ on $\mathbbm{R}\times S^2$ in presence of a $\mathrm{U}(1)$ monopole with charge $q$ which is kept at the centre of $S^2$. The equation of motion obeyed by $\psi$ in this system is,
\be
\label{eqn:eom-fermion}
\slashed{\mathcal{D}}\psi+\frac{\eta}{2} q\psi=0,\qquad \eta=\pm 1
\ee
The Dirac operator, $\slashed{\mathcal{D}}=\gamma^\tau \partial_\tau+ \slashed{\mathcal{D}}_S$, where $\slashed{\mathcal{D}}_S$ is the operator on  $S^2$, contains a $U(1)$ monopole solution with charge $q$. 
 The associated conserved current is,
\be
j^\mu=-i\psi^\g\gamma^\mu\psi\nn
\ee
which has the following conserved charge, 
\be
\label{oscillatorcharge}
Q= \,-i\int d\Omega \, \psi^\g\gamma^\tau\psi
\ee
Our goal is to find the normal ordering constant of \eqref{oscillatorcharge}. To solve for $\psi$ from \eqref{eqn:eom-fermion} we use the machinery of monopole spinor harmonics \cite{wu}. The explicit expressions of monopole spinor harmonics  and their properties are given in Appendix C of BKK. Monopole spinor harmonics are eigenfunctions of $\slashed{\mathcal{D}}_S$ and forms a basis on $S^2$ in the presence of a monopole.  Therefore we can expand $\psi(\tau,\Omega)$ in the monopole spinor harmonics basis as,
\be
\label{monopole-harmonics-basis}
\psi(\tau,\Omega)=\sum_m \psi_m(\tau)\Upsilon^0_{qm}(\Omega)+\sum_{jm\varepsilon}\psi^\varepsilon_{jm}(\tau)\Upsilon^\varepsilon_{qjm}(\Omega)
\ee
with $\varepsilon=\pm1$.
The time part gets separated because the monopole solution we are considering does not have a $\tau$ dependence.
$j$ is the total angular momentum  quantum number, taking values
\be
\textit{for $q\neq 0$}\quad
j=\frac{|q|-1}{2}, \frac{|q|+1}{2}, \frac{|q|+3}{2},...\qquad m=-j,-j+1,...,j\nn\\
\textit{for $q= 0$}\quad
j= \frac{|q|+1}{2}, \frac{|q|+3}{2},...\qquad m=-j,-j+1,...,j
\ee
Eigenvalue equations of $\slashed{\mathcal{D}}_S $,
\be
&&\slashed{\mathcal{D}}_S \Upsilon^0_{qm} =0 \qquad\qquad\qquad\,\, \textit{for}\,\, j=\frac{|q|-1}{2},\, q\neq 0\\
&&\slashed{\mathcal{D}}_S \Upsilon^\pm_{qjm} =i\Delta^\pm_{jq}\Upsilon^\pm_{qjm} \qquad\,\, \textit{for}\,\, j=\frac{|q|+1}{2},\frac{|q|+3}{2},..
\ee
where, $\Delta^{\pm}_{jq}=\pm\frac{1}{2}\sqrt{(2j+1)^2-q^2}$. $\Upsilon^0_{qm}$ is called the zero mode since it has zero eigenvalue. For the zero mode, $j=\frac{|q|-1}{2}$ whose  multiplicity is $m=|q|$.
\par Now putting the expansion of $\psi$ into \eqref{eqn:eom-fermion} one obtains the following first order differential equations,
\be
\label{eqn:psi_s2}
  \dot{\psi}_m = - \eta \frac{\abs{q}}{2} \psi_m
   ,\quad
 \frac{d{\psi}^+_{jm}}{d\tau}= (-i\Delta^- - \frac{q}{2})\,\psi^-_{jm},\quad
\frac{d{\psi}^-_{jm}}{d\tau}= (-i\Delta^+ - \frac{q}{2})\,\psi^+_{jm} 
\ee
\underline{Solving the equation of motion for $j=\frac{\abs q-1}{2}$},
\be
\label{zero_mode_energy}
{\psi_m(\tau)} =A\, e^{- \eta\frac{\abs{q}}{2} \tau},\quad \textit{$A$ is an integration constant}
\ee
this says that the zero modes have energy $E_j=\eta\frac{\abs{q}}{2}$.
\smallskip
\par
\underline{Solving the equation of motion for $j=\frac{\abs q-1}{2}+p, p\neq 0,\, \eta=1$}: 

Dividing the last two equations of \eqref{eqn:psi_s2} one obtains $\psi^-_{jm}$ in terms of $\psi^+_{jm}$ as,
\be
{\psi^-_{jm}} =\pm\sqrt{ \frac{i\Delta^+ + \frac{q}{2}}{i\Delta^- + \frac{q}{2}}}\,{\psi^+_{jm}}+C,\quad \textit{$C$ is an integration constant}\nn
\ee
Substituting this back in equation of motion we get,
\be
\psi^-_{jm}(\tau)= (P\, e^{ E_j{\tau}}+ Q\, e^{- E_j{\tau}})\nn
\ee
Similarly one obtains, 
\be
\psi^+_{jm}(\tau)= (R\, e^{ E_j{\tau}}+ S\, e^{- E_j{\tau}})\nn
\ee
where, $P,Q,R,S$ are integration constants and fixed by using normalisation of $\psi$ and canonical anti commutation relations of the operators,  $E_j=j+\frac{1}{2}$ is the energy of the corresponding state.
Quoting the  final solution from BKK,
\be
\label{solution_psi}
  \psi = \sum_m \Bigsbrk{ c_m \, u^0 \, e^{-\frac{\abs{q}}{2}\tau} + d^\dagger_m \, v^0 \, e^{\frac{\abs{q}}{2}\tau} } \Upsilon^0_m
       + \sum_{jm\varepsilon} \Bigsbrk{ c_{jm} \, u^\varepsilon_j \, e^{-E_j\tau} + d^\dagger_{jm} \, v^\varepsilon_j \, e^{E_j\tau} } \Upsilon^\varepsilon_{jm}
       \; \nn\\
        \psi^\g = \sum_m \Bigsbrk{ c^\g_m \, u^0 \, e^{\frac{\abs{q}}{2}\tau} + d_m \, v^0 \, e^{-\frac{\abs{q}}{2}\tau} } \Upsilon^{0\g}_m
       + \sum_{jm\varepsilon} \Bigsbrk{ c^\g_{jm} \, u^{\varepsilon\g}_j \, e^{E_j\tau} + d_{jm} \, v^{\varepsilon\g}_j \, e^{-E_j\tau} } \Upsilon^{\varepsilon\g}_{jm}
       \; 
\ee
(Note: In the expression of $\psi^\g$ we reversed sign of $\tau$ since we are working with Euclidean time.)
 The wave-functions for the, BPS case ($\eta=+1$),
\begin{align}
  u^0 = 1 \; , \;\;
  v^0 = 0 \; , \;\;
  u^+_j = v^+_j =  \tfrac{1}{\sqrt{2}} \; , \;\;
  u^-_j = - v^-_j = \tfrac{1}{\sqrt{2}} \Bigbrk{ \tfrac{q}{2j+1} + i \sqrt{1-\bigbrk{\tfrac{q}{2j+1}}^2} } \; 
\end{align}
anti-BPS case ($\eta=-1$),
\begin{align}
  u^0 = 0 \; , \;\;
  v^0 = 1 \; , \;\;
  u^+_j= -v^+_j  =  -\tfrac{1}{\sqrt{2}} \Bigbrk{ \tfrac{q}{2j+1} + i \sqrt{1-\bigbrk{\tfrac{q}{2j+1}}^2} } \; , \;\;
  u^-_j = - v^-_j = \tfrac{1}{\sqrt{2}} \; 
\end{align} 
Now, the $U(1)_R$ charge can be computed by using point splitting regularisation \cite{bor2}
\be 
\label{point-splitting}
  Q(\beta) = -\frac{i}{2} \int\,d\Omega\: \Bigsbrk{
             \psi^\dagger\bigbrk{\tau+\frac{\beta}{2}} \, \gamma^\tau \, \psi\bigbrk{\tau-\frac{\beta}{2}}
           - \psi\bigbrk{\tau+\frac{\beta}{2}} \, \gamma^\tau \, \psi^\dagger\bigbrk{\tau-\frac{\beta}{2}}
           } \; 
\ee
where $\beta>0$. In the end we take the limit $\beta\to 0$.
Now substituting \eqref{solution_psi} above and using properties of monopole spinor harmonics and canonical anti-commutation relations we get,

\be
 Q(\beta)
&=&\frac{1}{2}\sum_m \Bigsbrk{ c^\g_m  c_{m}  {u^0}^\g u^0 \, (e^{\frac{\abs{q}}{2}{\beta}}+ e^{-\frac{\abs{q}}{2}{\beta}}) -e^{-\frac{\abs{q}}{2}{\beta}}  {u^0}^\g u^0+{v^0}^\g v^0  e^{-\frac{\abs{q}}{2}{\beta}} -d^\g_m d_{m} {v^0}^\g v^0  (e^{-\frac{\abs{q}}{2}{\beta}} +  e^{\frac{\abs{q}}{2}{\beta}})}     \nn\\
&+&\frac{1}{2} \sum_{jm\varepsilon} \Bigsbrk{ c^\g_{jm}c_{jm}  {u^\varepsilon}^\g_j u^\varepsilon_{j} (e^{E_j\beta}+ e^{-E_j\beta})-{u^\varepsilon}^\g_j u^\varepsilon_{j} e^{-E_j\beta} +v^\varepsilon_{j} {v^\varepsilon}^\g_j e^{-E_j\beta} -d^\g_{jm} d_{jm} \, v^\varepsilon_{j} {v^\varepsilon}^\g_j (e^{-E_j\beta}+e^{E_j\beta})}\nn
\ee
whose normal ordered piece at $\beta=0$ is,
\be
Q_1(\beta=0)&=&\sum_m \Big( c^\g_m  c_{m}{u^0}^\g u^0 -d^\g_m d_m  {v^0}^\g v^0\Big)
+\sum_{jm\varepsilon} \Bigsbrk{ c^\g_{jm}c_{jm}  {u^\varepsilon}^\g_j u^\varepsilon_{j}  - d^\dagger_{jm} d_{jm}  \, v^\varepsilon_{j} {v^\varepsilon}^\g_j }
\ee
with a normal ordering constant,
\be
Q_0(\beta)
&=&- \half \sum_{jm\varepsilon} \Bigsbrk{ u^{\varepsilon\dagger}_j u^\Ev_j - v^{\Ev\dagger}_j v^\Ev_j } e^{-\beta E_j} 
\ee
where in the last sum  the zero modes with $j=\frac{\abs{q}-1}{2}$ is also included. Observing that $\sum_\varepsilon u^{\varepsilon\dagger}_j u^\varepsilon = 1$  for every positive energy state($\sim e^{-E_j\tau}$) and $\sum_\varepsilon v^{\varepsilon\dagger}_j v^\varepsilon = 1$ for every negative energy state($\sim e^{E_j\tau}$), we can write,
\be
  Q_0(\beta) = -\frac{1}{2} \sum_{\mathrm{states}} \sign(E) \, e^{-\beta \abs{E}} \; 
\ee
The above quantity would be zero when we have a symmetric spectrum with respect to $E=0$. But after turning on  the scalar fields one finds  that the energy corresponding to the zero mode is, $-\frac{|q|}{2}$  for BPS states and $\frac{|q|}{2}$  for anti-BPS states, i.e the zero mode energy spectrum is not symmetric\footnote{ The energy spectrum  plot can be found in BKK.}
 But for non zero modes both positive and negative energy states are present for a fixed value of $j$. Therefore the normal ordering constant is non zero only in the case of zero modes,
\be
\label{eqn:normal_order_u1}
Q_0=-\eta\frac{\abs q}{2}
\ee
factor $\abs{q}$ arises because of the sum over zero modes, which has multiplicity $|q|$. Bosonic fields do not contribute to the $U(1)_R$ charge because their spectrum is symmetric \cite{bor2}.
\subsection{Application to $\widehat{A}$-type quiver }
Now we can apply the above method to ${\superN}=3$ Yang-Mills deformed CS $\widehat{ADE}$ theories.  We  first compute the $SU(2)_R$ current and then extract the $U(1)_R$ part from that. Under infinitesimal $SU(2)$ transformation   fundamental and anti-fundamental $SU(2)_R$ indices transform as,
\be 
\label{eqn:SU2R-trafo}
  \delta \mathcal{X}^a = i \varepsilon^a_b \mathcal{X}^b
  ,\qquad
  \delta \mathcal{X}_a = - i \varepsilon^b_a \mathcal{X}^\dagger_b
  \; 
\ee
where $\mathcal{X}$ represents a generic field. Hence the fields in the vector multiplet transform as,
\be 
  \delta \lambda^{ab}_\m = i \varepsilon^a_c\, \lambda^{cb}_\m +  i \varepsilon^b_c\, \lambda^{ac}_\m
  ,\qquad
  \delta \phi^a_{b\m} =  i \varepsilon^a_c \, \phi^c_{b\m} -  i \varepsilon^c_b \, \phi^a_{c\m}
  \; 
\ee
The fields in the hypermultiplet transform as,
\begin{align}
\delta X^a_\m &= i \varepsilon^a_b X^b_\m
  \qquad
  \delta X^\g_{a\m} = - i \varepsilon^b_a X^\dagger_{b\m}
  \; &\nn\\
 \delta \xi^a_\m &= i \varepsilon^a_b \xi^b_\m
  \qquad
  \delta \xi^\g_{a\m} = - i \varepsilon^b_a \xi^\dagger_{b\m}&
  \; 
\end{align}
The conserved $SU(2)_R$ current, 
\be \label{eqn:SU2R-Noether-current}
 {(J^\mu)}^{~b}_a&=& \sum_{\m=1}^n \Tr \Big[
   i {X^\dagger_\m}_a \mathcal{D}_\mu X_\m^b - i \mathcal{D}_\mu {X^\g_\m}_a {X^b_\m}  
   -  {\xi^\dagger_a}_\m \gamma^\mu\xi^b_\m
   + \frac{i}{g^2} \mathcal{D}_\mu {\phi^c_a}_\m  {\phi^b_c}_\m\nn\\ &-&  \frac{i}{g^2} \mathcal{D}_\mu {\phi^b_c}_\m  {\phi^c_a}_\m
   + \frac{1}{2g^2} {\lambda_{ac}}_\m \gamma^\mu {\lambda_{bc}}_\m    + \frac{1}{2g^2} {\lambda_{ca}}_\m \gamma^\mu {\lambda_{cb}}_\m
   \Big]
\ee
where we have used $J^\mu={(J^\mu)}^{~b}_a \varepsilon_b^a$.
The $U(1)_R$ component of this current  is obtained by contracting ${(J^\mu)}^{~b}_a$  with $\varepsilon^a_b = (\sigma_3)^{~a}_b$ which gives,
\be \label{eqn:U1R-Noether-current-fermions}
  J^\mu &=& \sum_{\m=1}^n\Tr \Bigsbrk{ - \half \, \zeta^\dagger_\m \gamma^\mu \zeta_\m
                             - \half \, \omega^{\dagger }_\m \gamma^\mu \omega_\m
                             + \chi_{\sigma\m}^\dagger \gamma^\mu {\chi_\sigma}_\m
                           }
  \;
\ee
Now the expression of  charge in \eqref{oscillatorcharge} has become,
\be 
\label{eqn:U1R-charge}
  Q = - i \int \, d\Omega\: \sum_{\m=1}^n\Tr \Bigsbrk{ - \half \, \zeta^\dagger_\m \gamma^\tau \zeta_\m
                                          - \half \, \omega^{\dagger }_\m \gamma^\tau \omega_\m
                                          +\chi_{\sigma\m}^\dagger \gamma^\tau {\chi_\sigma}_\m}
\ee

The equations of motion of the fermions in the far UV limit $\tilde{g}\to 0$,
\begin{align} \label{eqn:eom-fluct-euclid}
  & \slashed{\mathcal{D}} \zeta_\m    + \frac{\eta}{2} \comm{H}{\zeta_\m} = 0 \;  &
  & \slashed{\mathcal{D}} {\chi_\sigma}_\m  + \frac{\eta}{2} \comm{H}{{\chi_\sigma}_\m} = 0 \;   &\nn  \\
  & \slashed{\mathcal{D}} \omega_\m   + \frac{\eta}{2} \comm{H}{\omega_\m} = 0 \;  &
  & \slashed{\mathcal{D}} {\chi_\phi}_\m    + \frac{\eta}{2} \comm{H}{{\chi_\phi}_\m} = 0 \; 
\end{align}
Now, the brilliant observation of BKK which enables us to use the abelian result of previous section in a non-abelian model is the following. For any $mr$-th entry of the  $N\times N$ matrix $\psi$($m,r$ are gauge indices) one can write,
\be 
\label{trick}
\comm{H}{\psi}_{mr} =H_{mp}\psi_{pr}-\psi_{mp}H_{pr} =q_m \delta_{mp} \, \psi_{pr} - \psi_{mp} \, q_p \delta_{pr} = ( q_m - q_r ) \psi_{mr} \; 
\ee
which happens because of the diagonal nature of $H$. Therefore we can treat each component in  \eqref{eqn:eom-fluct-euclid} separately, for example,
\be
 \slashed{\mathcal{D}} \zeta_{\m mr}    + \frac{\eta}{2} q_{mr}\zeta_{\m mr} = 0 \nn 
  \ee
where, $ q_{mr} \equiv q_m - q_r$.
comparing this with \eqref{eqn:eom-fermion} one sees that now  the effective monopole charge is $q_m-q_r$. 
Using the result \eqref{eqn:normal_order_u1} form the previous section and adding contributions from hyper and vector multiplet fermions, we get,
\be
Q^{mon}_R&=& \sum_{\m}^n \sum_{m,r=1}^N\Big[ -\half. \Big(-\eta\frac{\abs {q_{mr}}}{2}\Big)   -\half. \Big(-\eta\frac{\abs {q_{mr}}}{2}\Big)  +1. \Big(-\eta\frac{\abs {q_{mr}}}{2}\Big) \Big]\nn\\[2mm]
&=& 0
\ee
Therefore $U(1)_R$ charges of the monopoles of $\NN=3$ $\widehat{A}$-type quiver gauge theory is zero. This result is as anticipated as in BKK who obtained,
$Q^{mon}_R=\eta\Big(\frac{N_f}{2}-1 \Big)\sum_{m,r=1}^N \abs {q_m -q_r}$,
where $N_f$ is the number of hypermultiplets between two nodes. Therefore for $N_f=2$ which is ABJM we get $Q^{mon}_R=0$. Our result indeed matches with ABJM(i.e $n=2$) as it should.
\subsection{\label{u1_dn}$\widehat{D}$-type quiver }
This case is almost similar to the previous case except for  the external edges. The $U(1)_R$ component of the Noether current is,
\be \label{eqn:U1R-Noether-current-fermions_dn}
  J^\mu &=&\Tr \sum_{\m=1}^n \Bigsbrk{ - \half \, \zeta^\dagger_\m \gamma^\mu \zeta_\m
                             - \half \, \omega^{\dagger }_\m \gamma^\mu \omega_\m}+
                           \Tr \sum_{\m=1}^{n+1}  \chi_{\sigma\m}^\dagger \gamma^\mu {\chi_\sigma}_\m
\ee
The equation of motions in the far UV limit of the hypermultiplet fermions,
\begin{align} \label{eqn:eom-fermion_dn}
\m=5,..,n:\quad   & \slashed{\mathcal{D}} \zeta_\m    + \frac{\eta}{2} \comm{H^{(2)}}{\zeta_\m} = 0 \;  &
\;\;\;\; \m=1,...,4:\quad  & \slashed{\mathcal{D}} \zeta_\m    + \frac{\eta}{2}\big( H^{(1)}{\zeta_\m} -  {\zeta_\m} H^{(2)}\big) = 0  \;   &\nn  \\
  & \slashed{\mathcal{D}} \omega_\m   + \frac{\eta}{2} \comm{H^{(2)}}{\omega_\m} = 0 \;  &
  & \slashed{\mathcal{D}} \omega_\m    + \frac{\eta}{2}\big( H^{(1)}{\omega_\m} -  {\omega_\m} H^{(2)}\big) = 0 \; 
\end{align}
equations of motions of the vector multiplet fermions,
\begin{align} 
\m=5,..,n+1\quad   & \slashed{\mathcal{D}} \chi_\sigma   + \frac{\eta}{2} \comm{H^{(2)}}{\chi_{\sigma\m}} = 0 \;  &
 \m=1,..,4\quad  & \slashed{\mathcal{D}} {\chi_\sigma}_\m  + \frac{\eta}{2} \comm{H^{(1)}}{{\chi_\sigma}_\m} = 0 \;   &\nn  \\
  & \slashed{\mathcal{D}} \chi_{\phi\m}   + \frac{\eta}{2} \comm{H^{(2)}}{\chi_{\phi\m}} = 0 \;  &
  & \slashed{\mathcal{D}} {\chi_\phi}_\m    + \frac{\eta}{2} \comm{H^{(1)}}{{\chi_\phi}_\m} = 0 \; 
\end{align}
It is straight forward to generalise \eqref{trick} for the internal edges, i.e $\m=5,..., n$ except now we have $q_m=q_r$. This implies that the fermions associated to the internal edges do not interact with the monopole  hence do not contribute to the $U(1)_R$ charge.  For the external edges, 
\begin{eqnarray}
\label{trick_dn}
\big(H^{(1)}{\zeta_\m} -  {\zeta_\m} H^{(2)}\big)_{r\hat{s}}=H^{(1)}_{rm}\,{\zeta_\m}_{m\hat{s}} -  {\zeta_\m}_{r\hat{p}} H^{(2)}_{\hat{p}\hat{s}}&=&q\, \delta_{rm}\,{\zeta_\m}_{m\hat{s}} - q\, {\zeta_\m}_{r\hat{p}} \delta_{\hat{p}\hat{s}}\nn\\
&=&q\,\,{\zeta_\m}_{r\hat{s}} - q\, {\zeta_\m}_{r\hat{s}}=0 
\end{eqnarray}
 where the hatted gauge indices imply that $\zeta_\1$ is an $N\times 2N$ matrix and so on. Therefore the hyperinos associated with the  external  edges also do not contribute to the charge. This happens because of our choice of the gauge ansatz in \eqref{gauge_ansatz_dn}. 
Therefore we find that the $U(1)_R$ charges of the monopole operators in the $\widehat{D}$ case do not receive any quantum corrections. This readily implies that $U(1)_R$ charges of the monopole operators in the $\widehat{E}_6$ case also do not receive any quantum corrections.
\bigskip
\section{\label{sec:su2_charge}$\su_R$ charges of the monopole operators}
In this section we compute the quantised $\su_R$ charges of the monopole operators which is the main goal of this note. In the previous sections we have shown that  
the Yang-Mills deformations  to the superconformal CS theories under $\widehat{ADE}$ classification    preserve $\NN=3$ supersymmetry. Also we have  explicitly solved for the hypermultiplet scalars, which implies that monopole solution exists for more than two gauge groups.  Therefore it is legitimate to use BKK method, in $\widehat{ADE}$ quiver theories as well to compute the $SU(2)_R$ charges. We report in this note that, the smallest possible representation of the $SU(2)_R$  charge is zero. This  result was  anticipated in \cite{bkk} for the $\widehat{A}$-type quiver and  here it  is verified by explicit calculations. 
\smallskip
\par
Let us first briefly describe the method used in BKK which is collective co-ordinate quantisation method  to obtain the ${SU}(2)_R$ charge of the BPS background. 
The first step in obtaining the charges is to generalise the BPS background in \eqref{monopole_solution}(and their analogs for $\widehat{DE}$) to arbitrary $SU(2)_R$ orientation, 
\be
\phi_{i\,(j)} = - \frac{{H}}{2}\,n_i
\ee
where, $n_i$ is the unit vector on  two sphere $\su_R/{U}(1)_R$. It can be checked that these are the bosonic zero modes by computing the equation of motion for $\phi_\m$.  In the previous section $\phi_{i\m}$ was chosen in the $\mathbf{3}$ direction with $n_i=\delta_{i3}$ . Therefore the conserved quantity with that background was $U(1)_R$ charge. 
 Now, to specify the  the collective co-ordinates one makes the unit vector $n_i$ time dependent,
\be
\phi_{i\,(j)} = - \frac{H}{2}\,n_i(\tau)
\ee
$n_i(\tau)$ is the collective co-ordinate of the BPS background. Now the global $\su_R$ symmetry can act on this background whose action is to  rotate $n_i(\tau)$ on the two sphere. Since the collective co-ordinate is interacting with the fermions of the theory the motion is not free. The effect of these interactions are obtained by calculating the effective action of the collective co-ordinate, i.e by integrating out the fermions. From the effective Lagrangian  we compute the conserved quantity by Noether's procedure. The conserved quantity will be the angular momentum since $SU(2)_R$ acts as rotational symmetry. After computing the angular momentum we compute its quantized values which are the $SU(2)_R$ charges of the BPS-monopole operators in $\widehat{ADE}$ theories.

\subsection{Quantum effective action of the collective co-ordinate}
\smallskip
In this section we present the details for obtaining the effective action of the collective co-ordinates. Following BKK  we first consider a simple model with one fermion in an abelian gauge theory. Then we will generalise the results of the former to Yang-Mills deformed quiver theories. 
\smallskip
\par Let us consider a fermion $\psi^a(\tau)$ in the fundamental representation of $SU(2)$,  on $\mathbbm{R}\times {S}^2$ with the action,
\be
\label{spatial_dependence}
\mathcal{S}=\int\, d\tau\, d\Omega \big(-i\psi^\g_a \slashed{\mathcal{D}} \psi^a- \frac{iq}{2}n_i(\tau)\psi^\g_a\,(\sigma_i)^a_{\,\,b}\, \psi^b  \big)
\ee
where the operator,
\be
 \slashed{\mathcal{D}}=\gamma^\tau\partial_\tau+\slashed{\mathcal{D}}_\mathrm{S} =\gamma^\tau\partial_\tau+\gamma^\theta\partial_\theta+\gamma^\varphi \nabla_\varphi+\gamma^\varphi A_\varphi\nn
 \ee 
 on $\mathbbm{R}\times S^2$ contains the abelian generalisation of the monopole background \eqref{monopole_solution} with monopole charge $q$.
Now to compute the effective action, we expand $\psi(\tau,\Omega)$ in monopole spinor harmonics basis as \eqref{monopole-harmonics-basis}.
The full action after substituting \eqref{monopole-harmonics-basis} in \eqref{spatial_dependence} and using properties monopole harmonics,
\be
\mathcal{S}
&=& 
\sum_{m}\int\, d\tau\,  \Bigg(-i \,\psi^\g_{am}\partial_\tau {\psi}^a_m
-\frac{iq}{2}\,\sign(q)\, n_i(\tau)\, \, \psi^\g_{am}.(\sigma_i)^a_{\,\,b}.  \psi^b_m(\tau)\Bigg)\nn\\
&+&\sum_{j m \varepsilon} \int\, d\tau\,  \Bigg(-i{\psi^\varepsilon}^\g_{aj m}\partial_\tau  {\psi}^{a\varepsilon}_{jm}\,+ \, \Delta^\varepsilon_{jq}\,{\psi^{-\varepsilon}}^\g_{aj m}\psi^{a\varepsilon}_{jm}
-\frac{iq}{2}n_i(\tau) {\psi^{-\varepsilon}}^\g_{aj m}.(\sigma_i)^a_{\,\,b}. \psi^{b\varepsilon}_{jm} \Bigg)
\ee
 The orthogonality property of monopole harmonics ensures that modes with different $(jm)$ values  and zero modes do not couple to each other. Therefore, the effective action can be computed easily for each $(jm)$ separately. 

The effective action for this system,
\be
\label{effective_action}
e^{-\Gamma(\vec{n})}&=&\int\, [d\psi^\g_m]\,[d\psi_m]\, e^{-\mathcal{S}[\psi_m,\psi^\g_m]}\int\, [d\psi^\g_{jm}]\,[d\psi_{jm}]\, e^{-\mathcal{S}[\psi_{jm},\psi^\g_{jm}]}\nn\\
&=& \det \big(i\, \delta^b_a \partial_\tau\,  - \frac{iq}{2}n_i(\tau)\,(\sigma_i)_a^{\,\,b}\,   \big).\det \begin{pmatrix}\vspace{.5cm} i\delta^b_a\partial_\tau & -\Delta^--\frac{iq}{2}n_i(\tau)\,(\sigma_i)_a^{\,\,b}\\ -\Delta^+-\frac{iq}{2}n_i(\tau)\,(\sigma_i)_a^{\,\,b} & i\delta^b_a\partial_\tau \end{pmatrix}\nn\\[4mm]
\implies \Gamma(\vec{n})&=& -\ln \det \big(i\, \delta^b_a \partial_\tau\,  - \frac{iq}{2}n_i(\tau)\,(\sigma_i)_a^{\,\,b}\,   \big)-\ln \det \begin{pmatrix}\vspace{.5cm} i\delta^b_a\partial_\tau & -\Delta^--\frac{iq}{2}n_i(\tau)\,(\sigma_i)_a^{\,\,b}\\ -\Delta^+-\frac{iq}{2}n_i(\tau)\,(\sigma_i)_a^{\,\,b} & i\delta^b_a\partial_\tau \end{pmatrix}\nn\\
\ee

To evaluate this functional determinant  first we write the general form of effective action using derivative expansion in $n_i(\tau)$,
\be
\label{effective_action_general}
\Gamma(\vec{n})=\int \, d\tau\Big( -V_{eff}(\vec{n}) + i\dot{n}_i A_i(\vec{n})+\frac{1}{2}\dot{n}_i\dot{n}_j B_{ij}(\vec{n})+...\Big)
\ee

Now we will expand both \eqref{effective_action} and \eqref{effective_action_general} and compare to find $A_i(\vec{n})$. We are keeping terms upto first order derivative in $n_i$, because the higher orders will be suppressed(in the far UV limit $\tilde{g}\to 0$) by a term proportional to $\frac{1}{\tilde{g}^2}\dot{n}_i^2$  which comes from the bosonic kinetic term in the action.
\smallskip
\par Now, to expand the effective action  we write,
\be
n_i(\tau)=\mathring{n}_i+\tilde{n}_i(\tau)
\ee
where, $\mathring{n}_i$ is a constant, satisfying $\mathring{n}^2=1$ and $\tilde{n}_i(\tau)$ is a small fluctuation. Expanding \eqref{effective_action_general} around $\mathring{n}_i$,
\be
\label{eff_acn_expansion}
\Gamma(\vec{n})
&=& \int \, d\tau\Big( -V_{eff}(\mathring{\vec{n}})- \tilde{n}_i \partial_i V_{eff}(\mathring{\vec{n}}) -\frac{1}{2} \tilde{n}_i \tilde{n}_j \partial_i\partial_j V_{eff}(\mathring{\vec{n}}) + {i\dot{\tilde{n}}_i A_i(\mathring{\vec{n}}) + i\dot{\tilde{n}}_i {\tilde{n}}_j \partial_j A_i(\mathring{\vec{n}})}\nn\\
&+&\frac{1}{2}\dot{\tilde{n}}_i \dot{\tilde{n}}_j B_{ij}(\mathring{\vec{n}})+...\Big)
\ee
 The above says that to  determine $A_i$ we will have to look at the terms with two powers of  $\tilde{n}_i$ with one derivative. Such term is denoted by $\Gamma_{(2,1)}(\vec{n})$. We present the final result, details of which can be found in BKK.
 \be
\Gamma_{(2,1)}(\vec{n})
&=&-\frac{i}{4}  \int d\tau \epsilon_{ijk} \dot{\tilde{n}}_i.\tilde{n}_j.  \mathring{n}_k\, \frac{1}{ |\vec{\mathring{n}}|^3} \nn\\
&+&\int d\tau\,\int 
\frac{d\omega}{2\pi} \,\, \frac{2i\epsilon_{ijk} \dot{\tilde{m}}_i\tilde{m}_j\mathring{m}_k (\Delta^+ +\Delta^-)\Big( 2\omega^2
+ {\Delta^+}^2+{\Delta^-}^2+2\mathring{m}^2\Big)\omega}{\big(\omega^4+2(\Delta^+\Delta^--\mathring{m}^2) \omega^2+{\Delta^+}^2{\Delta^-}^2+({\Delta^+}^2+{\Delta^-}^2)\mathring{m}^2+ \mathring{m}^4\big)^2}\nn\\
\ee
where, $\frac{q}{2}\mathring{n}_i:= \mathring{m}_i$, $\frac{q}{2}\tilde{n}_i:= \tilde{m}_i$, $\omega$ is the energy.  The contribution to the effective action from non-zero modes cancels  because of the fact that $\Delta^+=-\Delta^-$.

comparing the above with, \eqref{eff_acn_expansion} we get,
\be
\partial_i A_j(\vec{n})-  \partial_j A_i(\vec{n})= \frac{|q|}{2}\,  \epsilon_{ijk}   \, \frac{\vec{n}_k}{ |\vec{n}|^3}
\ee
The factor $|q|$ occurs because of the sum over zero modes which has multiplicity $m=2j+1=|q|$.

\subsection{Application to $\widehat{A}$-type quiver}
 Now we are ready to generalise the previous result in our case.  The relevant part of the action in \eqref{eqn:radial_action_an} and \eqref{eqn:radial_action_int_an}  for computing effective action is,
\begin{align}
\mathcal{S}
 = \int d\tau d\Omega\,\,\sum_{\m=1}^n \mathrm{tr} \Bigg(\,
   - i \xi^\dagger_\m \slashed{\mathcal{D}} \xi_\m
   + \frac{i}{2} {\lambda^{ab}}_\m &\slashed{\mathcal{D}} {\lambda_{ab}}_\m -
 \frac{i}{2} n_i(\tau) {\xi^\dagger_a}_\m {(\sigma_i)}^a_{\,\,\,b} [H, \xi^b_\m ]\nn\\& -  \frac{i}{4} n_i(\tau) {\lambda_{ab}}_\m{(\sigma_i)}_c^{\,\,\,b} \comm{{ H}}{\lambda^{ac}_\m} \Bigg)&
\end{align}

  Now, to take care of the gauge indices we apply the same trick as \eqref{trick} and
rewriting the action with gauge indices,
\be
  \mathcal{S}
 = \int d\tau d\Omega\,\,\sum_{\m=1}^n\sum_{m,r=1}^N  \Bigg(\,
   &-& i \xi^\dagger_{\m rm} \slashed{\mathcal{D}} \xi_{\m mr}
 -  \frac{i}{2} n_i(\tau)  ( q_m - q_r ) {\xi^\dagger_a}_{\m rm} {(\sigma_i)}^a_{\,\,\,b}\xi^b_{\m mr}\nn\\
  &-&  i (\lambda_{1a\m})_{rm} \slashed{\mathcal{D}} (\lambda^{1a}_\m)_{mr}   +\frac{i}{2} n_i(\tau) (q_m-q_r)({\lambda^\g_{1a}}_\m)_{rm}  {(\sigma_i)}^{a}_{\,\,\,b} ({\lambda^{1b}_\m})_{mr} \Bigg)\nn\\
\end{eqnarray}
To write the vector-multiplet fermion action in the above form so that we can use the result from the abelian case, we have defined the following similarly as BKK,
\be
\lambda_{11}=-{\lambda^\g_{11}},\,\lambda_{22}=\lambda^{11},\,\lambda_{21}=-\lambda^{12},\,\lambda^{21}=\lambda^\g_{12},\,\lambda_{12}=-\lambda^\g_{12},\,\lambda^{22}=-\lambda^\g_{11}\nn
\ee
The above action is similar to \eqref{spatial_dependence} except for the vector multiplet where the sign of interaction term is changed. 
Now we can treat each matrix elements of the fields  as abelian fields and use the   result from previous section to compute the effective action. 
The only modification we have to make is to put a negative sign in the final result for vector multiplet fermions. We obtain,
 \be
\left.\partial_i A_j(\vec{n})-  \partial_j A_i(\vec{n})\right|_{\mathrm{hyper}}+\left.\partial_i A_j(\vec{n})-  \partial_j A_i(\vec{n})\right|_{\mathrm{vector}}&=&n\sum_{m,r=1}^N\big( \frac{|q_m-q_r|}{2} -\frac{|q_m-q_r|}{2}\big)\epsilon_{ijk}   \, \frac{n_k}{ |\vec{n}|^3}\nn\\[3mm]
&=&0
\ee
which means the induced monopole charge, in the $SU(2)_R$ moduli space, due to fermionic interaction is zero. This happens because of the field configuration of $\widehat{A}$-quiver. Quoting the result of BKK who obtained,
 \be
\left.\partial_i A_j(\vec{n})-  \partial_j A_i(\vec{n})\right|_{\mathrm{hyper}}+\left.\partial_i A_j(\vec{n})-  \partial_j A_i(\vec{n})\right|_{\mathrm{vector}}=(N_f-2)\sum_{m,r=1}^N  \big( \frac{|q_m-q_r|}{2}\big)\epsilon_{ijk}   \, \frac{n_k}{ |\vec{n}|^3}
\ee
which is zero for ABJM where $N_f=2$ and hence our result is consistent.
\smallskip
\par 
The effective action adding all contributions from bosons and fermions,
\be
\label{eff-action}
\Gamma(\vec{n})=\int \, d\tau \Big( \frac{1}{2}M\, \dot{n}^2_i+\lambda(n^2_i-1)  \Big)
\ee
where,
\be
M=\frac{n}{2\tilde{g}^2}\,\,\mathrm{tr} (H^2)=\frac{n}{2{g}^2}e^{-2\tau}\Tr H^2
\ee
The above action can be thought of as a free particle of mass $M$ moving on a unit sphere due to the presence of  Lagrange multiplier(last term) in \eqref{eff-action}. The conserved angular momentum,
\be
L= iM\vec{n}\times \dot{\vec{n}}
\ee
whose quantized values are $l=0,1,2...$. Now, to find the conformal dimension of the monopole operators we solve the Schrodinger equation and read off the scaling dimension from the wave function by using state operator correspondence at IR. It can be checked that to get the
correct behaviour of the wave function one has to include  second order correction($\Gamma_{2,2}(\vec{n})$) in the  effective action.  By doing so we recover the correct exponential behaviour of the wave function and  read off the conformal dimension of the monopole operator  by using state operator correspondence.  This  also verifies that the lowest energy state $l=0$ is BPS in our case and hence the lowest possible value of the conformal dimension is zero.

\subsection{Application to $\widehat{D}$-type quiver}
\smallskip
It is easy to see now that in $\widehat{D}$ case everything from the previous section follows for the internal edges. For the external edges also one can generalise the results as follows.
\smallskip
\par 
The relevant part of the action for charge computation is,
\begin{eqnarray}
\mathcal{S}  &=& \int d\tau\, d\Omega\,\, \Tr \Bigg( \sum_{\m=1}^{n+1}
 \frac{i}{2} \lambda^{ab}_\m\, \slashed{\mathcal{D}} {\lambda_{ab}}_\m   + \sum_{\m=1}^{n+1}\frac{i}{2}{\lambda_{ab}}_\m[{\phi^b_c}_\m,\lambda^{ac}_\m]\nn\\
&+& \sum_{\m=3}^{4}\Big[
-i    \xi^\g_\m \slashed{\mathcal{D}} \xi_\m 
 +i {\xi^\g_a}_\m {\phi^a_b}_\m \xi^b_\m-i {\xi^a}_\m {\phi^b_a}_{(n+1)} {\xi^\g_b}_\m\Big]\nn \\
&+& \sum_{\m=1}^{2}\Big[-i\xi^\g_\m \slashed{\mathcal{D}} \xi_\m -i \xi^a_\m {\phi^b_a}_\5 {\xi^\g_b}_\m +i {\xi^\g_a}_\m {\phi^a_b}_\m {\xi^b}_\m \Big]\nn\\
&+& \sum_{\m=5}^{n}\Big[ -i \xi^\g_\m \slashed{\mathcal{D}} \xi_\m -i \xi^a_\m {\phi^b_a}_{\p} {\xi^\g_b}_\m +i {\xi^\g_a}_\m {\phi^a_b}_\m \xi^b_\m   \Big]\Bigg) 
\end{eqnarray}
The last summation is same as $\widehat{A}$ case. Rest of the interaction terms for the hypermultiplets after substituting the monopole solution for the gauge fields,
\begin{eqnarray}
&&\int d\tau\, d\Omega \,\, \Tr \Bigg( 
 \sum_{\m=1}^{4}\Big[
-\frac{i}{2} n_i(\tau){\xi^\g_a}_\m H^{(1)} (\sigma_i)_{\,\,b}^a  \xi^b_\m + \frac{i}{2} n_i(\tau) {\xi^\g_a}_\m (\sigma_i)_{\,\,b}^a{\xi^b}_\m H^{(2)}  \Big] \Bigg)\nn\\
&=& \sum_{\m=1}^{4}\sum_{s=1}^{N}\sum_{\hat{r}=1}^{2N}\Big(-\frac{iq}{2} n_i(\tau){\xi^\g_a}_{\m \hat{r}s} \, (\sigma_i)_{\,\,b}^a  \xi^b_{\m s\hat{r}} + \frac{iq}{2} n_i(\tau) {\xi^\g_a}_{\m \hat{r}s} (\sigma_i)_{\,\,b}^a{\xi^b}_{\m s\hat{r}}  \Big)=0
\end{eqnarray}
Therefore the external hypermultiplet fermions do not contribute to the charge. By doing same manipulations as above it is easy to see that 
the hypermultiplets associated to the internal edges also do not contribute to the charge. This can be anticipated from the $\widehat{A}$ case  if we set $q_m=q_r$. Similarly the action for the vector multiplet fermions  after substituting the BPS background is,
\be
\mathcal{S}&=& \int d\tau d\Omega\,\,\Tr\Bigg(\sum_{\m=1}^4  \,
 -  i \lambda_{1a\m} \slashed{\mathcal{D}} \lambda^{1a}_\m   +\frac{i}{2} n_i(\tau) {\lambda^\g_{1a}}_\m  {(\sigma_i)}^{a}_{\,\,\,b} \comm{{ H^{(1)}}}{\lambda^{1b}_\m} \nn\\
 &+&\sum_{\m=5}^{n+1} 
 -  i \lambda_{1a\m} \slashed{\mathcal{D}} \lambda^{1a}_\m   +\frac{i}{2} n_i(\tau) {\lambda^\g_{1a}}_\m  {(\sigma_i)}^{a}_{\,\,\,b} \comm{{ H^{(2)}}}{\lambda^{1b}_\m} \Bigg)
\ee
Now the commutator in above expression,
\begin{eqnarray*}
\comm{{ H^{(2)}}}{\lambda^{1b}_\m}_{rs}=  H^{(2)}_{rp}\lambda^{1b}_{\m ps} - \lambda^{1b}_{\m rp} H^{(2)}_{ps}=
q\delta_{rp}\lambda^{1b}_{\m ps} -q \lambda^{1b}_{\m rp} \delta_{ps}=q\lambda^{1b}_{\m rs} -q \lambda^{1b}_{\m rs}=0 
\end{eqnarray*}

which implies that they  do not contribute to the $SU(2)_R$ charge. Therefore,
\be
\partial_i A_j(\vec{n})-  \partial_j A_i(\vec{n})|_\mathrm{hyper}+ \partial_i A_j(\vec{n})-  \partial_j A_i(\vec{n})|_\mathrm{vector}=0
\ee

The same logic applies in the $\widehat{E}_6$ case as well and we get same  result as above.

\section{Discussion} 
 
To summarise our results,

\smallskip
\begin{itemize}
\item In this note we have constructed actions and  supersymmetry variations of three dimensional $\NN=3$ Yang-Mills deformed CS quiver gauge theories with $\widehat{A}_{n-1}, \widehat{D}_n, \widehat{E}_6$ quiver diagrams. These theories flow to a conformal fixed point in the IR via RG flow.

\item We have obtained $\frac{1}{3}$  BPS and anti-BPS monopole solutions in the $\widehat{A}_{n-1}$ quiver  case  with  equal ranks of all gauge groups. 

\item  We have obtained  $\frac{1}{3}$  BPS and anti-BPS monopole solutions in the  $\widehat{D}_n$ quiver theory where we choose the gauge group to be $U(N)^4\times U(2N)^{n-3}$ which has been studied extensively in the context of matrix models and have a dual  M-theory description. Similarly in the $\widehat{E}_6$ case we find BPS and anti-BPS monopole solution with a gauge group $U(N)^3\times U(2N)^3\times U(3N)$.
\item We find that the quantum corrections to the $U(1)_R$ charges  are zero for each quiver. In the $\widehat{A}$ case this happens because of the field content of the theory, i.e contribution to the $U(1)_R$ charge coming from hypermultiplet fermions precisely cancel the contribution from vector multiplet fermions.  In the $\widehat{DE}$ case this happens because of our choice of gauge ansatz which is equally charged under all $U(1)$ factors of the gauge groups.
\item  We find that the lowest possible value of quantised  $SU(2)_R$ charge for each quiver theory is zero. In the $\widehat{A}$ case, it is observed by calculating the  path integral for the adjoint fermions that they give negative contribution to the R-charge and therefore cancelling the positive contribution from hypermultiplet fermions and making net  quantum correction zero.  The contribution from the adjoint fermions wouldn't have been captured in the IR theory, since thy are not dynamical in the IR. In the $\widehat{DE}$ case we find $SU(2)_R$ charges to be zero for the same reasons explained in the previous point.

\end{itemize}

\smallskip

\par 
Our result is similar to ABJM theory which has monopole operator of zero conformal dimension. In fact these monopole operators  are the ones needed to match the spectrum of ABJM theory with dual gravity theory.
\smallskip

\par Some of the interesting questions that maybe worth exploring are as follows:
\begin{itemize}

\item
It is well known that in ABJM theory there is a supersymmetry enhancement from $\NN=6$ to $\NN=8$ for $k=1,2$. In this phenomena monopole operators   played an important role. It will be interesting to check
  if there is any supersymmetry enhancement in $\widehat{ADE}$ theories as well. 

  \item To the best of our knowledge the brane construction for exceptional quivers is not well understood \cite{kap_dn}. One can try to see if the results derived here are useful for brane engineering\cite{vafa}.

\item In interacting CFT's, monopole operators are usually studied via state operator correspondence, where the monopole operators become states on $\mathbbm{R}\times S^2$(considering three dimensional theories), which provides quantized flux through $S^2$ due to a monopole  kept at the centre of $S^2$\cite{kapustin}.  Then one quantizes the theory in the monopole background and find several quantities like scaling dimensions, superconformal index. 
 A recent study on monopole operators in CS matter theories in \cite{assel2018} proposes a prescription to describe the monopole operators as local operators directly on $\mathbbm{R}^3$. In a $\NN=2$ abelian SQED with single charged chiral multiplet and a CS term with level $k$, it can be done  by giving a singular profile to the bosonic and fermionic fields in the theory along with the singular gauge field at the insertion point, keeping in mind that they should be consistent with the equations of motion and Gauss law constraints.  They find  $\frac{1}{4}$-BPS monopoles on $\mathbbm{R}^3$ and compute their dimensions.  As suggested in \cite{assel2018} that this method is applicable to continuous deformation of ABJM theory as it won't affect the discrete global charges of the monopole operators in the theory. Therefore it would be nice to apply this method in $\widehat{ADE}$ theories as a consistency check. 
 \end{itemize}


\begin{center}
\textbf{Acknowledgements}
\end{center}
I am grateful to my advisor Chethan N. Gowdigere for suggesting me the problem. I would like to thank Bobby Ezhuthachan and Chethan N. Gowdigere for collaboration at the initial stage, many helpful discussions and reading the manuscript.
I would like to thank Ashoke Sen and Palash Dubey for helpful discussions. 
I would like to thank RKMVERI and HRI for the warm hospitality where some part of the work was done.

\appendix
\section{Notations and conventions}
\label{appendix:A}
We use  similar conventions as   BKK. $\alpha,\beta=1,2$ are spinor indices raised and lowered  from the left, $\psi^\alpha = \epsilon^{\alpha\beta} \psi_\beta$ and $\psi_\alpha = \epsilon_{\alpha\beta} \psi^\beta$, with $\epsilon^{12} = -\epsilon_{12} = 1$.  $a,b,c,d=1,2$ are R-symmetry indices which are raised and lowered by $\su_R$ metric $\epsilon^{ab}$, with $\epsilon^{12}=\epsilon_{21}=+1$. To contract the spinor indices we use NW-SE convention.
 Inner products on superspace in three dimension
$\theta^2=\theta^\alpha\theta_\alpha,
\theta\bar{\theta}=\theta^\alpha\bar{\theta}_\alpha,
\bar{\theta}^2=\bar{\theta}^\alpha\bar{\theta}_\alpha, \theta \gamma^\mu\bar{\theta}=\theta^\alpha {(\gamma^\mu)}_\alpha^{\hspace{.1cm}\beta}\theta_\beta$. For doing the superspace integral we use,
\begin{eqnarray}
\theta^\alpha\theta^\beta=-\frac{1}{2}\epsilon^{\alpha\beta}\theta^2,\quad
\bar{\theta}^\alpha\bar{\theta}^\beta=-\frac{1}{2}\epsilon^{\alpha\beta}\bar{\theta}^2
\end{eqnarray}
The super covariant derivatives in $x$ basis,
\begin{eqnarray}
D_\alpha &=& \frac{\partial}{\partial\theta^\alpha} -i  (\gamma^\mu)_\alpha^{\,\,\beta}\, \bar{\theta}_{\beta}\frac{\partial}{\partial x^\mu}\nn\\
\bar{D}_\alpha &=& -\frac{\partial}{\partial\bar{\theta}^\alpha} +i  (\gamma^\mu)_\alpha^{\,\,\beta}\, \theta_{\beta}\frac{\partial}{\partial x^\mu}
\end{eqnarray}
The Fierz identities are,
\be
 (\psi_1\psi_2)(\psi_3\psi_4)&=&-\frac{1}{2}(\psi_1\psi_4)(\psi_3\psi_2)-\frac{1}{2}(\psi_1\gamma^\mu\psi_4)(\psi_3\gamma_\mu\psi_2)\nn\\
 (\psi_1\psi_2)(\psi_3\gamma^\mu\psi_4)&=&-\frac{1}{2}(\psi_1\gamma^\mu\psi_4)(\psi_3\psi_2)-\frac{1}{2}(\psi_1\psi_4)(\psi_3\gamma^\mu\psi_2)-\frac{1}{2}\epsilon^{\mu\nu\rho}(\psi_1\gamma_\nu\psi_4)(\psi_3\gamma_\rho\psi_2)\nn\\
(\psi_1\gamma^\mu\psi_2)(\psi_3\gamma^\nu\psi_4)&=&  -\frac{1}{2}g^{\mu\nu}(\psi_1\psi_4)(\psi_3\psi_2)+\frac{1}{2}g^{\mu\nu}(\psi_1\gamma^\rho\psi_4)(\psi_3\gamma_\rho\psi_2) -(\psi_1\gamma^{(\mu}\psi_4)(\psi_3\gamma^{\nu)}\psi_2)\nn\\&-&\frac{1}{2}\epsilon^{\mu\nu\rho}\Big[(\psi_1\gamma_\rho\psi_4)(\psi_3\psi_2)-(\psi_1\psi_4)(\psi_3\gamma_\rho\psi_2)\Big]
 \ee
\smallskip
\textbf{On $\boldsymbol{\mathbbm{R}^{1,2}}$:}\\
 We list here the conventions used in  the   Minkowski  space $\mathbbm{R}^{1,2}$ with metric $ds^2=-(dx^0)^2+(dx^1)^2+(dx^2)^2$.  $\mu,\nu=0,1,2$ to denote  space-time indices.  Choices of $\gamma$ matrices are ${(\gamma^\mu)}_\alpha^{\hspace{.1cm}\beta}=(i\sigma^2,  \sigma^1, \sigma^3)$ which satisfy  $\gamma^\mu \gamma^\nu = \eta^{\mu\nu} + \epsilon^{\mu\nu\rho} \gamma_\rho$. Notice that $(\gamma^\mu)_{\alpha\beta}=(-\mathbb{1}, -\sigma^3, \sigma^1)$ is symmetric.
 
The Killing spinor equation,
\be
\mathcal{D}_\mu\varepsilon=0 
\ee

 \smallskip

\par \textbf{On $\boldsymbol{\mathbbm{R}\times S^2}$:}\\
The metric on $\mathbbm{R}\times S^2$ is $ds^2=g_{mn}dx^m \,dx^n=d\tau^2+(d\theta^2+   \sin^2\theta \, d\varphi^2)$. $k,m,m=1,2,3$ are space time(Euclidean) indices.
Choices for gamma matrices, in the tangent frame${(\gamma^A)}_\alpha^{\hspace{.1cm}\beta}=(-\sigma^2,  \sigma^1, \sigma^3)$, which satisfy $\gamma^A \gamma^B = \delta^{AB} + i\epsilon^{ABC} \gamma^C$, where $A=1,2,3$ are flat indices in the tangent frame. The gamma matrices $\gamma^m= e^m_{\,\,\,A} \gamma^A$ satisfy the Clifford algebra $\gamma^m\gamma^n+\gamma^n\gamma^m=2g^{mn}$, where $e^m_{\,\, A}$ are the veirbeins, taking values $e^\theta_{\,\,\,1}=1, e^\varphi_{\,\,\,2}=\frac{1}{\sin\theta}$. The covariant derivative of a spinor $\psi$,
\be
\nabla_m\,\psi &=&(\partial_m+\omega_m)\psi,\quad where\,\,\textit{$\omega_m$ is the spin connection}\nn\\
\omega_m &=&\frac{1}{4}\omega_{mAB}\gamma^{AB},\quad \gamma^{AB}=\frac{1}{2}[\gamma^A,\gamma^B]
\ee
where, $\omega_{\varphi 21}=-\omega_{\varphi 12}=\cos\theta$. The covariant derivatives of a vector $\A^n$,
\be
\nabla_m \A^n=\partial_m \A^n+\Gamma^n_{mp} \A^p
\ee
where the non zero components of the  Christoffel connection are $\Gamma^\theta_{\varphi\varphi}=-\sin\theta\cos\theta, \Gamma^\varphi_{\theta \varphi}=  \frac{\cos\theta}{\sin\theta}$.
 \\
\section{Component action computation}
\label{appendix:B}
We use dimensionally reduced $\NN=2$ multiplet of four dimension which are again written in terms of $\NN=1$ superfields.
The component expansion of the $\superN=2$ superfields are given as follows. In the gauge multiplet we have a vector superfield,
\be
  \VV_\m = 2i \, \theta \bar{\theta} \, \sigma_\m(x)
    - 2 \, \theta\gamma^m\bar{\theta} \, {A_m}_\m(x)
    + \sqrt{2} i \, \theta^2 \, \bar{\theta} {\chi_\sigma^\dagger}_\m(x)
    - \sqrt{2} i \, \bar{\theta}^2 \, \theta {\chi_\sigma}_\m(x)
    + \theta^2 \, \bar{\theta}^2 \, D_\m(x)\nn
  \; 
\ee
and an adjoint chiral superfield,
\be
  \Phi_\m             & =& \phi_\m(x_L)               + \sqrt{2} \, \theta       {\chi_\phi}_\m(x_L)                 + \theta^2       \, {F_\phi}_\m(x_L) \nn  \\
  \bar{\Phi}_\m       & = &\phi_\m^\dagger(x_R)       - \sqrt{2} \, \bar{\theta} {\chi^\dagger_\phi}_\m(x_R)       - \bar{\theta}^2 \, {F^\dagger_\phi}_\m(x_R) \;\nonumber
\ee
In the hypermultiplet we have two bifundamental chiral superfields,
\begin{align}
\label{eqn:bifundamental}
  \ZZ_\m       & = Z_\m(x_L)         + \sqrt{2} \, \theta\zeta_\m(x_L)                + \theta^2       \, F_i(x_L) \;  &
  \bar{\ZZ}_\m & = Z_\m^\dagger(x_R) - \sqrt{2} \, \bar{\theta}\zeta_\m^\dagger(x_R)  - \bar{\theta}^2 \, F_\m^\dagger(x_R) \; \nonumber \\\nonumber
  \WW _\m      & = W_\m(x_L)         + \sqrt{2} \, \theta\omega_\m(x_L)               + \theta^2       \, G_\m(x_L) \;  &
  \bar{\WW}_\m & = W_\m^\dagger(x_R) - \sqrt{2} \, \bar{\theta}\omega_\m^\dagger(x_R) - \bar{\theta}^2 \, G_\m^\dagger(x_R) \;\nonumber
\end{align}
where, $x_L^m = x^m - i\theta^\alpha(\gamma^m)^{\beta}_{\alpha}\bar{\theta}_\beta$ and $x_R^m = x^m + i\theta^\alpha(\gamma^m)^{\beta}_{\alpha}\bar{\theta}_\beta$.

\subsection{\label{appendix:an_action}$\widehat{A}$-type quiver}
\begin{eqnarray}
\label{eqn:cs_an}
\Action_{\mathrm{CS}}=\int d^3x\,\,\sum_{\m=1}^n \Tr\big[-2\kappa_\m \sigma_\m D_\m+\kappa_\m\epsilon^{\mu\nu\lambda}({A_\mu}_\m\partial_\nu {A_\lambda}_\m+\frac{2i}{3}{A_\mu}_\m {A_\nu}_\m {A_\lambda}_\m)\nonumber\\+\frac{i{\kappa}_\m}{2}{\chi_\sigma}_\m{\chi^\dagger_\sigma}_\m+\frac{i\kappa_\m}{2}{\chi^\dagger_\sigma}_\m{\chi_\sigma}_\m\big]
\end{eqnarray}
where, $\kappa_\m=\frac{k_\m}{4\pi}$.
The Yang-Mills part,
\begin{eqnarray}
\label{eqn:ym_an}
  \Action_{\mathrm{YM}}=\int d^3x\,\,\Tr\sum_{\m=1}^n \Big(-\frac{1}{2g^2}{F_{\mu\nu}}_\m F^{\mu\nu}_\m + \frac{i}{g^2}{\chi_\sigma}_\m\slashed{\mathcal{D}}{\chi_\sigma^\g}_\m &-& \frac{1}{g^2}(\mathcal{D}_m\sigma_\m)(\mathcal{D}^m\sigma_\m) +\frac{1}{g^2}D_\m^2\nonumber\\ &+&\frac{i}{g^2}{\chi_\sigma}_\m[\sigma_\m,{\chi^{\dagger}_{\sigma}}_\m]\Big)
\end{eqnarray}
The component expression of $\Action_{\mathrm{adj}} $,
\begin{eqnarray}
\label{eqn:adj}
\Action_{\mathrm{adj}} &=& \int d^3x\,\, \Tr\sum_{\m=1}^n \Big(-\frac{1}{g^2}(\mathcal{D}_m \phi^\dagger_\m)(\mathcal{D}^m \phi_\m)+ \frac{i}{g^2}{\chi_\phi}_\m \slashed{\mathcal{D}}{\chi_\phi^\g}_\m -\frac{i}{g^2}\phi^\g_\m[{\chi_\sigma}_\m,{\chi_\phi}_\m]\nn\\
& +&\frac{i}{g^2}{\chi_\phi^\g}_\m[\sigma_\m,{\chi_\phi}_\m]+\frac{i}{g^2}{\chi_\phi^\g}_\m[{\chi_\sigma^\g}_\m,\phi_\m] +\frac{1}{g^2}[\sigma_\m, \phi_\m][\sigma_\m,\phi^\g_\m] -\frac{1}{g^2}\phi^\g_\m \phi_\m D_\m\nn\\ 
&+&\frac{1}{g^2} \phi^\g_\m  D_\m\phi_\m+\frac{1}{g^2}{F_\phi^\g}_\m {F_\phi}_\m
\end{eqnarray}
The matter action,
\be
\Action_{\mathrm{mat}}&=&\int d^3x \hspace{.1cm}  \sum_{\m=1}^n\Tr\Big[-\mathcal{D}_m Z_\m.\mathcal{D}^m Z^\g_\m+i\zeta^\g_\m\slashed{\mathcal{D}} \zeta_\m -Z^\g_\m Z_\m{D}_\p +Z^\g_\m D_\m Z_\m\nn\\[1mm]\hspace{17mm}
&-& iZ^\g_\m\big(\zeta_\m {\chi_\sigma}_\p-{\chi_\sigma}_\m \zeta_\m\big) - i\zeta^\g_\m\big(Z_\m{\chi^\g_\sigma}_\p -{\chi^\g_\sigma}_\m Z_\m \big)\nn\\[2mm]\hspace{17mm}
&-& i \zeta^\g_\m\big(\sigma_\m\zeta_\m-\zeta_\m{\sigma}_\p\big)- Z^\g_\m Z_\m{\sigma}_\p^2 +  2 Z^\g_\m\sigma_\m Z_\m{\sigma}_\p\nn\\[2mm]\hspace{17mm}
&-& Z^\g_\m \sigma^2_\m Z_\m+F_\m^\g F_\m -\mathcal{D}_m W_\m.\mathcal{D}^m W^\g_\m +i\omega_\m\slashed{\mathcal{D}} \omega^\g_\m -W^\g_\m W_\m D_\m\nn\\[2mm]\hspace{17mm} &+&W^\g_\m D_\p W_\m -iW^\g_\m \big(\omega_\m {\chi_\sigma}_\m-{\chi_\sigma}_\p \omega_\m \big) - i\omega^\g_\m \big(W_\m{\chi^\g_\sigma}_\m -{\chi^\g_\sigma}_\p W_\m \big)\nn\\[2mm]\hspace{17mm} &+& i \omega^\g_\m\big(\omega_\m\sigma_\m-\sigma_\p\omega_\m \big)- W^\g_\m W_\m\sigma^2_\m +  2 W^\g_\m{\sigma}_\p W_\m{\sigma}_\m\nn\\[2mm]\hspace{17mm}
&-& W^\g_\m {\sigma}^2_\p W_\m +G^\g_\m G_\m\Big] 
\ee
The component action for $\Action_{\mathrm{pot}}$,
\begin{eqnarray}
\label{eqn:pot}
\mathcal{S}_\mathrm{pot}&=&\int d^3x\;\Tr\Big[ \phi_\m Z_\m G_\m-\phi_\m\zeta_\m\omega_\m+\phi_\m F_\m W_\m-{\chi_\phi}_\m Z_\m\omega_\m\nn\\[2mm]\hspace{17mm}
&-&{\chi_\phi}_\m\zeta_\m W_\m + {F_\phi}_\m Z_\m W_\m
-{\phi}_\m W_\n F_\n+\phi_\m\omega_\n\zeta_\n
\nn\\[2mm]\hspace{17mm}
&-&\phi_\m G_\n Z_\n+{\chi_\phi}_\m \omega_\n Z_\n+{\chi_\phi}_\m W_\n\zeta_\n - {F_\phi}_\m W_\n Z_\n \nn \\[2mm]\hspace{17mm}
&+&\phi^\g_\m W^\g_\m F^\g_\m+ \phi^\g_\m\omega^\g_\m \zeta^\g_\m+ \phi^\g_\m G^\g_\m Z^\g_\m+{\chi^\g_\phi}_\m  W^\g_\m \zeta^\g_\m+{\chi^\g_\phi}_\m\omega^\g_\m Z^\g_\m\nn\\[2mm]\hspace{17mm}
&+&{F^\g_\phi}_\m W^\g_\m Z^\g_\m-{\phi}^\g_\m Z^\g_\n G^\g_\n 
-{\phi}^\g_\m\omega^\g_\n\zeta^\g_\n -{\phi}^\g_\m F^\g_\n W^\g_\n\nn\\[2mm]\hspace{17mm}
&-&{\chi_\phi}^\g_\m Z^\g_\n \omega^\g_\m
-{\chi_\phi}^\g_\m \zeta^\g_\n W^\g_\n
 -{F^\g_\phi}_\m Z^\g_\n W^\g_\n \Big]\nonumber\\[2mm]\hspace{17mm}
&+&\int d^3x\;\frac{\kappa_\m}{2}tr \Big[2\phi_\m {F_\phi}_\m-{\chi_\phi}_\m{\chi_\phi}_\m+2 \phi^\g_\m {F^\g_\phi}_\m +{\chi^\g_\phi}_\m {\chi^\g_\phi}_\m  \Big]
\end{eqnarray}

Auxiliary fields are eliminated by using their equation of motions,
\be
D_\m &=&\frac{g^2}{2}\big(2\kappa_\m\sigma_\m-\frac{1}{g^2}\comm{\phi_\m}{\phi^\dagger_\m}-Z_\m Z_\m^\dagger + W^\g_\m W_\m+Z^\g_\n Z_\n - W_\n W^\g_\n\big)\nn\\[2mm]\hspace{17mm}
{F_{\phi}}_\m &=&-g^2\big(W_\m^\g Z^\g_\m -Z^\g_\n W^\g_\n+\kappa_\m\phi^\g_\m\big)\nn\\[2mm]\hspace{17mm}
F_\m &=&-\big( \phi^\g_\m W^\g_\m-W^\g_\m\phi^\g_\p\big)\nn\\[2mm]\hspace{17mm}
G_\m&=&-\big( Z^\g_\m\phi^\g_\m -\phi^\g_\p Z^\g_\m\big)\nn\\[2mm]\hspace{17mm}
{F_{\phi}^\dagger}_\m&=&-g^2\big( Z_\m W_\m-W_\n Z_\n+\kappa_\m\phi_\m\big)\nn\\[2mm]\hspace{17mm}
F^\dagger_\m &=&-\big(  W_\m\phi_\m-\phi_\p W_\m\big)\nn\\[2mm]\hspace{17mm}
G^\dagger_\m &=&-\big( \phi_\m Z_\m -Z_\m\phi_\p\big)
\ee
After eliminating the auxiliary fields,
\begin{align}
&{(\mathcal{L}_\mathrm{aux})}&\nonumber\\&
=-\frac{g^2}{2}\Big( (Z_\m Z^\dagger_\m)^2+(W_\m W^\dagger_\m)^2+Z_\m Z^\dagger_\m W^\dagger_\m W_\m+Z^\dagger_\m Z_\m W_\m W^\dagger_\m) \Big)&\nonumber\\[1mm]\hspace{17mm}&
+\frac{g^2}{2}(Z_\m Z^\dagger_\m Z^\dagger_\n Z_\n-Z_\m Z^\dagger_\m W_\n W^\dagger_\n-W^\dagger_\m W_\m Z^\dagger_\n Z_\n &\nonumber\\[2mm]\hspace{17mm}&+W^\dagger_\m W_\m W_\n W^\dagger_\n)+g^2W^\dagger_\m Z^\g_\m W_\n Z_\m+g^2Z^\dagger_\n W^\dagger_\n Z_\m W_\m\nn\\[2mm]\hspace{17mm}
&-\frac{1}{4g^2}[\phi_\m,\phi^\g_\m]^2+\kappa_\m\sigma_\m[\phi_\m,\phi^\g_\m]-\kappa^2_\m g^2 \sigma^2_\m+\kappa_\m g^2 \sigma_\m(Z_\m Z^\g_\m- Z^\g_\n Z_\n &\nonumber\\[2mm]\hspace{17mm}& + W_\n W^\g_\n -W^\g_\m W_\m)+\kappa_\m g^2 \phi_\m (Z^\g_\n W^\g_\n- W^\g_\m Z^\g_\m)&\nonumber\\[2mm]\hspace{17mm}&+\kappa_\m g^2 \phi^\g_\m (W_\n Z_\n- Z_\m W_\m)-\kappa^2_\m g^2 \phi^\g_\m \phi_\m &\nonumber\\[2mm]\hspace{17mm}&-\frac{1}{2}(\phi_\m \phi^\g_\m+\phi^\g_\m\phi_\m)(Z_\m Z^\g_\m+W^\g_\m W_\m + Z^\g_\n Z_\n +W_\n W^\g_\n)
\end{align}
\subsection{$\widehat{D}$-type quiver}
The  Euler-Lagrange equations of motion of the auxiliary fields,
\be
D_\m &=&\frac{g^2}{2}\big(2\kappa_\m\sigma_\m-\frac{1}{g^2}\comm{\phi_\m}{\phi^\g_\m}-Z_\m Z_p^\g + W^\dagger_\m W_\m\big),\qquad \textit{for \m=1,2,3,4.}  \nn\\
D_\5&=&\frac{g^2}{2}\Big(2\kappa_\5\sigma_\5-\frac{1}{g^2}\comm{\phi_\5}{\phi^\g_\5}+\sum_{\m=1}^2 (Z_\m^\g Z_\m  -W_\m W^\g_\m )- Z_\5 Z^\g_\5+ W^\g_\5 W_\5\Big)\nn\\
D_{(n+1)}&=&\frac{g^2}{2}\Big(2\kappa_{(n+1)}\sigma_{(n+1)}-\frac{1}{g^2}\comm{\phi_{(n+1)}}{\phi^\g_{(n+1)}}+\sum_{\m=3,4,n} (Z_\m^\g Z_\m  -W_\m W^\g_\m )\Big)\nn\\
D_\m&=&\frac{g^2}{2}\Big(2\kappa_\m \sigma_\m -\frac{1}{g^2}\comm{\phi_\m}{\phi^\g_\m}+ Z_{\n}^\g Z_{\n}  -W_{\n} W^\g_{\n}-Z_{\m} Z^\g_{\m}  +W^\g_{\m} W_{\m} )\Big) \nn\\&&\hspace{12cm} \textit{for \m=6 to n.} \nn\\[2mm]
{F_{\phi}}_\m &=&-g^2\Big(W^\g_\m Z^\g_\m +\kappa_\m \phi^\g_\m \Big)\qquad (\textit{for \m=1,2,3,4.}  \nn\\[2mm]
{F_{\phi}}_\5&=&-g^2\big(W^\g_\5 Z^\g_\5 +\kappa_\5 \phi^\g_\5- Z^\g_\1 W^\g_\1- Z^\g_\2 W^\g_\2 \big)\nn\\
{F_{\phi}}_{(n+1)}&=&-g^2\Big(\kappa_{(n+1)} \phi^\g_{(n+1)}-\sum_{\m=3,4,n}Z^\g_\m W^\g_\m \Big)\nn\\[2mm]
{F_{\phi}}_\m &=&-g^2\big(\kappa_\m \phi^\g_\m + W^\g_\m Z^\g_\m -  Z^\g_{\n} W^\g_{\n} \big)\quad (\textit{for \m=6 to n})\nn\\[2mm]
\ee
\be
&&F_\m=-\big( \phi^\g_\m W^\g_\m- W^\g_\m \phi^\g_\5\big)\quad (\textit{for \m=1,2})\nn\\[2mm]
&&F_\m=-\big( \phi^\g_\m W^\g_\m- W^\g_\m \phi^\g_{(n+1)}\big)\quad (\textit{for \m=3,4})\nn\\[2mm]
&&F_\m=-\big( \phi^\g_\m W^\g_\m- W^\g_\m \phi^\g_{\p}\big)\quad (\textit{for \m=5 to n})\nn\\[2mm]
&&G_\m=-\big( Z^\g_\m\phi^\g_\m -\phi^\dagger_\5 Z^\g_\m\big)\quad (\textit{for \m=1,2})\nn\\[2mm]
&&G_\m=-\big( Z^\g_\m\phi^\g_\m -\phi^\dagger_{(n+1)} Z^\g_\m\big)\quad (\textit{for \m=3,4})\nn\\[2mm]
&&G_\m=-\big( Z^\g_\m\phi^\g_\m -\phi^\dagger_{\p} Z^\g_\m\big)\quad (\textit{for \m=5 to n})
\ee

We present the component action for only the  $\mathcal{L}_{\mathrm{pot}}$ and $\mathcal{L}_{\mathrm{mat}}$ rest of the part is straight forward from the previous section.

\begin{eqnarray}
\label{eqn:pot_dn}
\mathcal{L}_\mathrm{pot}&=&\sum_{\m=1}^n\Tr\Big[ \phi_\m Z_\m G_\m-{\phi_\m\zeta_\m\omega_\m}+ \phi_\m  F_\m W_\m  -{\chi_\phi}_\m Z_\m\omega_\m -{\chi_\phi}_\m\zeta_\m W_\m+ {F_\phi}_\m Z_\m W_\m \nn\\ &+& \phi^\g_\m W^\dagger_\m F^\dagger_\m+ \phi^\dagger_\m \omega^\dagger_\m \zeta^\dagger_\m+ \phi^\g_\m G^\g_\m Z^\dagger_\m+{\chi^\dagger_\phi}_\m  W^\dagger_\m \zeta^\dagger_\m
+{\chi^\dagger_\phi}_\m\omega^\dagger_\m Z^\dagger_\m+{F^\g_\phi}_\m W^\dagger_\m Z^\dagger_\m\Big]
\nonumber\\ 
&+&\sum_{\m=1}^2\Big[ - {\phi}_\5 W_\m F_\m+{\phi_\5 \omega_\m\zeta_\m} -{\phi}_\5 G_\m Z_\m+{\chi_\phi}_\5 \omega_\m Z_\m 
+ {\chi_\phi}_\5 W_\m\zeta_\m - {F_\phi}_\5 W_\m Z_\m\nn\\ &-&{\phi}^\dagger_\5 Z^\dagger_\m G^\g_\m -{{\phi}^\g_5\zeta^\dagger_\m \omega^\dagger_\m} -{\phi}^\dagger_\5 F^\dagger_\m W^\dagger_\m -{\chi_\phi}^\dagger_\5 Z^\dagger_\m \omega^\dagger_\m -{\chi_\phi}^\dagger_5 \zeta^\dagger_\m W^\dagger_\m -{F^\dagger_\phi}_\5 Z^\dagger_\m W^\dagger_\m\Big] \nonumber \\
&+&\sum_{\m=3}^4\Big[ - {\phi}_{(n+1)}W_\m F_\m+\phi_{(n+1)}\omega_\m\zeta_\m -{\phi}_{(n+1)} G_\m Z_\m+{\chi_\phi}_{(n+1)} \omega_\m Z_\m + {\chi_\phi}_{(n+1)} W_\m\zeta_\m\nn\\ 
&-&  {F_\phi}_{(n+1)} W_\m Z_\m -{\phi}^\dagger_{(n+1)} Z^\dagger_\m G^\dagger_\m -{\phi}^\dagger_{(n+1)}\omega^\dagger_\m\zeta^\dagger_\m -{\phi}^\dagger_{(n+1)} F^\dagger_\m W^\dagger_\m- {\chi_\phi}^\dagger_{(n+1)} Z^\dagger_\m \omega^\dagger_\m \nn\\&-& {\chi_\phi}^\dagger_{(n+1)} \zeta^\dagger_\m W^\dagger_\m -{F^\dagger_\phi}_{(n+1)} Z^\dagger_\m W^\dagger_\m\Big] + \sum_{\m=5}^n\Big[ - {\phi}_{\p}W_\m F_\m+\phi_{\p}\omega_\m\zeta_\m- {\phi}_{\p} G_\m Z_\m  \nonumber\\
&+&{\chi_\phi}_{\p} \omega_\m Z_\m+{\chi_\phi}_{\m+1} W_\m\zeta_\m - {F_\phi}_{\p} W_\m Z_\m 
- {\phi}^\dagger_{\p} Z^\dagger_\m G^\dagger_\m -{\phi}^\g_{\p}\omega^\dagger_\m\zeta^\dagger_\m\nonumber\\
& -&{\phi}^\dagger_{\p} F^\dagger_\m W^\dagger_\m -{\chi_\phi}^\dagger_{\p} Z^\dagger_\m \omega^\dagger_\m -{\chi_\phi}^\dagger_{\p} \zeta^\dagger_\m W^\dagger_\m -{F^\dagger_\phi}_{\p} Z^\dagger_\m W^\dagger_\m\Big] \nonumber \\
&+&\int d^3x\;\sum_{\m=1}^{n+1}\frac{\kappa_\m}{2}\Tr\Big[2{\phi_\m {F_\phi}_\m-{\chi_\phi}_\m{\chi_\phi}_\m+2 \phi^\dagger_\m {F^\dagger_\phi}_\m +{\chi^\dagger_\phi}_\m {\chi^\dagger_\phi}_\m} \Big]
\end{eqnarray}

\begin{eqnarray}
\label{eqn:matZ_dn}
\mathcal{L}_{\mathrm{mat}}&=&
\sum_{\m=1}^2 \Tr\Big[-\mathcal{D}_m Z_{\m}\mathcal{D}^m Z^\dagger_{\m}+i\zeta^\dagger_{\m}\slashed{\mathcal{D}} \zeta_{\m} -Z^\g_{\m} Z_{\m}D_\5 +Z^\g_{\m} D_{\m} Z_{\m} -iZ^\g_{\m}\big(\zeta_{\m} {\chi_\sigma}_\5\nn\\[2mm]\hspace{17mm}
&-&{\chi_\sigma}_{\m} \zeta_{\m}\big) -{ i\zeta^\g_{\m}\big(Z_{\m}{\chi^\g_\sigma}_\5} -{\chi^\g_\sigma}_{\m} Z_{\m}\big) - i \zeta^\g_{\m}\big(\sigma_{\m}\zeta_{\m}-\zeta_{\m}{\sigma}_\5\big)- Z^\g_{\m} Z_{\m}\sigma^2_{\5} \nn \\[2mm]\hspace{17mm}
&+&  2 Z^\g_{\m}\sigma_{\m} Z_{\m}{\sigma}_5 -Z^\g_{\m} \sigma^2_{\m} Z_{\m}+F_{\m}^\g F_{\m}\Big]
-(\mathcal{D}_m W_{\m})(\mathcal{D}^m W^\g_{\m})+i\omega_{\m}\slashed{\mathcal{D}} \omega^\g_{\m}\nn\\[2mm]\hspace{17mm}
&-&W^\g_{\m} W_{\m} D_{\m}
 +W^\g_{\m} {D}_\5 W_{\m} + i {\omega^\g_{\m}\big(\omega_{\m}\sigma_{\m}}-{\sigma}_\5\omega_{\m}\big)-iW^\g_{\m}\big(\omega_{\m} {\chi_\sigma}_{\m}-{\chi_\sigma}_{\5} \omega_{\m}\big) \nonumber\\ [2mm]\hspace{17mm}
 &-&   {i\omega^\g_{\m}(W_{\m}{\chi^\g_\sigma}_{\m}} -{\chi^\g_\sigma}_\5 W_{\m}\big)-W^\g_{\m} W_{\m}\sigma^2_{\m}
+   2 W^\g_{\m}{\sigma}_\5 W_{\m}{\sigma}_{\m} -W^\g_{\m} \sigma^2_\5 W_{\m} +G^\g_{\m} G_{\m}\Big]\nn\\[2mm]\hspace{17mm}
&+& \sum_{\m=3}^4 \Tr\Big[-\mathcal{D}_m Z_{\m}\mathcal{D}^m Z^\g_{\m}+i\zeta^\g_{\m}\slashed{\mathcal{D}} \zeta_{\m} -Z^\g_{\m} Z_{\m}D_{(n+1)}
+ Z^\g_{\m} D_{\m} Z_{\m}\nn\\
&-& iZ^\g_{\m}\big(\zeta_{\m} {\chi_\sigma}_{n+1}-{\chi_\sigma}_{\m} \zeta_{\m}\big)- i\zeta^\g_{\m}\big(Z_{\m}{\chi^\g_\sigma}_{(n+1)} -{\chi^\g_\sigma}_{\m} Z_{\m}\big) - {i \zeta^\g_{\m}\big(\sigma_{\m}\zeta_{\m}}-\zeta_{\m}{\sigma}_{(n+1)}\big)\nn \\ [2mm]\hspace{17mm}&-& Z^\g_{\m} Z_{\m}\sigma^2_{(n+1)} +  2 Z^\g_{\m}\sigma_{\m} Z_{\m}{\sigma}_{(n+1)} -Z^\g_{\m} \sigma^2_{\m} Z_{\m}+F_{\m}^\g F_{\m}\Big]-(\mathcal{D}_m W_{\m})(\mathcal{D}^m W^\g_{\m})\nn\\[2mm]\hspace{17mm}
&+&i\omega_{\m}\slashed{\mathcal{D}} \omega^\g_{\m} -W^\g_{\m} W_{\m} D_{\m}+ W^\g_{\m} {D}_{(n+1)} W_{\m} + i \omega^\g_{\m}\big(\omega_{\m}\sigma_{\m}-{\sigma}_{(n+1)}\omega_{\m}\big) \nonumber\\[1mm]\hspace{17mm} &-& iW^\g_{\m}\big(\omega_{\m} {\chi_\sigma}_{\m}-\chi_{\sigma(n+1)} \omega_{\m}\big) - i\omega^\g_{\m}\big(W_{\m}{\chi^\g_\sigma}_{\m} -\chi^\g_{\sigma(n+1)} W_{\m}\big)-W^\g_{\m} W_{\m}\sigma^2_{\m} \nn\\[1mm]\hspace{17mm}
&+&   2 W^\g_{\m}{\sigma}_{(n+1)} W_{\m}{\sigma}_{\m} -W^\g_{\m} {\sigma}^2_{(n+1)} W_{\m} +G^\g_{\m} G_{\m}\Big]+\sum_{\m=5}^n \Tr\Big[-\mathcal{D}_m Z_{\m}\mathcal{D}^m Z^\g_{\m}\nn\\
&+& i\zeta^\g_{\m}\slashed{\mathcal{D}} \zeta_{\m} -Z^\g_{\m} Z_{\m}D_{\m+1}+ Z^\g_{\m} D_{\m} Z_{\m} -iZ^\g_{\m}\big(\zeta_{\m} {\chi_\sigma}_{\p}-{\chi_\sigma}_{\m} \zeta_{\m}\big)\nn\\[1mm]\hspace{17mm} &-& i\zeta^\g_{\m}\big(Z_{\m}{\chi^\g_\sigma}_{\p} -{\chi^\g_\sigma}_{\m} Z_{\m}\big) - i \zeta^\g_{\m}\big(\sigma_{\m}\zeta_{\m}-\zeta_{\m}{\sigma}_{\p}\big)- Z^\g_{\m} Z_{\m}{\sigma^2}_{\p} \nn \\[1mm]\hspace{17mm} &+&  2 Z^\g_{\m}\sigma_{\m} Z_{\m}{\sigma}_{\p} -Z^\g_{\m} \sigma^2_{\m} Z_{\m}+F_{\m}^\g F_{\m}\Big]
- (\mathcal{D}_m W_{\m})(\mathcal{D}^m W^\g_{\m})+i\omega_{\m}\slashed{\mathcal{D}} \omega^\g_{\m} \nonumber\\ &-& -W^\g_{\m} W_{\m} D_{\m}+W^\g_{\m} {D}_{\p} W_{\m} +i \omega^\g_{\m}\big(\omega_{\m}\sigma_{\m}-{\sigma}_{\p}\omega_{\m}\big) \nonumber\\ &-&  iW^\g_{\m}\big(\omega_{\m} {\chi_\sigma}_{\m}-{\chi_\sigma}_{\p} \omega_{\m}\big) - i\omega^\g_{\m}\big(W_{\m}{\chi^\g_\sigma}_{\m} -{\chi^\g_\sigma}_{\p} W_{\m}\big)-W^\g_{\m} W_{\m}\sigma^2_{\m} \nn\\&+&  2 W^\g_{\m}{\sigma}_{\p} W_{\m}{\sigma}_{\m} -W^\g_{\m} {\sigma}^2_{\p} W_{\m} +G^\g_{\m} G_{\m}\Big]
\end{eqnarray}

\subsection{$\widehat{E}_6$ quiver}
The component action in this case is similar to $\widehat{D}_n$ case.
The auxiliary fields are eliminated using the following equations,
\be
&&D_\m=\frac{g^2}{2}\big(2\kappa_\m\sigma_\m-\frac{1}{g^2}\comm{\phi_\m}{\phi^\g_\m}-Z_\m Z_\m^\g + W^\dagger_\m W_\m\big)\qquad\qquad\qquad (\textit{for \m=1,3,5.}  )\nn\\
&&D_\7=\frac{g^2}{2}\Big(2\kappa_\7\sigma_\7-\frac{1}{g^2}\comm{\phi_\7}{\phi^\g_\7}+\sum_{\m=2,4,6} (Z_\m^\g Z_\m  -W_\m W^\g_\m )\Big)\nn\\
&&D_\m=\frac{g^2}{2}\big(2\kappa_\m \sigma_\m -\frac{1}{g^2}\comm{\phi_\m}{\phi^\g_\m}+Z_{\n}^\g Z_{\n}  -W_{\n} W^\g_{\n}-Z_{\m} Z^\g_{\m}  +W^\g_{\m} W_{\m} )\big) \nn\\&&\hspace{12cm} (\textit{for $\m=2,4,6$}  )\nn\\
&&{F_{\phi}}_\m=-g^2\big(W^\g_\m Z^\g_\m +\kappa_\m \phi^\g_\m \big)\qquad (\textit{for \m=1,3,5.}  )\nn\\[2mm]
&&{F_{\phi}}_\7=-g^2\Big(\kappa_\7 \phi^\g_\7-\sum_{\m=2,4,6} Z^\g_\m W^\g_\m \Big)\nn\\[2mm]
&&{F_{\phi}}_\m=-g^2\big(\kappa_\m \phi^\g_\m + W^\g_\m Z^\g_\m -  Z^\g_{\n} W^\g_{\n} \big),\qquad (\textit{for \m=2,4,6 })\nn\\[2mm]
&&F_\m=-\big( \phi^\g_\m W^\g_\m- W^\g_\m \phi^\g_\7\big)\qquad\hspace{.5cm} (\textit{for \m= 2, 4, 6})\nn\\[2mm]
&&F_\m=-\big( \phi^\g_\m W^\g_\m- W^\g_\m \phi^\g_{\p}\big),\qquad (\textit{for \m=1,3,5})\nn\\[2mm]
&&G_\m=-\big( Z^\g_\m\phi^\g_\m -\phi^\dagger_\7 Z^\g_\m\big)\qquad\hspace{.7cm}(\textit{for \m=2, 4, 6})\nn\\[2mm]
&&G_\m=-\big( Z^\g_\m\phi^\g_\m -\phi^\dagger_{\p} Z^\g_\m\big)\qquad\hspace{.5cm} (\textit{for \m=1,3,5})
\ee

\bigskip
\section{\label{app:susy_variation_check} Checking supersymmetry variation of $\widehat{D}_n$}

We present some steps of   checking the  supersymmetry variation of $\widehat{D}_n$ action. We vary all the fields in \eqref{eqn:min_action_int_dn} and \eqref{eqn:min_action_kin_dn}  simultaneously and substitute \eqref{eqn:min_var_vec_dn} and \eqref{eqn:min_var_hyper_dn}. The terms which combine and cancel are labelled by same alphabet. We do not write $\Tr$ in front  of every term, which implies the  trace of gauge indices.
\smallskip

\subsection{Variation of  the kinetic part}

We now explicitly show the term by term variation of the kinetic part of the action,
\ben
\label{1}
(1)\qquad -\frac{1}{g^2}\sum_{\m=1}^{n+1}\Tr(\delta F_{\mu\nu})_\m F^{\mu\nu}_\m=\sum_{\m=1}^{n+1}\underbrace{\frac{i}{g^2} \Ev_{ab} \gamma_\mu \lambda^{ab}_\m\mathcal{D}_\nu F^{\mu\nu}_\m}_B
\en

\ben
\label{2}
(2)\sum_{\m=1}^{n+1}\kappa_\m  \epsilon^{\mu\nu\lambda} \delta \big({A_\mu}_\m \partial_\nu {A_\lambda}_\m + \frac{2i}{3} {A_\mu}_\m {A_\nu}_\m {A_\lambda}_\m \big)
=\sum_{\m=1}^{n+1}\underbrace{-\frac{i\kappa_\m}{2}\epsilon^{\mu\nu\lambda}\Tr \Big(\varepsilon_{ab}\gamma_\mu\lambda^{ab}F_{\nu\lambda\m}}_C\Big)
\en

\ben
(3)\label{3}
&&\sum_{\m=1}^{n}- \delta\big(\mathcal{D}_\mu X^\g_\m.\, \mathcal{D}^\mu X_\m\big)\\
&=&\sum_{\m=1}^{n}\underbrace{(- i  \xi^{\g}_{b\m} \Ev^a_b).\,\mathcal{D}_\mu \mathcal{D}^\mu X_\m}_D  +\sum_{\m=1}^{n} \underbrace{ (\frac{1}{2}\Ev_{ab}\gamma_\mu \lambda^{ab}_\m).\,( \mathcal{D}^\mu X_\m .X^\g_\m - X_\m\mathcal{D}^\mu X^\g_\m)}_E 
\\&+&\sum_{\m=1}^{n}\underbrace{ (- i \Ev^a_b \xi^b_\m).\,\mathcal{D}_\mu\mathcal{D}^\mu X^\g_\m }_{D^*} +\sum_{\m=1}^{2} (\underbrace{\frac{1}{2}\Ev_{ab}\gamma_\mu \lambda^{ab}_\5).\,(\mathcal{D}^\mu X^\g_\m. X_\m- X^\g_\m \mathcal{D}^\mu X_\m)}_{E_5}\\
&+&\sum_{\m=3}^{4}(\underbrace{\frac{1}{2}\Ev_{ab}\gamma_\mu \lambda^{ab}_{(n+1)}).\,(\mathcal{D}^\mu X^\g_\m. X_\m- X^\g_\m \mathcal{D}^\mu X_\m)}_{E_{n+1}}
+\sum_{\m=5}^{n} \underbrace{(\frac{1}{2}\Ev_{ab}\gamma_\mu \lambda^{ab}_\p).\,(\mathcal{D}^\mu X^\g_\m. X_\m}_F\nn\\
&-& \underbrace{X^\g_\m \mathcal{D}^\mu X_\m)}_F
\en

\ben
\label{4}
(4)&&\sum_{\m=1}^{n}i \delta  (\xi^\g_\m\, \slashed{\mathcal{D}} \xi_\m)\\
&=&\sum_{\m=5}^{n}\Bigg(\underbrace{i(- \phi^b_{a\p} \Ev^c_b X^\g_{c\m}) \slashed{\mathcal{D}} \xi_\m
+i(- X^c_\m \Ev^b_c \phi^a_{b\p}). \slashed{\mathcal{D}}\xi^\g_\m}_G    + \underbrace{  \xi^\g_\m.\gamma^\mu \xi_\m (
-\frac{i}{2} \Ev_{ab} \gamma_\mu \lambda^{ab}_\p)}_I\Bigg)\\
&+&\sum_{\m=1}^{2}\Bigg(\underbrace{i (- \phi^b_{a\5} \Ev^c_b X^{\g}_{c\m}  ) \slashed{\mathcal{D}} \xi_\m
+i(- X^c_\m \Ev^b_c \phi^a_{b\5}). \slashed{\mathcal{D}}\xi^\g_\m }_{G_5}   +\underbrace{  \xi^\g_\m.\gamma^\mu \xi_\m (
-\frac{i}{2} \Ev_{ab} \gamma_\mu \lambda^{ab}_\5)}_I\Bigg)\\
&+&\sum_{\m=3}^{4}\Bigg(i\underbrace{(- \phi^b_{a(n+1)} \Ev^c_b X^\g_{c\m}) \slashed{\mathcal{D}} \xi_\m
+i(- X^c_\m \Ev^b_c \phi^a_{b(n+1)}  ). \slashed{\mathcal{D}}\xi^\g_\m   }_{G_{n+1}} +\underbrace{  \xi^\g_\m.\gamma^\mu \xi_\m (
-\frac{i}{2} \Ev_{ab} \gamma_\mu \lambda^{ab}_{(n+1)}}_I)\Bigg)\\
&+&\sum_{\m=1}^n\Bigg( \underbrace{-i \Ev^b_a\slashed{\mathcal{D}} X^\g_{b\m} \slashed{\mathcal{D}} \xi_\m}_{D^*}  -\underbrace{i \Ev^a_b\slashed{\mathcal{D}} X^b_{\m} \slashed{\mathcal{D}} \xi^\g_\m}_D \\
&-& \underbrace{ \xi^\g_\m.\gamma^\mu  (
-\frac{i}{2} \Ev_{ab} \gamma_\mu \lambda^{ab}_\m) .\xi_\m}_I+ \underbrace{iX^\dagger_{c\m} \Ev^c_b \phi^b_{a\m} \slashed{\mathcal{D}} \xi_\m +i \phi^a_{b\m} \Ev^b_c X^c_\m . \slashed{\mathcal{D}}\xi^\g_\m}_H\Bigg)\\
\en

\ben
\label{5}
(5)&&\sum_{\m=1}^{n+1}   - \frac{1}{2g^2}\delta \big(\mathcal{D}_\mu {\phi^a_b}_\m\big) \mathcal{D}^\mu {\phi^b_a}_\m  - \frac{1}{2g^2}\mathcal{D}_\mu {\phi^a_b}_\m \delta \big(\mathcal{D}^\mu {\phi^b_a}_\m\big)\\
&=&   \underbrace{\frac{1}{g^2} \Ev_{cb} \mathcal{D}_\mu\lambda^{ca}_\m \mathcal{D}^\mu {\phi^b_a}_\m}_J
-\frac{1}{2g^2} \cancelto{0} {\Ev_{cd} \mathcal{D}_\mu\lambda^{cd}_\m \mathcal{D}^\mu {\phi^a_a}_\m} -\underbrace{ \frac{1}{2g^2} \Ev_{cd} \gamma_\mu [\lambda^{cd}_\m ,{\phi^a_b}_\m] \mathcal{D}^\mu {\phi^b_a}_\m }_K
\en

\ben
\label{6}
(6)\sum_{\m=1}^{n+1}- \frac{1}{2} \kappa_\m^2 g^2 \delta \Big(\phi^a_{b\m} \phi^b_{a\m}\Big)
=\underbrace{\kappa_\m^2 g^2 \Ev_{cb} \lambda^{ca}_\m {\phi^b_a}_\m}_L
\en

\ben
\label{7}
(7)&&\sum_{\m=1}^{n+1} - \frac{i}{2g^2}\delta( \lambda^{ab}_\m \slashed{\mathcal{D}} \lambda_{ab\m})\\
  &=& - \frac{i}{g^2}\sum_{\m=1}^{n}\Big(\underbrace{ i g^2 \, X^a_\m {X^\g_c}_\m \Ev^{cb}  - \frac{ig^2}{2} (X X^\g)_\m \Ev^{ab} \Big) \slashed{\mathcal{D}} \lambda_{ab\m}}_E\\
&-&\underbrace{\frac{i}{g^2}\Big(- \Ev^{bc} g^2 \, i  \sum_{\m=1}^2{X^\g_c}_\m X^a_\m  + \frac{ig^2}{2}\sum_{\m=1}^2(X^\g X)_\m  \Ev^{ab}\Big) \slashed{\mathcal{D}} \lambda_{ab\5}}_{E_5}\\
&-&\underbrace{\frac{i}{g^2} \sum_{\m=6}^{n}\Big( - g^2 \, i \Ev^{bc}{X^\g_c}_{\m-1} X^a_{\m-1}  + \frac{ig^2}{2} (X^\g X)_{\m-1}  \Ev^{ab}\Big) \slashed{\mathcal{D}} \lambda_{ab\m}}_{E}\\ 
&-&\underbrace{\frac{i}{g^2}\Big( - ig^2 \,    \Ev^{bc}\sum_{\m=3,4,n}{X^\g_c}_\m X^a_\m + \frac{ig^2}{2} \sum_{\m=3,4,n}(X^\g X)_\m  \Ev^{ab} \Big)\slashed{\mathcal{D}} \lambda_{ab(n+1)} }_{E_{n+1}}
 \\
 &+&\sum_{\m=1}^{n+1}\Bigg(\underbrace{ \frac{1}{2g^2} \lambda^{ab}_\m\gamma^\mu[ -\frac{i}{2} \Ev_{ab} \gamma_\mu \lambda^{ab}_\m.\,\lambda_{ab\m}]}_M +\underbrace{\frac{i}{2g^2}  \epsilon^{\mu\nu\lambda} {F_{\mu\nu}}_\m \Ev^{ab}\gamma_\lambda \slashed{\mathcal{D}} \lambda_{ab\m}}_B + \underbrace{\frac{1}{g^2} \Ev^{ac} \slashed{\deriD} \phi^b_{c\m}\slashed{\mathcal{D}} \lambda_{ab\m}}_J\\
 &+&\underbrace{\frac{1}{2g^2} [\phi^b_{c\m},\phi^c_{d\m}] \Ev^{ad}\slashed{\mathcal{D}} \lambda_{ab\m}}_K+ \underbrace{\kappa_\m  \phi^b_{c\m} \Ev^{ac}\slashed{\mathcal{D}} \lambda_{ab\m}}_N\Bigg)   
 \en

\ben
\label{8}
 (8)&-&\sum_{\m=1}^{n+1} \frac{\kappa_\m}{2}  i\delta\Big( {\lambda^{ab}}_\m {\lambda_{ba}}_\m\Big) \\
&=&-i\sum_{\m=1}^{n}\kappa_\m \Big( \underbrace{+i g^2 \, X^a_\m {X^\dagger_c}_\m \Ev^{cb}  - \frac{ig^2}{2} (X X^\dagger)_\m \Ev^{ab}}_O \Big) {\lambda_{ba}}_\m\\
&-&i\kappa_\5 \Big(\underbrace{ - g^2 \, i \Ev^{bc}\sum_{\m=1}^2{X^\dagger_c}_\m X^a_\m  + \frac{ig^2}{2} \sum_{\m=1}^2(X^\dagger X)_\m  \Ev^{ab}}_O \Big) {\lambda_{ba}}_\5\\
  &-&i\sum_{\m=6}^{n}\kappa_\m \Big(\underbrace{ -i g^2 \,  \Ev^{bc}{X^\dagger_c}_{\m-1} X^a_{\m-1} + \frac{ig^2}{2}(X^\dagger X)_{\m-1} \Ev^{ab}}_O\Big) {\lambda_{ba}}_\m\\ 
  &-&i\kappa_{(n+1)} \Big(\underbrace{- ig^2 \,    \Ev^{bc}\sum_{\m=3,4,n}{X^\dagger_c}_\m X^a_\m + \frac{ig^2}{2} \sum_{\m=3,4,n}(X^\dagger X)_\m  \Ev^{ab}  }_O\Big){\lambda_{ba}}_{(n+1)}\\
  &-&i\sum_{\m=1}^{n+1}\kappa_\m \Big(\underbrace {\half \epsilon^{\mu\nu\lambda} {F_{\mu\nu}}_\m \gamma_\lambda \Ev^{ab}}_C-\underbrace{ i \slashed{\deriD} \phi^b_{c\m} \Ev^{ac}}_N+ \underbrace{\frac{i}{2} [\phi^b_{c\m},\phi^c_{d\m}] \Ev^{ad}}_P+\underbrace{ \kappa_\m g^2 \, i \phi^b_{c\m} \Ev^{ac}}_L\Big)\lambda_{ba\m}
\en

\subsection{Variation of the interaction part}
We now explicitly show the term by term variation of the interaction part of the action,
\ben
\label{9}
(9)&-& i\sum_{\m=1}^n \delta \big(\xi^\g_{a\m}\phi^a_{b\m} \xi^{b}_\m\big) \\
&=&- i\sum_{\m=1}^2 \delta \xi^\g_{a\m}.\phi^a_{b\m} \xi^{b}_\m - i\sum_{\m=3}^4 \delta \xi^\g_{a\m}.\phi^a_{b\m} \xi^{b}_\m - i\sum_{\m=5}^n \delta \xi^\g_{a\m}.\phi^a_{b\m} \xi^{b}_\m
- i\sum_{\m=1}^n \xi^\g_{a\m}.\delta \phi^a_{b\m}. \xi^{b}_\m\\
 &-& i\sum_{\m=1}^2 \xi^\g_{a\m}\phi^a_{b\m} \delta \xi^{b}_\m- i\sum_{\m=3}^4 \xi^\g_{a\m}\phi^a_{b\m} \delta \xi^{b}_\m- i\sum_{\m=5}^n \xi^\g_{a\m}\phi^a_{b\m} \delta \xi^{b}_\m
 \en
substituting transformation equations in each terms we get the following terms,\\
 \ben
 \label{9.1}
(9.1)&-& i\sum_{\m=1}^2\Big( \big( \underbrace{ -\Ev^b_a\slashed{\mathcal{D}} X^\g_{b\m}}_H  - \underbrace{\phi^b_{a\5} \Ev^c_b X^\g_{c\m}+ X^\g_{c\m} \Ev^c_b \phi^b_{a\m}}_Q\big).\phi^a_{d\m} \xi^{d}_\m \\
& +& \xi^\g_{a\m}\phi^a_{b\m} \big( \underbrace{\slashed{\mathcal{D}} X^{c}_\m \Ev^b_c}_H +\underbrace{ \phi^b_{d\m} \Ev^d_c X^c_\m - X^c_\m \Ev^d_c \phi^b_{d\5}\big)}_Q\Big)
\en 
 
\ben
 \label{9.2}
( 9.2)&-& i\sum_{\m=3}^4\Big( \big( \underbrace{-\Ev^b_a\slashed{\mathcal{D}} X^\g_{b\m}}_H  -\underbrace{ \phi^b_{a(n+1)} \Ev^c_b X^\g_{c\m}+ X^\g_{c\m} \Ev^c_b \phi^b_{a\m}}_Q \big).\phi^a_{m\m} \xi^{m}_\m\\
& +&\xi^\g_{m\m}\phi^m_{a\m} \big( \underbrace{ \slashed{\mathcal{D}} X^{b}_\m \Ev^a_b}_H + \underbrace{\phi^a_{b\m} \Ev^b_c X^c_\m - X^c_\m \Ev^b_c \phi^a_{b(n+1)}\big)}_Q\Big) 
\en
 
 \ben
 \label{9.3}
(9.3)& -& i\sum_{\m=5}^n \Big(\big( \underbrace{-\Ev^b_a \slashed{\mathcal{D}} X^\g_{b\m}}_H  - \underbrace{\phi^b_{a\p} \Ev^c_b X^\g_{c\m}+ X^\dagger_{c\m} \Ev^c_b \phi^b_{a\m}}_Q).\phi^a_{m\m} \xi^{m}_\m\\
& +& \xi^\g_{m\m}\phi^m_{a\m} \big( \underbrace{\slashed{\mathcal{D}} X^b_\m \Ev^a_b }_H+\underbrace{ \phi^a_{b\m} \Ev^b_c X^c_\m - X^c_\m \Ev^b_c \phi^a_{b\p}\big)}_Q\Big)
\en
\ben
\label{9.4}
(9.4)\underbrace{ i\sum_{\m=1}^n \xi^\g_{a\m}.(- \Ev_{cb} \lambda^{ca}_\m+ \half\, \delta^a_b\, \Ev_{cd}\, \lambda^{cd}_\m). \xi^{b}_\m}_I
\en

\ben
\label{10}
(10)&&\sum_{\m=1}^n\epsilon_{ac}\delta \big(\lambda^{cb}_\m\, X^a_\m\,\xi^\dagger_{b\m}\,  \big)\\
&=&\epsilon_{ac}\sum_{\m=1}^n \Big(\underbrace{-\half \epsilon^{\mu\nu\lambda} {F_{\mu\nu}}_\m \Ev^{cb}\gamma_\lambda}_D +  \underbrace{i \Ev^{cd} \slashed{\deriD} \phi^b_{d\m}}_H + \underbrace{\frac{i}{2} [\phi^b_{m\m},\phi^m_{d\m}] \Ev^{cd}}_Q+\underbrace{ \kappa_\m g^2 \, i \phi^b_{d\m} \Ev^{cd}}_R \\
&+& \underbrace{i g^2 \, X^c_\m {X^\g_d}_\m \Ev^{db}  - \frac{ig^2}{2} (X X^\g)_\m \Ev^{cb}}_S \Big).\, X^a_\m\,\xi^\dagger_{b\m}\\ 
&+&\epsilon_{ac}\Big(\underbrace{- g^2 \, i  \Ev^{bd}\sum_{\m=1}^2{X^\g_d}_\m X^c_\m \big) + \frac{ig^2}{2} \sum_{\m=1}^2(X^\g X)_\m  \Ev^{cb} }_S \Big).\, X^a_\5\,\xi^\dagger_{b\5}\\
&+&\epsilon_{ac}\sum_{\m=6}^n \Big( \underbrace{ - g^2 \, i \Ev^{bd}{X^\g_d}_{\n} X^c_{\n}  + \frac{ig^2}{2}(X^\g X)_{\n}  \Ev^{cb}}_S\Big).\, X^a_\m\,\xi^\dagger_{b\m}\\ 
 &+&\epsilon_{am}\sum_{\m=1}^2\underbrace{\lambda^{mb}_\m\, X^a_\m\,( - \phi^d_{b\5} \Ev^c_d X^\g_{c\m})
+\epsilon_{ac}\sum_{\m=3}^4\lambda^{cb}_\m\, X^a_\m\,(- \phi^m_{b(n+1)} \Ev^d_m X^\g_{d\m} )}_T\\
&+&\epsilon_{mn}\sum_{\m=5}^n\underbrace{\lambda^{na}_\m\, X^m_\m\,(  - \phi^b_{a\p} \Ev^c_b X^\g_{c\m})}_T
+\underbrace{\epsilon_{ac}\sum_{\m=1}^n \lambda^{cb}_\m.\,(- i \Ev^a_d \xi^d_\m).\,\xi^\dagger_{b\m}}_I\\
&+&\epsilon_{am}\sum_{\m=1}^n\lambda^{mb}_\m\, X^a_\m\,\Big( \underbrace{\slashed{\mathcal{D}} X^\g_{c\m} \Ev^c_b}_E+\underbrace{ X^\g_{c\m} \Ev^c_d \phi^d_{b\m}}_T \Big)
\en

\ben
\label{11}
 (11) &&\sum_{\m=1}^{n}\delta
  \Big[-\epsilon^{ac}{\lambda_{cb}}_\m\xi^b_\m{X^\g_a}_\m  \Big]\\
&=& -\sum_{\m=1}^{n}
  \epsilon^{ac}\epsilon_{cm}\epsilon_{bn} \Big(\underbrace{-\half \epsilon^{\mu\nu\lambda} {F_{\mu\nu}}_\m \Ev^{mn}\gamma_\lambda}_{D^*} + \underbrace{ i\Ev^{md} \slashed{\deriD} \phi^n_{d\m}}_H + \underbrace{\frac{i}{2} [\phi^n_{r\m},\phi^r_{d\m}] \Ev^{md}}_Q+\underbrace{ \kappa_\m g^2 \, i \phi^n_{d\m} \Ev^{md}}_R \\
  &+&i g^2\underbrace{ \, X^m_\m {X^\g_d}_\m \Ev^{dn}  - \frac{ig^2}{2} (X X^\g)_\m \Ev^{mn}}_S \Big).\xi^b_\m{X^\g_a}_\m\\
&-&\epsilon^{ac}\epsilon_{cm}\epsilon_{bn}\Big(\underbrace{- g^2 \, i  \Ev^{nd}\sum_{\m=1}^2{X^\g_d}_\m X^m_\m + \frac{ig^2}{2} \sum_{\m=1}^2(X^\g X)_\m \Ev^{ab}}_S \Big).\xi^b_\5{X^\g_a}_\5 \\
&-& \epsilon^{ac}\epsilon_{cm}\epsilon_{bn} \sum_{\m=6}^{n}\Big(\underbrace{ + g^2 \, i \big( - \Ev^{nr}{X^\g_r}_{\n} X^m_{\n} \big) + \frac{ig^2}{2} \big((X^\g X)_{\n} \big) \Ev^{mn}}_S\Big).\xi^b_\m{X^\g_a}_\m \\
&-&\sum_{\m=1}^{2}\epsilon^{ac}\underbrace{{\lambda_{cb}}_\m.(- X^m_\m \Ev^d_m \phi^b_{d\5}). X^\g_{a\m}  
   -\sum_{\m=3}^{4}\epsilon^{mc}\lambda_{ca\m}. (  - X^c_\m \Ev^b_c \phi^a_{b(n+1)}). X^\g_{m\m}}_T
\\
&-&\sum_{\m=5}^{n}\underbrace{\epsilon^{mn}{\lambda_{na}}_\m. (- X^c_\m \Ev^b_c \phi^a_{b\p}). X^\g_{m\m} }_T
   -\underbrace{\sum_{\m=1}^{n}\epsilon^{ac}{\lambda_{cb}}_\m\xi^b_\m. (- i \xi^\g_{d\m} \Ev^d_a)}_I\\
 &-&\sum_{\m=1}^{n}\epsilon^{ac}{\lambda_{cb}}_\m.\Big(\underbrace{\slashed{\mathcal{D}} X^{d}_\m \Ev^b_d}_E + \underbrace{\phi^b_{d\m} \Ev^d_m X^m_\m }_T\Big). X^\g_{a\m}   
 \en

\ben
\label{12}
(12)&& \sum_{\m=1}^n-\kappa_\m g^2 \,\delta( X^\dagger_{a\m} \phi^a_{b\m} X^{b}_\m)
 \\
   &=& \underbrace{- \sum_{\m=1}^n\kappa_\m g^2 \,(- i \xi^\g_{c\m} \Ev^c_a ). \phi^a_{b\m} X^{b}_\m}_R -  \sum_{\m=1}^n\underbrace{\kappa_\m g^2 \, X^\dagger_{a\m} (- \Ev_{pb} \lambda^{pa}_\m
+ \half \delta^a_b \Ev_{pq} \lambda^{pq}_\m). X^{b}_\m }_O\\ &&\underbrace{- \sum_{\m=1}^n\kappa_\m g^2 \, {X^\dagger_{a\m}} \phi^a_{b\m}. (- i \Ev^b_c \xi^{c}_\m)}_R
\en

\ben
\label{13}
 (13)&& \sum_{\m=1}^{n}\delta
  \Big[-\frac{1}{2} X_\m X^\g_\m{\phi^a_b}_\m {\phi^b_a}_\m \Big]\\
&=&\underbrace{\frac{i}{2} (\Ev^c_d \xi^{d}_\m {X^\dagger_c}_\m) {\phi^a_b}_\m {\phi^b_a}_\m+\frac{i}{2} (X_\m\xi^\dagger_{{d}_\m} \Ev^d_c ) {\phi^a_b}_\m {\phi^b_a}_\m}_Q\\
&-& \underbrace{\frac{1}{2} (X_\m X^\dagger_\m)( \Ev_{cb} \lambda^{ca}_\m {\phi^b_a}_\m}_T \,
+ \half\cancelto{0} {\delta^a_b \Ev_{cd} \lambda^{cd}_\m {\phi^b_a}_\m})\\
&-& \underbrace{\frac{1}{2} (X_\m X^\dagger_\m)( - \Ev_{ca} {\phi^a_b}_\m\lambda^{cb}_\m}_T\hspace{.1cm}
+ \half\cancelto{0} {\delta^b_a \Ev_{cd} {\phi^a_b}_\m\lambda^{cd}_\m} )\Bigg)
\en

\ben
\label{14}
(14)&&  \sum_{\m=1}^{n}  \delta\Big[
 -\frac{g^2}{4} (X_\m \sigma_i X^\g_\m) (X_\m \sigma_i X_\m^\g) \Big]\\
&=&\frac{ig^2}{2}\sum_{\m=1}^{n}\Bigg(\underbrace{(X_\m \sigma_i X^\dagger_\m) \Ev^a_b \xi^{b}_\m \sigma_i\ X^\dagger_\m+(X_\m \sigma_i X^\dagger_\m) X_\m \sigma_i \xi^\dagger_{{b}_\m} \Ev^b_a}_S\Bigg)
 \en

\ben
\label{15}
(15)&&\frac{g^2}{2}   \sum_{\m=1}^{n}\delta\Big[- (X^\g_\m \sigma_i X_\m)(X^\g_\m \sigma_i X_\m)    \Big]\\
&=&\frac{ig^2}{2}\sum_{\m=1}^{n} \Bigg(\underbrace{(X^\dagger_\m \sigma_i X_\m) \xi^\dagger_{{b}_\m} \Ev^b_a \sigma_i X_\m+ (X^\dagger_\m\sigma_i X_\m) X^\dagger_\m \sigma_i  \Ev^a_b \xi^{b}_\m}_S  \Bigg)
\en

  \ben
  \label{16}
(16)&&\sum_{\m=1}^2-\epsilon_{ac}\delta \big(\lambda^{cb}_\5\, \xi^\dagger_{b\m}\, X^a_\m \big)\\[4mm]
&=&-\epsilon_{ac}\Bigg( \underbrace{-\half \epsilon^{\mu\nu\lambda} F_{\mu\nu_\5}\Ev^{cb} \gamma_\lambda}_D +\underbrace{ i\Ev^{cd} \slashed{\deriD} \phi^b_{d\5}}_{G_5}\,\, +\underbrace{ \frac{i}{2} [\phi^b_{p\5},\phi^p_{d\5}] \Ev^{cd}}_Q+\underbrace{ \kappa_\5 g^2 \, i \phi^b_{d\5} \Ev^{cd}}_R \\
&+& g^2 \, \underbrace{i \big(X^c_\5 X^\g_{d\5} \Ev^{db} - \Ev^{bd}\sum_{\m=1}^2{X^\g_d}_\m X^c_\m \big) + \frac{ig^2}{2} \sum_{\m=1}^2\big((X^\dagger X)_\m -(X X^\g)_\5\big) \Ev^{cb}}_S\Bigg)\,.\sum_{\m=1}^2 \xi^\dagger_{b\m}\, X^a_\m \\
&-&\epsilon_{ac} \lambda^{cb}_\5\,\sum_{\m=1}^2\Bigg( \big( \underbrace{\slashed{\mathcal{D}} X^\g_{d\m} \Ev^d_b}_{E_5}\, -\underbrace{ \phi^d_{b\5} \Ev^p_d X^\g_{p\m}+ X^\dagger_{d\m} \Ev^d_p\phi^p_{b\m}\big)\,. X^a_\m}_T -\underbrace{ \xi^\dagger_{b\m}\,(- i \Ev^a_d \xi^d_\m)}_I \Bigg)
\en

\ben
\label{17}
(17) && \sum_{\m=1}^{2}\delta\Big[+ \epsilon^{ac}{\lambda_{cb}}_\5{X^\g_a}_\m\xi^b_\m  \Big]\\
&=&\epsilon^{ac}\epsilon_{cm}\epsilon_{bn} \Big(\underbrace{- \half \epsilon^{\mu\nu\lambda} F_{\mu\nu_\5}  \Ev^{mn}\gamma_\lambda}_{D^*}+ \underbrace{i\Ev^{md} \slashed{\deriD} \phi^n_{d\5}}_{G_5}\,\, +\underbrace{ \frac{i}{2} [\phi^n_{r\5},\phi^r_{d\5}] \Ev^{md}}_Q+\underbrace{ \kappa_\5 g^2 \, i \phi^n_{d\5} \Ev^{md}}_R \\
&+& g^2 \, i \big( \underbrace{X^m_\5 X^\g_{d\5} \Ev^{dn} - \Ev^{nd}\sum_{\m=1}^2{X^\g_d}_\m X^m_\m \big) + \frac{ig^2}{2} (\sum_{\m=1}^2(X^\g X)_\m -(X X^\g)_\5) \Ev^{mn}}_S  \Big).\sum_{\m=1}^{2}{X^\g_a}_\m\xi^b_\m \\
&+&\epsilon^{ac}{\lambda_{cb}}_\5 \sum_{\m=1}^{2}\Bigg(  \, (\underbrace{- i \xi^\g_{d\m} \Ev^d_a).\xi^b_\m}_I + {X^\g_a}_\m. \big(\underbrace{ \slashed{\mathcal{D}} X^{c}_\m \Ev^b_c}_{E_5} +\underbrace{ \phi^b_{d\m} \Ev^d_c X^c_\m - X^c_\m \Ev^d_c \phi^b_{d\5}}_T\big)  \Bigg)
\en

 \ben
 \label{18}
(18)&&\sum_{\m=1}^{2}\delta\Big[+i {\xi^a}_\m {\phi^b_a}_\5 {\xi^\g_b}_\m\Big]\\
  &=&  i\sum_{\m=1}^{2}\Bigg(  \big( \underbrace{-\Ev^a_c\slashed{\mathcal{D}} X^{c}_\m }_{G_5} +\underbrace{ \phi^a_{d\m} \Ev^d_c X^c_\m - X^c_\m \Ev^d_c \phi^a_{d\5}}_Q\big) {\phi^b_a}_\5 {\xi^\g_b}_\m \\
&+&\underbrace{\xi^a_\m.\, \big( - \Ev_{cb} \lambda^{ca}_\5
+ \half\, \delta^a_b\, \Ev_{cd}\, \lambda^{cd}_\5 \big). {\xi^\g_b}_\m}_I + \xi^d_\m {\phi^a_d}_\5 .\big(\underbrace{\slashed{\mathcal{D}} X^\g_{b\m} \Ev^b_a}_{G_5} - \underbrace{\phi^b_{a\5} \Ev^c_b X^\g_{c\m}+ X^\g_{c\m} \Ev^c_b \phi^b_{a\m}}_T \big)\Bigg)
\en

 \ben
 \label{19}
(19)&&   \sum_{\m=1}^{2}\delta \Big[-\frac{1}{2}(X^\g_\m X_\m){\phi^a_b}_\5 {\phi^b_a}_\5\Big]\\
  &=&-\frac{1}{2}\sum_{\m=1}^{2}\Bigg( \underbrace{ (- i \xi^\g_{d\m} \Ev^d_c) X^c_\m.{\phi^a_b}_\5 {\phi^b_a}_\5  -i  X^\g_\m . \Ev^a_d \xi^d_\m.{\phi^a_b}_\5 {\phi^b_a}_\5}_Q\\
  &-& \underbrace{X^\g_\m X_\m. \Ev_{cb} \lambda^{ca}_\5 {\phi^b_a}_\5 
+ X^\g_\m X_\m.{\phi^a_b}_\5 .\, \Ev_{cb} \lambda^{ca}_\5}_T\Bigg)
\en

\ben
\label{20}
(20)&&  \sum_{\m=1}^{2}\delta\Big[ {X^\g_\m}_a{\phi^b_c}_\m X^a_\m {\phi^c_b}_5 \Big]\\
&=&\sum_{\m=1}^{2} \Bigg(\underbrace{(- i \xi^\g_{d\m} \Ev^d_a)\,{\phi^b_c}_\m X^a_\m {\phi^c_b}_5}_Q+{X^\g_\m}_a.\, (\underbrace{- \Ev_{dc} \lambda^{db}_\m)X^a_\m {\phi^c_b}_\5}_T-\underbrace{i {X^\g_\m}_a{\phi^b_c}_\m  \Ev^a_d \xi^d_\m. {\phi^c_b}_5}_Q\\
&+&\underbrace{ {X^\g_\m}_a{\phi^b_c}_\m X^a_\m (- \Ev_{db} \lambda^{dc}_\5)}_T \Bigg)
\en

\ben
\label{21}
(21)&& \sum_{\m=1}^{2}\delta \Big[  \kappa_\m g^2 X^a_\m {(\phi^b_a)}_5 {X^\g_b}_\m \Big]\\
 &=&   \kappa_\m g^2\sum_{\m=1}^{2} \Bigg(\underbrace{ - i \Ev^a_b \xi^b_\m \phi^b_{a\5} {X^\g_b}_\m-  X^a_\m  \Ev_{ca} \lambda^{cb}_\5 {X^\g_b}_\m
+\half X^b_\m \, \Ev_{cd}\, \lambda^{cd}_\5 {X^\g_b}_\m+  X^a_\m \phi^b_{a\5}.\, (- i \xi^\g_{d\m} \Ev^d_b)}_R \Bigg)
\en

\ben
\label{22}
(22)&&  \sum_{\m=1}^{2}\delta \Big[   \frac{g^2}{2}(X_5 \sigma_i X^\g_5)(X^\g_{\m}\sigma_i X_{\m})\Big]\\
& =&  \frac{g^2}{2} \sum_{\m=1}^{2} \Bigg( \underbrace{\Big( (- i \Ev^a_b \xi^b_\5)\sigma_i X^\g_5+ X_5 \sigma_i.\, (- i \xi^\g_{b\5} \Ev^b_a)\Big).(X^\g_{\m}\sigma_i X_{\m}}_S)\\
&+&\underbrace{(  X_5 \sigma_i X^\g_5).\Big(  (- i \xi^\g_{b\m} \Ev^b_a) \,\sigma_i X_{\m}+X^\g_{\m}\sigma_i (- i \Ev^a_b \xi^b_\m)}_S \Big)\Bigg)
\en

\ben
\label{23}
(23)&&\sum_{\m=3}^4-\epsilon_{ac}\delta \big(\lambda^{cb}_{(n+1)}\, \xi^\dagger_{b\m}\, X^a_\m \big)\\
&=&\sum_{\m=3}^4\Bigg(-\epsilon_{ac} \Big(\underbrace{-\half \epsilon^{\mu\nu\lambda} F_{\mu\nu(n+1)}\Ev^{cb} \gamma_\lambda}_D + \underbrace{i\Ev^{cd} \slashed{\deriD} \phi^b_{d(n+1)}}_{G_{n+1}} + \underbrace{\frac{i}{2} [\phi^b_{p(n+1)},\phi^p_{d(n+1)}] \Ev^{cd}}_Q+ \underbrace{\kappa_{(n+1)} g^2 \, i \phi^b_{d(n+1)} \Ev^{cd}}_R \\
&-& g^2 \, i\underbrace{  \Ev^{bd}\sum_{\m=3,4,n}{X^\g_d}_\m X^c_\m + \frac{ig^2}{2} \sum_{\m=3,4,n}(X^\dagger X)_\m  \Ev^{cb}}_S\Big)\,.\, \xi^\dagger_{b\m}\, X^a_\m  \\
&-&\epsilon_{ac} \lambda^{cb}_{(n+1)}\,\big(\underbrace{  \slashed{\mathcal{D}} X^\g_{d\m} \Ev^d_b}_{E_{n+1}} -\underbrace{ \phi^d_{b(n+1)} \Ev^p_d X^\g_{p\m}}_{G_{n+1}}+\underbrace{ X^\dagger_{d\m} \Ev^d_p\phi^p_{b\m}}_T\big)\, X^a_\m-\underbrace{\epsilon_{ac}\lambda^{cb}_{(n+1)}\, \xi^\dagger_{b\m}\,(- i \Ev^a_d \xi^d_\m)}_I\Bigg)
\en

   \ben
   \label{24}
(24)&&\sum_{\m=3}^{4}\delta\Big[ \epsilon^{ac}{\lambda_{cb}}_{(n+1)}{X^\g_a}_\m\xi^b_\m  \Big]\\
&=&\sum_{\m=3}^{4}\Bigg( \epsilon_{bc} \Big(\underbrace{-\half \epsilon^{\mu\nu\lambda} F_{\mu\nu(n+1)}\Ev^{ac} \gamma_\lambda}_{D^*} + \underbrace{i\Ev^{ad} \slashed{\deriD} \phi^c_{d(n+1)}}_{G_{n+1}}\,\,\, +\underbrace{ \frac{i}{2} [\phi^c_{p(n+1)},\phi^p_{d(n+1)}] \Ev^{ad}}_Q+ \underbrace{\kappa_{(n+1)} g^2 \, i \phi^c_{d(n+1)} \Ev^{ad}}_R \\
&-& g^2 \, i \underbrace{ \Ev^{cd}\sum_{\m=3,4,n}{X^\g_d}_\m X^a_\m + \frac{ig^2}{2} \sum_{\m=3,4,n}(X^\dagger X)_\m  \Ev^{cd}}_S\Big).{X^\g_a}_\m\xi^b_\m +\underbrace{\epsilon^{ac}{\lambda_{cb}}_{(n+1)}.\, (- i \xi^\g_{d\m} \Ev^d_a)\,.\xi^b_\m}_I\\
&+&\epsilon^{ac}{\lambda_{cb}}_{(n+1)}.{X^\g_a}_\m. ( \underbrace{\slashed{\mathcal{D}} X^{m}_\m \Ev^b_m}_{E_{n+1}} + \underbrace{\phi^b_{m\m} \Ev^m_d X^d_\m - X^d_\m \Ev^m_d \phi^b_{m(n+1)})}_T \Bigg)
\en

\ben
\label{25}
(25)&+&\sum_{\m=3}^{4}\delta\Big[i \xi^a_\m {\phi^b_a}_{(n+1)} {\xi^\g_b}_\m \Big]\\
&=& i\sum_{\m=3}^{4}\Bigg( \big(\underbrace{-\Ev^a_b \slashed{\mathcal{D}} X^{b}_\m}_{G_{n+1}}  +\underbrace{ \phi^a_{b\m} \Ev^b_c X^c_\m - X^c_\m \Ev^b_c \phi^a_{b(n+1)}}_Q\big) {\phi^m_a}_{(n+1)} {\xi^\g_m}_\m
 + \underbrace{\xi^a_\m \big(- \Ev_{ca} \lambda^{cb}_{(n+1)}}_I\\
&+& \underbrace{\half\, \delta^b_a\, \Ev_{cd}\, \lambda^{cd}_{(n+1)}\big) {\xi^\g_b}_\m }_I
+ \xi^m_\m {\phi^a_m}_{(n+1)} \big(\underbrace{\slashed{\mathcal{D}} X^\g_{b\m} \Ev^b_a}_{G_{n+1}} -\underbrace{ \phi^b_{a(n+1)} \Ev^c_b X^\g_{c\m}+ X^\g_{c\m} \Ev^c_b \phi^b_{a\m}}_Q \big) \Bigg)
\en

\ben
\label{26}
(26)&&\sum_{\m=3}^{4}\delta \Big[
   -\frac{1}{2}(X^\g_\m X_\m){\phi^a_b}_{(n+1)} {\phi^b_a}_{(n+1)} \Big]\\
&=&-\frac{1}{2}\sum_{\m=3}^{4} \Bigg(
 \underbrace{ (- i \xi^\g_{d\m} \Ev^d_a)\,. X_\m{\phi^a_b}_{(n+1)} {\phi^b_a}_{(n+1)}+  X^\g_\m. (- i \Ev^a_d \xi^d_\m){\phi^a_b}_{(n+1)} {\phi^b_a}_{(n+1)}}_Q\\
  &+& \underbrace{ X^\g_\m. X_\m (- \Ev_{cb} \lambda^{ca}_{(n+1)}) {\phi^b_a}_{(n+1)}
+  X^\g_\m. X_\m{\phi^a_b}_{(n+1)} (- \Ev_{ca} \lambda^{cb}_{(n+1)})}_T \Bigg)
\en

\ben
\label{27}
(27)&&\sum_{\m=3}^{4}\delta \Big[
   + {X^\g_\m}_a {\phi^b_c}_\m X^a_\m{\phi^c_b}_{(n+1)}\Big]\\
&=&\sum_{\m=3}^{4} \Bigg(
 \underbrace{ (- i \xi^\g_{d\m} \Ev^d_a)\,\, {\phi^b_c}_\m X^a_\m{\phi^c_b}_{(n+1)}}_Q + \underbrace{{X^\g_\m}_a (- \Ev_{dc} \lambda^{db}_\m) X^a_\m{\phi^c_b}_{(n+1)}}_T\\
 &+&\underbrace{{X^\g_\m}_a {\phi^b_c}_\m (- i \Ev^a_d \xi^d_\m). {\phi^c_b}_{(n+1)}}_Q
+\underbrace{{X^\g_\m}_a {\phi^b_c}_\m X^a_\m (- \Ev_{db} \lambda^{dc}_{(n+1)})}_T\Bigg)
   \en

  \ben
  \label{28}
(28)&&  \sum_{\m=3}^{4} \delta\Big[
   \kappa_\m g^2 X^a_\m {(\phi^b_a)}_{n+1} {X^\g_b}_\m    \Big]\\
 &=& \kappa_\m g^2\sum_{\m=3}^{4} \Bigg(
( \underbrace{- i \Ev^a_d \xi^d_\m){\phi^b_a}_{(n+1)} {X^\g_b}_\m }_R   +  \underbrace{X^a_\m.\, (- \Ev_{ca} \lambda^{cb}_{(n+1)}
+ \half\, \delta^b_a\, \Ev_{cd}\, \lambda^{cd}_{(n+1)}) {X^\g_b}_\m}_O\\
&+& \underbrace{X^a_\m {(\phi^b_a)}_{(n+1)}  (- i \xi^\g_{d\m} \Ev^d_b}_R)\Bigg)
     \en

 \ben
 \label{29}
(29)&& \sum_{\m=3}^{4}\delta \Big[ - \frac{g^2}{2}(X^\g_\m \sigma_i X_\m)(X^\g_{(n)} \sigma_i X_{(n)}) \Big]\\
 &=&- \frac{g^2}{2} \sum_{\m=3}^{4} \Bigg(\big(  \underbrace{(- i \xi^\g_{b\m} \Ev^b_a)\, \sigma_i X_\m+X^\g_\m \sigma_i (- i \Ev^a_b \xi^b_\m)\big )(X^\g_{(n)} \sigma_i X_{(n)})}_S\\
&+&\underbrace{(X^\g_\m \sigma_i X_\m)\big(  (- i \xi^\g_{b(n)} \Ev^b_a)\, \sigma_i X_{(n)}+X^\g_{(n)} \sigma_i (- i \Ev^a_b \xi^b_{(n)})}_S \big) \Bigg)
   \en

\ben
\label{30}
(30)&+&\sum_{\m=5}^n-\epsilon_{ac}\delta \big(\lambda^{cb}_\p\, \xi^\dagger_{b\m}\, X^a_\m \big)\\
&=&-\epsilon_{ac}\sum_{\m=6}^{n}\Big(\underbrace{-\half \epsilon^{\mu\nu\lambda} {F_{\mu\nu}}_\m \Ev^{cb}\gamma_\lambda}_D + \underbrace{i \Ev^{cd}\slashed{\deriD} \phi^b_{d\m}}_G +\underbrace{ \frac{i}{2} [\phi^b_{m\m},\phi^m_{d\m}] \Ev^{cd}}_Q+ \underbrace{\kappa_\m g^2 \, i \phi^b_{m\m} \Ev^{cm}}_R \\
&+& \underbrace{g^2 \, i \big(X^c_\m {X^\g_m}_\m \Ev^{mb} - \Ev^{bd}{X^\g_d}_{\n} X^c_{\n} \big) + \frac{ig^2}{2} \big((X^\g X)_{\n} -(X X^\g)_\m\big) \Ev^{cb}}_S\Big)\, \xi^\dagger_{b\n}\, X^a_{\n}\\
&-&\epsilon_{ac}\Big(\underbrace{-\half \epsilon^{\mu\nu\lambda} {F_{\mu\nu}}_{(n+1)} \Ev^{cb}\gamma_\lambda}_D + \underbrace{i \Ev^{cd}\slashed{\deriD} \phi^b_{d{(n+1)}}}_G +\underbrace{ \frac{i}{2} [\phi^b_{m{(n+1)}},\phi^m_{d{(n+1)}}] \Ev^{cd}}_Q+ \underbrace{\kappa_{(n+1)} g^2 \, i \phi^b_{m\m} \Ev^{cm}}_R \\
&+& \underbrace{g^2 \, i \sum_{\m=3,4,n }\big( - \Ev^{bd}{X^\g_d}_{\m} X^c_{\m} \big) +\sum_{\m=3,4,n } \frac{ig^2}{2} \big((X^\g X)_{\n} \big) \Ev^{cb}}_S\Big) \xi^\dagger_{bn}\, X^a_{(n)}\\
&-&\epsilon_{ac}\sum_{\m=6}^{n+1} \Big(\underbrace{\lambda^{cb}_\m\, \big( \slashed{\mathcal{D}} X^\g_{d\n} \Ev^d_b}_F -\underbrace{ \phi^d_{b\m} \Ev^m_d X^\g_{m\n}}_T\\
&+& \underbrace{X^\g_{d\n} \Ev^d_m \phi^m_{b\n}\big) \, X^a_{\n}}_T+\underbrace{\lambda^{cb}_\m\, \xi^\dagger_{b\n}\, \big( - i \Ev^a_d\, \xi^d_\n\big)}_I \Big)
\en

\ben
\label{31}
(31) &&    \sum_{\m=5}^{n}\delta \Big[ \epsilon^{ac}{\lambda_{cb}}_{\p}{X^\g_a}_\m\xi^b_\m   \Big]\\
&=&\epsilon_{bn}\sum_{\m=6}^{n+1} \Big( \underbrace{-\half \epsilon^{\mu\nu\lambda} {F_{\mu\nu}}_\m \Ev^{an} \gamma_\lambda}_{D^*} + \underbrace{i \Ev^{ad}\slashed{\deriD} \phi^n_{d\m} }_G+\underbrace{ \frac{i}{2} [\phi^n_{r\m},\phi^r_{d\m}] \Ev^{ad}}_Q+ \underbrace{\kappa_\m g^2 \, i \phi^n_{d\m} \Ev^{ad}}_R\Bigg){X^\g_a}_{\n}\xi^b_{\n} \\
&+&\epsilon_{bn} \sum_{\m=6}^{n} \Bigg[g^2 \, i\underbrace{ \big(X^a_\m {X^\g_d}_\m \Ev^{dn} - \Ev^{nd}{X^\g_d}_{\n} X^a_{\n} \big) + \frac{ig^2}{2} \big((X^\g X)_{\n} -(X X^\g)_\m}_S\big) \Ev^{an} \Big)\Bigg] {X^\g_a}_{\n}\xi^b_{\n} \\
&+&\epsilon_{bn}\Big(\underbrace{ - ig^2 \,    \Ev^{nc}\sum_{\m=3,4,n}{X^\g_c}_\m X^a_\m + \frac{ig^2}{2} \sum_{\m=3,4,n}(X^\g X)_\m  \Ev^{an}}_S \Big){X^\g_a}_{(n)}\xi^b_{(n)}\\
&+&\underbrace{\epsilon^{ac}\sum_{\m=5}^{n} \Bigg( {\lambda_{cb}}_{\p}.\,
     (- i \xi^\g_{d\n} \Ev^d_a)\xi^b_\m}_I\Bigg)\\
&+&\epsilon^{mn}\sum_{\m=5}^{n} \Bigg(  {\lambda_{na}}_{\p}{X^\g_m}_\m. \big(\underbrace{\slashed{\mathcal{D}} X^b_\m \Ev^a_b }_F+\underbrace{ \phi^a_{b\m} \Ev^b_c X^c_\m - X^c_\m \Ev^b_c \phi^a_{b\p}}_T\big)  \Bigg)  
   \en

 \ben
 \label{32}
(32)   &&\sum_{\m=5}^{n}\delta\Big[i {\xi^a}_\m {\phi^b_a}_{\p} {\xi^\g_b}_\m \Big]\\
   &=& i\sum_{\m=5}^{n}\Bigg((\underbrace{-\Ev^a_b\slashed{\mathcal{D}} X^b_\m }_G +\underbrace{ \phi^a_{b\m} \Ev^b_c X^c_\m - X^c_\m \Ev^b_c \phi^a_{b\p}}_Q). {\phi^b_a}_{\p} \xi^\g_{b\m}\\
   &+&\xi^a_\m  \underbrace{(- \Ev_{ca} \lambda^{cb}_\p+ \half\, \delta^a_b\, \Ev_{cd}\, \lambda^{cd}_\p). \xi^\g_{b\m}}_I\\
   &+&\xi^m_\m {\phi^a_m}_{\p}. \big(\underbrace{ \slashed{\mathcal{D}} X^\g_{b\m} \Ev^b_a}_G -\underbrace{ \phi^b_{a\p} \Ev^c_b X^\g_{c\m}+ X^\dagger_{c\m} \Ev^c_b \phi^b_{a\m}}_Q\big) \Bigg)
\en

 \ben
 \label{33}
(33)&& \sum_{\m=5}^{n} \delta \Big[
   + g^2\kappa_{\p} {X^a_\m} {\phi^b_a}_{\p} {X^\g_b}_\m   \Big]\\
&=& g^2\kappa_{\p} \sum_{\m=5}^{n}  \Bigg(
    (\underbrace{- i \Ev^a_c \xi^c_\m). {\phi^b_a}_{\p} {X^\g_b}_\m}_R + \underbrace{{X^a_\m}   (- \Ev_{ca} \lambda^{cb}_\p
+ \half\, \delta^b_a\, \Ev_{cd}\, \lambda^{cd}_\p). {X^\g_b}_\m}_O\\
& +& \underbrace{{X^a_\m} {\phi^b_a}_{\p}. ( - i \xi^\g_{c\m} \Ev^c_b }_R)   \Bigg)
\en

\ben
\label{34}
(34)&& \sum_{\m=5}^{n}\delta \Big[
   -\frac{1}{2}(X^\g_\m X_\m){(\phi^a_b)}_{\p}{(\phi^b_a)}_{\p} \Big]\\
&=&-\frac{1}{2} \sum_{\m=5}^{n} \Bigg(
  \underbrace{ (  - i \xi^\g_{{b}_\m} \Ev^b_a) X_\m\phi^a_{b\p}  \phi^a_{b\p} 
+X^\g_\m (- i \Ev^a_d \xi^d_\m)\phi^a_{b\p}  \phi^a_{b\p}}_Q\\
&+&\underbrace{ (X^\g_\m X_\m) (- \Ev_{cb} \lambda^{ca}_\p) \phi^b_{a\p} 
+(X^\g_\m X_\m)\phi^a_{b\p} (- \Ev_{ca} \lambda^{cb}_\p)}_T \Bigg)
\en

\ben
\label{35}
(35)&&    \sum_{\m=5}^{n} \delta\Big[   X^\g_{\m a} \phi^b_{c\m} X^a_\m  \phi^c_{b\p}   \Big]\\
  &=&  \sum_{\m=5}^{n} \Bigg(  \underbrace{ ( - i \xi^\g_{{b}_\m} \Ev^b_a) \phi^b_{c\m} X^a_\m  \phi^c_{b\p}}_Q +\underbrace{ X^\g_{\m a}  (- \Ev_{dc} \lambda^{db}_\m) X^a_\m  \phi^c_{b\p}}_T \\
  &+& \underbrace{X^\g_{\m a} \phi^b_{c\m}  (- i \Ev^a_d \xi^d_\m)    \phi^c_{b\p}}_Q
    + \underbrace{X^\g_{\m a} \phi^b_{c\m} X^a_\m \big(- \Ev_{db} \lambda^{dc}_\p}_T\big)
\Bigg)
\en

\ben
\label{36}
(36)&& \sum_{\m=1}^{n+1} \delta \Big[\frac{1}{8g^2}[{\phi^a_b}_\m,{\phi^c_d}_\m][{\phi^b_a}_\m,{\phi^d_c}_\m]\Big]\\
 &=&\underbrace{-\frac{1}{4g^2} \Ev_{pb} [\lambda^{pa}_\m,{\phi^c_d}_\m] \comm{{\phi^b_a}_\m}{{\phi^d_c}_\m}
 -\frac{1}{4g^2} \Ev_{pd} [{\phi^a_b}_\m,\lambda^{pc}_\m]\comm{{\phi^b_a}_\m}{{\phi^d_c}_\m}}_U
\en

\ben 
\label{37}
(37) \sum_{\m=1}^{n+1}\delta\Big[ -\frac{\kappa_\m}{6}{\phi^a_b}_\m[{\phi^b_c}_\m{\phi^c_a}_\m]\Big]=\frac{\kappa_\m}{2}\Big(\underbrace{- \Ev_{cb} \lambda^{ca}_\m+ \half \delta^a_b \Ev_{cd} \lambda^{cd}_\m\Big)\comm{\phi^b_{m\m}}{\phi^m_{a\m}}}_P
\en

\ben
\label{38}
(38)&&\sum_{\m=1}^{n+1}\delta \Big[-\frac{i}{2g^2}{\lambda_{ab}}_\m[{\phi^b_c}_\m,{\lambda^{ac}}_\m]\Big]\\
&=& -\frac{i}{2g^2}\sum_{\m=1}^{n+1} \Bigg(\lambda_{ab\m}  [ \delta \phi^b_{c\m},\lambda^{ac}_\m]\Bigg) -\frac{i}{2g^2}\sum_{\m=1}^{4} \Bigg(\delta \lambda_{ab\m} [\phi^b_{c\m},\lambda^{ac}_\m]+\lambda_{ab\m} [\phi^b_{c\m}, \delta \lambda^{ac}_\m]\Bigg)\\
 &-&\frac{i}{2g^2} \Bigg(\delta \lambda_{ab\5} [\phi^b_{c\5},\lambda^{ac}_\5]+\lambda_{ab\5} [\phi^b_{c\m}, \delta \lambda^{ac}_\5]\Bigg)\\
 &-&\frac{i}{2g^2}\sum_{\m=6}^{n} \Bigg(\delta \lambda_{ab\m} [\phi^b_{c\m},\lambda^{ac}_\m]+\lambda_{ab\m} [\phi^b_{c\m}, \delta \lambda^{ac}_\m]\Bigg)\\
 &-&\frac{i}{2g^2}\Bigg(\delta \lambda_{ab (n+1)} [\phi^b_{c(n+1)},\lambda^{ac}_{(n+1)}]+\lambda_{ab(n+1)} [\phi^b_{c(n+1)}, \delta \lambda^{ac}_{(n+1)}]\Bigg)
\en
Terms from the above,\\

\ben
\label{38.1}
(38.1)\quad -\frac{i}{2g^2}\sum_{\m=1}^{n+1} \Bigg(\lambda_{ab\m}  [ \delta \phi^b_{c\m},\lambda^{ac}_\m]\Bigg)= \underbrace{ -\frac{i}{2g^2}\sum_{\m=1}^{n+1} \lambda_{ab\m} \Bigg[- \Ev_{dc} \lambda^{db}_\m
+ \half\, \delta^b_c\, \Ev_{mn}\, \lambda^{mn}_\m,\lambda^{ac}_\m\Bigg]}_M
\en

\ben
\label{38.2}
(38.2)\quad &-&\frac{i}{2g^2}\sum_{\m=1}^{4} \Bigg(\delta \lambda_{ab\m} [\phi^b_{c\m},\lambda^{ac}_\m]+\lambda_{ab\p} [\phi^b_{c\m}, \delta \lambda^{ac}_\m]\Bigg)\\
& =& -\frac{i}{2g^2}\epsilon_{am}\epsilon_{bn}\sum_{\m=1}^{4}  \Big(\underbrace{- \half \epsilon^{\mu\nu\lambda} {F_{\mu\nu}}_\m \Ev^{mn} \gamma_\lambda}_J+\underbrace{ i \Ev^{mr}\slashed{\deriD} \phi^n_{r\m}}_K +\underbrace{ \frac{i}{2} [\phi^n_{r\m},\phi^r_{d\m}] \Ev^{md}}_U+ \underbrace{\kappa_\m g^2 \, i \phi^n_{d\m} \Ev^{md} }_P     \\& +&
i g^2 \underbrace{\, X^m_\m {X^\g_d}_\m \Ev^{dn}  - \frac{ig^2}{2} (X X^\g)_\m \Ev^{mn}}_T \Big)[\phi^b_{c\m},\lambda^{ac}_\m]\\& -&\frac{i}{2g^2}\sum_{\m=1}^{4} \lambda_{ab\m} \Bigg[\phi^b_{c\m}, \underbrace{\half \epsilon^{\mu\nu\lambda} {F_{\mu\nu}}_\m \gamma_\lambda \Ev^{ac}}_J-\underbrace{ i \slashed{\deriD} \phi^c_{d\m} \Ev^{ad}}_K+ \underbrace{\frac{i}{2} [\phi^c_{r\m},\phi^r_{d\m}] \Ev^{ad}}_U+\underbrace{ \kappa_\m g^2 \, i \phi^c_{d\m} \Ev^{ad}}_P\\& +&i g^2 \,\underbrace{ X^a_\m {X^\g_d}_\m \Ev^{dc}  - \frac{ig^2}{2} (X X^\g)_\m \Ev^{ac}}_T \Bigg]  
\en

\ben
\label{38.3}
(38.3)\quad &-&\frac{i}{2g^2}\epsilon_{am}\epsilon_{bn} \Big(\underbrace{- \half \epsilon^{\mu\nu\lambda} F_{\mu\nu_\5}\Ev^{mn} \gamma_\lambda}_J +\underbrace{ i \Ev^{md} \slashed{\deriD} \phi^n_{d\5}}_K+\underbrace{ \frac{i}{2} [\phi^n_{r\5},\phi^r_{d\5}] \Ev^{md}}_U+ \underbrace{\kappa_\5 g^2 \, i \phi^n_{d\5} \Ev^{md}}_P \\
& +& g^2 \,\underbrace{ i \big(X^m_\5 X^\g_{d\5} \Ev^{dn} - \Ev^{nd}\sum_{\m=1}^2{X^\g_d}_\m X^m_\m \big) + \frac{ig^2}{2} (\sum_{\m=1}^2(X^\g X)_\m -(X X^\g)_\5) \Ev^{mn}}_T  \Big) [\phi^b_{c\5},\lambda^{ac}_\5]\\
& -&\frac{i}{2g^2} \lambda_{ab\5} \Bigg[\phi^b_{c\5},  \underbrace{\half \epsilon^{\mu\nu\lambda} {F_{\mu\nu}}_\5 \gamma_\lambda \Ev^{ac}}_J- \underbrace{i \slashed{\deriD} \phi^c_{d\5} \Ev^{ad}}_K+\underbrace{ \frac{i}{2} [\phi^c_{r\5},\phi^r_{d\5}] \Ev^{ad}}_U+\underbrace{ \kappa_\5 g^2 \, i \phi^c_{d\5} \Ev^{ad}}_P\\
& +& g^2 \, i \underbrace{\big(X^a_\5 X^\g_{d\5} \Ev^{dc} - \Ev^{cd}\sum_{\m=1}^2{X^\g_d}_\m X^a_\m \big) + \frac{ig^2}{2} (\sum_{\m=1}^2(X^\g X)_\m -(X X^\g)_\5) \Ev^{ac}}_T\Bigg]
\en

\ben
\label{38.4}
(38.4)\quad & -&\frac{i}{2g^2}\sum_{\m=6}^{n} \Bigg(\delta \lambda_{ab\m} [\phi^b_{c\m},\lambda^{ac}_\m]+\lambda_{ab\m} [\phi^b_{c\m}, \delta \lambda^{ac}_\m]\Bigg)\\
& =& -\frac{i}{2g^2}\epsilon_{am}\epsilon_{bn}\sum_{\m=6}^{n} \Big( \underbrace{-\half \epsilon^{\mu\nu\lambda} {F_{\mu\nu}}_\m \Ev^{mn}\gamma_\lambda}_J +\underbrace{ i \Ev^{mr}\slashed{\deriD} \phi^n_{r\m}}_K +\underbrace{ \frac{i}{2} [\phi^n_{r\m},\phi^r_{d\m}] \Ev^{md}}_U+\underbrace{ \kappa_\m g^2 \, i \phi^n_{d\m} \Ev^{md}}_P\\
& +& g^2 \, \underbrace{i \big(X^m_\m {X^\g_d}_\m \Ev^{dn} - \Ev^{nd}{X^\g_d}_{\n} X^m_{\n} \big) + \frac{ig^2}{2} \big((X^\g X)_{\n} -(X X^\g)_\m\big) \Ev^{mn}}_T \Big)   [\phi^b_{c\m},\lambda^{ac}_\m]\nn\\
& -&\frac{i}{2g^2}\sum_{\m=6}^{n}\lambda_{ab\m} \Bigg[\phi^b_{c\m}, \underbrace{\half \epsilon^{\mu\nu\lambda} {F_{\mu\nu}}_\m \gamma_\lambda \Ev^{ac}}_J-\underbrace{ i \slashed{\deriD} \phi^c_{d\m} \Ev^{ad}}_K+ \underbrace{\frac{i}{2} [\phi^c_{r\m},\phi^r_{d\m}] \Ev^{ad}}_U+\underbrace{ \kappa_\m g^2 \, i \phi^c_{d\m} \Ev^{ad}}_P \\
& +& g^2 \underbrace{\, i \big(X^a_\m {X^\g_d}_\m \Ev^{dc} - \Ev^{cd}{X^\g_d}_{\n} X^a_{\n} \big) + \frac{ig^2}{2} \big((X^\g X)_{\n} -(X X^\g)_\m}_T \big) \Ev^{ac}\Bigg]
\en

\ben
\label{38.5}
(38.5)& -&\frac{i}{2g^2}\epsilon_{am}\epsilon_{bn}  \Big(\underbrace{-\half \epsilon^{\mu\nu\lambda} {F_{\mu\nu}}_{(n+1)} \Ev^{mn}\gamma_\lambda}_J + \underbrace{i \Ev^{mc}\slashed{\deriD} \phi^n_{c (n+1)}}_K + \underbrace{\frac{i}{2} [\phi^n_{c(n+1)},\phi^c_{d(n+1)}] \Ev^{md}}_U+ \underbrace{\kappa_{(n+1)} g^2 \, i \phi^n_{c (n+1)} \Ev^{mc}} \\
& -& ig^2 \,   \underbrace{ \Ev^{nc}\sum_{\m=3,4,n}{X^\g_c}_\m X^m_\m + \frac{ig^2}{2} \sum_{\m=3,4,n}(X^\g X)_\m  \Ev^{mn}}_T\Big) [\phi^b_{r(n+1)},\lambda^{ar}_{(n+1)}]\\
& -&\frac{i}{2g^2}\lambda_{ab(n+1)} \Bigg[\phi^b_{c(n+1)}, \underbrace{\half \epsilon^{\mu\nu\lambda} {F_{\mu\nu}}_{(n+1)} \gamma_\lambda \Ev^{ac}}_J- \underbrace{i \slashed{\deriD} \phi^c_{d (n+1)} \Ev^{ad}}_K+\underbrace{ \frac{i}{2} [\phi^c_{m(n+1)},\phi^m_{d(n+1)}] \Ev^{ad}}_U\\
&+&\underbrace{ \kappa_{(n+1)} g^2 \, i \phi^c_{d (n+1)} \Ev^{ad}}_P
 - ig^2 \,    \underbrace{\Ev^{cd}\sum_{\m=3,4,n}{X^\g_d}_\m X^a_\m + \frac{ig^2}{2} \sum_{\m=3,4,n}(X^\g X)_\m  \Ev^{ac}}_T\Bigg]
\en

\ben
\label{39}
(39) && \frac{g^2}{2}\sum_{\m=6}^n \delta  \Big((X_\m \sigma_i X^\g_\m)(X^\g_{\n}\sigma_i X_{\m-1})\Big)\\
&=&\underbrace{- \frac{ig^2}{2} \Big( \Ev^a_b \xi^{b}_\m \sigma_i X^\g_\m (X^\dagger_{\n} \sigma_i X_{\n})\Big) - \frac{ig^2}{2} \Big( X_\m \sigma_i\xi^\dagger_{{b}_\m} \Ev^b_a (X^\dagger_{\n} \sigma_i X_{\n})}_S\Big)\\
&&\underbrace{- \frac{ig^2}{2}\Big(( X_\m \sigma_i X^\dagger_\m) \xi^\dagger_{{b}_{\n}} \Ev^b_a \sigma_i X_{\m-1}\Big) - \frac{ig^2}{2}\Big(( X_\m \sigma_i X^\dagger_\m) X^\dagger_{\n} \sigma_i \Ev^a_b \xi^{b}_{\n}\Big)}_S 
\en

\ben
\label{40}
(40)&& \frac{g^2}{2}\delta\Bigg(-(X^\g_1 \sigma_i X_1)(X^\g_2 \sigma_i X_2)\Bigg)\\
 &=& - \frac{g^2}{2}\Bigg( \underbrace{\big (  ( - i \xi^\g_{{b}_1} \Ev^b_a) \sigma_i X_1 + X^\g_1 \sigma_i (- i \Ev^a_b \xi^b_1) \big)(X^\g_2 \sigma_i X_2) + (X^\g_1 \sigma_i X_1) \big (  ( - i \xi^\g_{{b}_2} \Ev^b_a) \sigma_i X_2 + X^\g_2 \sigma_i  (- i \Ev^a_b \xi^b_2)  \big)}_S\Bigg)
\en

\ben
\label{41}
(41)&&\frac{g^2}{2}\delta\Bigg(-(X^\g_\3 \sigma_i X_\3)(X^\g_\4 \sigma_i X_\4)\Bigg)\\
&=& - \frac{g^2}{2}\Bigg((\delta X^\g_\3 \sigma_i X_\3 + X^\g_\3 \sigma_i \delta X_\3)(X^\g_\4 \sigma_i X_\4) + (X^\g_\3 \sigma_i X_\3)(\delta X^\g_\4 \sigma_i X_\4 + X^\g_\4 \sigma_i\delta X_\4)\Bigg)\\
  &=& - \frac{g^2}{2}\Bigg(\big( \underbrace{  ( - i \xi^\g_{{b}_\3} \Ev^b_a) \sigma_i X_\3 - i \Ev^a_b X^\g_\3 \sigma_i   \xi^b_\3 \big)(X^\g_\4 \sigma_i X_\4) + (X^\g_\3 \sigma_i X_\3)\big( (- i \xi^\g_{{b}_\4} \Ev^b_a) \sigma_i X_\4 + X^\g_\4 \sigma_i  (- i \Ev^a_b \xi^b_\4) }_S \big)\Bigg)
\en

\bigskip

\section{\label{app:d4_monopole_soln} Monopole solution of $\hat{D}_4$}
 In this section we will consider a simple example with gauge group $U(N)^4\times U(2N)$. For $\hat{D}_4$,  $5$-th and $n+1$-th nodes coincide. The constraint on CS levels,
\be
\label{constraint_cs level_d4}
\kappa_\1+ \kappa_\2+\kappa_\3+\kappa_\4+ 2\kappa_\5=0
\ee
Equation \eqref{phi_corners} and \eqref{phi_5} reduces to,
\be
 {\phi_i}_\m  &=&      -\frac{1}{2\kappa_\m}\Big(   X_\m \sigma_i {X^\dagger}_\m \Big), \quad for\,\, \m=1,...4\nn\\
\phi_{i\5}  &=& \frac{1}{2\kappa_\5}\Big(   \sum_{\m=1}^4{X^\dagger}_\m  \sigma_i X_\m\Big)
\ee

The above equations   when explicitly written in terms of fields  give the following set of equations,
\smallskip
\par \textbf{For $\boldsymbol{\m=1,...,4}$:}

\be
\label{eqn:phi_fork_d4}
{\phi_1}_\m=0\implies  Z_\m W_\m+ W^\g_\m Z^\g_\m &=& 0\nn\\
{\phi_2}_\m=0 \implies  Z_\m W_\m - W^\g_\m Z^\g_\m&=& 0\nn\\
{\phi_3}_\m=-\eta \frac{H^{(1)}}{2}\implies Z_\m Z^\g_p - W^\g_\m W_\m &=&\eta H^{(1)}\kappa_\m 
\ee
\smallskip
\par \textbf{For $\boldsymbol{\m=5}$:}
\be
\label{eqn:phi_line_d4}
\phi_{1\5} &=& 0 \implies    \sum_{\m=1}^4(Z^\g_\m W^\g_\m  +W_\m Z_\m ) =0\nn\\
\phi_{2\5}  &=&0 \implies    \sum_{\m=1}^4(Z^\g_\m W^\g_\m  -W_\m Z_\m ) =0\nn\\
{\phi_3}_\5 &=& \sum_{\m=1}^4\Big( Z^\g_\m Z_\m - W_\m W^\g_\m\Big) = -\eta H^{(2)}\kappa_\5
\ee

For a simple solution, there can be the following possibilities which solve the above  equations.\\
(i)$Z_\m=0, \quad for \,\, \m=1,2,...,4$\\
(ii)$W_\m=0, \quad for \,\, \m=1,2,..., 4$\\
(iii)$W_\1=Z_\2=W_\3= Z_{(4)}=0$ or $Z_\1=W_\2=Z_\3= W_\4=0$\\
 We need to solve the above three equations for the three cases written above. 
For  positive semi-definite $H^{(1)}, H^{(2)}$, we can solve the several constraints on the hyper multiplet scalars for (anti)BPS solutions in the following ways.
\smallskip
\par \textbf{ Case(i)\quad$\boldsymbol{Z_\m=0},\quad\forall\,\, \m$}
\smallskip
\par \underline{{BPS $\eta=+1$}}

\be
 - W^\g_\m W_\m   = H^{(1)}\kappa_\m,\quad \sum_{\m=1}^4  W_\m W^\g_\m   = H^{(2)}\kappa_\5\nn
 \ee
which can be solved by choosing,
\be
 W_\m = A_\m\, e^{- \tau/2} \ , \hspace{3mm}  W^\g_\m = A_\m^\g\, e^{\tau/2}\nn
\ee
 where $A_\m$ is an $2N\times N$ matrix such that, 
\be
  A^\g_\m A_\m= -H^{(1)}\kappa_\m,\qquad \sum_{\m=1}^4 A_\m A^\g_\m = H^{(2)}\kappa_\5
\ee
Therefore for a positive semi definite solution $\kappa_\m\leq 0$ for $\m=1,...4$ and $\kappa_\5\geq 0$ which is compatible with \eqref{constraint_cs level_d4}. \\
 \smallskip
\underline{{anti-BPS $\eta=-1$}:}
Similarly the anti-BPS case is solved by,
\be
 W_\m = A_\m\, e^{ \tau/2} \ , \hspace{3mm}  W^\g_\m =  A_\m^\g\, e^{-\tau/2}  
\ee
such that, $
  A^\g_\m A_\m= H^{(1)}\kappa_\m
,\,\,\sum_{\m=1}^4 A_\m A^\g_\m =- H^{(2)}\kappa_\5$. Therefore for a positive semi definite solution $\kappa_\m\geq 0 for\,\, \m=1,..., 4$ and $\kappa_\5\leq 0$, which is again compatible with \eqref{constraint_cs level_d4}. Solving case (ii) is straight forward now.\\
\smallskip
\par 
\textbf{ Case (iii)\quad {$\boldsymbol{W_\1=Z_\2=W_\3= Z_{(4)}= W_\5=0}$}}\\
\underline{BPS $\eta=+1$}

\be
Z_\1 Z^\g_\1  &=& H^{(1)}\kappa_\1,\quad
 W^\g_\2 W_\2 =- H^{(1)}\kappa_\2 ,\nn\\
 Z_\3 Z^\g_\3  &=& H^{(1)}\kappa_\3 ,\quad
W^\g_\4 W_\4 =- H^{(1)}\kappa_\4 \nn\\
\ee
\smallskip
\par \textbf{For $\boldsymbol{\m=5}$:}
\be
\Big( Z^\g_\1 Z_\1  - W_\2 W^\g_\2 +Z^\g_\3 Z_\3  - W_\4 W^\g_\4\Big) = -H^{(2)}\kappa_\5
\ee
Solving the above,
\be
 Z_\1 = A_\1\, e^{ -\tau/2} \ , \hspace{5mm} Z^\g_\1 = A_\1^\g\, e^{\tau/2} \nn  \\
 W_\2 = B_\2\, e^{ -\tau/2} \ , \hspace{3mm}  W^\g_\2 = B_\2^\g\, e^{\tau/2}  \nn \\
Z_\3 = C_\3\, e^{- \tau/2} \ , \hspace{5mm} Z^\g_\3 = B_\3^\g\, e^{\tau/2}\nn\\
 W_\4 = D_\4\, e^{ -\tau/2} \ , \hspace{3mm}  W^\g_\4 = D_\4^\g\, e^{\tau/2}
\ee
such that, $A_\1 A^\g_\1= H^{(1)}\kappa_\1, B^\g_\2 B_\2=- H^{(1)}\kappa_\2, C_\3 C^\g_\3= H^{(1)}\kappa_\3, D^\g_\2 D_\2= -H^{(1)}\kappa_\4$, $\Big( A^\g_\1 A_\1  - B_\2 B^\g_\2 +C^\g_\3 C_\3  - D_\4 D^\g_\4\Big) = -H^{(2)}\kappa_\5$. 
For a positive semi-definite solution, $\kappa_\1\geq 0, \kappa_\2\leq 0, \kappa_\3\geq 0, \kappa_\4\leq 0, \kappa_\5\leq 0$. The anti-BPS case can be similarly solved by reversing the signs of the CS levels.

\end{document}